\documentclass[12pt]{article}
 
\usepackage[superscript,sort]{cite}

\usepackage{ifthen,ifpdf}
\ifpdf
	\usepackage[pdftex]{hyperref}	\hypersetup{colorlinks=true,linkcolor=black,citecolor=black,urlcolor=blue,backref=page,breaklinks=true,plainpages=false }
	\usepackage[pdftex]{graphicx}
	\usepackage[usenames,pdftex]{color}
\else
	\usepackage[colorlinks=true,breaklinks=true,plainpages=false]{hyperref}
	\usepackage{graphicx}
	\usepackage[usenames]{color}
\fi

\usepackage{amsthm,amscd,amsxtra,amsfonts,amsmath,amssymb,multirow}
\usepackage{wrapfig}
\usepackage{algorithm,algorithmic,extarrows}

\begin{document}

\title{Persistent homology  analysis of protein structure, flexibility and folding
}

\author{
Kelin Xia$^{1,2}$
Guo-Wei Wei$^{1,2,3,4}$ \footnote{Address correspondences  to Guo-Wei Wei. E-mail:wei@math.msu.edu}\\
$^1$Department of Mathematics \\
Michigan State University, MI 48824, USA\\
$^2$Center for Mathematical Molecular Biosciences\\
Michigan State University, MI 48824, USA \\
$^3$Department of Electrical and Computer Engineering \\
Michigan State University, MI 48824, USA \\
$^4$Department of Biochemistry and Molecular Biology\\
Michigan State University, MI 48824, USA \\
}

\date{\today}
\maketitle

\begin{abstract}
Proteins are the most important biomolecules for living organisms. The understanding of protein structure, function, dynamics and transport is one of most challenging tasks in biological science. In the present work, persistent homology is, for the first time, introduced for extracting molecular topological fingerprints (MTFs) based  on the persistence of molecular topological invariants.   MTFs are utilized  for protein characterization, identification and classification. The method of slicing is proposed to track the geometric origin of protein topological invariants. Both all-atom and coarse-grained representations of MTFs are constructed. A new   cutoff-like filtration is proposed to  shed light on the optimal cutoff distance in elastic network models. Based on the correlation between  protein compactness,   rigidity and connectivity, we propose an accumulated bar length generated from persistent topological invariants for the quantitative  modeling of protein flexibility. To this end, a correlation matrix based filtration is developed.  This approach gives rise to an accurate prediction of the optimal characteristic distance used in protein B-factor analysis.  Finally, MTFs are employed to characterize protein topological evolution during protein folding and  quantitatively predict the protein folding stability.  An excellent consistence between our persistent homology prediction and molecular dynamics simulation is found.  {This work reveals the topology-function relationship of proteins.} 

\end{abstract}

\newpage

{\setcounter{tocdepth}{5} \tableofcontents}

\newpage

\section{Introduction}

Proteins are of paramount importance to living systems not only because of their role in providing the structural stiffness and rigidity to define the distinct shape of each living being, but also due to their functions in catalyzing cellular chemical reactions,   immune systems, signaling and signal transduction. It is commonly believed that protein functions are determined by protein structures, including primary amino acid sequences, secondary  alpha helices and beta sheets, and the associated tertiary structures. 
Rigidity and flexibility are part of protein functions. Since  65-90\% of human cell mass is water,   structural proteins, such as keratin, elastin and  collagen,    provide stiffness and rigidity to prevent biological material from flowing around. The prediction of protein flexibility and rigidity is not only crucial to structural proteins, but also important to membrane and globular proteins due to the correlation of flexibility to many other protein functions. Protein functions are well known to correlate with protein folding, a process in which random coiled polypeptides assume their three-dimensional structures. Although Anfinsen's dogma \cite{Anfinsen:1973} has been challenged due to  the existence of prions and amyloids, most functional proteins are well folded. The folding funnel hypothesis associates each folded protein structure with the global minimum of the Gibbs free energy. Unfolded conformations have higher energies and are thermodynamically unstable, albeit they can be kinetically favored.

The understanding  of the structure-function relationship of proteins is a central issue in experimental biology and is regarded by many to be the holy grail of computational biophysics. Numerous attempts have been made to unveil the structure-function relationship in proteins. One  approach  to  this  problem  is  to  design experiments from the evolutionary point of view to understand how evolutionary processes have led to various protein functions that strengthen the sustainability of live beings.  The past decade  has witnessed a rapid growth in gene sequencing. Vast sequence databases are readily available for  entire genomes of many bacteria, archaea and eukaryotes. New  genomes are updated on a daily basis. The Protein Data Bank (PDB) has also accumulated  near one hundred thousand tertiary structures. The availability of these structural data enables the comparative  study of evolutionary processes, which has a potential to decrypt the structure-function relationship. Another approach is to utilize  abundant protein  sequence  and  structural information  at  hand to set up theoretical models for relationships between protein structures and functions. An ultimate goal is to predict protein functions from known protein structures, which  is one of the most challenging tasks  in biological sciences. 

Theoretical study of the structure-functions relationship of proteins is usually based on fundamental laws of physics, i.e., quantum mechanics (QM), molecular mechanism (MM), statistical mechanics, thermodynamics, etc. QM methods are indispensable for chemical reactions and protein degradations. MM approaches are able to elucidate the conformational landscapes and flexibility patterns of proteins \cite{McCammon:1977}. However, the all-electron or all-atom representations and long-time integrations lead to such an excessively large number of degrees of freedom that their application to real-time scale protein dynamics becomes prohibitively expensive. One way to reduce the number of degrees of freedom is to pursue time-independent formulations, such as normal mode analysis (NMA)  \cite{Go:1983,Tasumi:1982,Brooks:1983,Levitt:1985},  elastic network model (ENM) \cite{Tirion:1996}, including Gaussian network model (GNM)   \cite{Flory:1976, Bahar:1997,Bahar:1998} and  anisotropic network model (ANM) \cite{Atilgan:2001}.  Multiscale methods are some of the most popular approaches for studying protein  structure, function, dynamics and transport \cite{Cui:2002, YZhang:2009a,WFTian:2014,Geng:2011}.  Recently, we have introduced differential geometry based multiscale models for biomolecular structure, solvation, and transport \cite{Wei:2009,Wei:2012,DuanChen:2012a,Wei:2013}. A new interaction free approach, called flexibility-rigidity index (FRI), has also been proposed for the estimation of the shear modulus in the theory of continuum elasticity with atomic rigidity (CEWAR) for biomolecules \cite{KLXia:2013d}.

A common feature of the above mentioned models for the study of structure-functions relationship of proteins  is that they are structure or geometry based approaches \cite{XFeng:2012a,QZheng:2012}. Mathematically, these approaches  make use of local  geometric information,  i.e., coordinates, distances, angles, surfaces \cite{Bates:2008,Bates:2009,QZheng:2012}  and sometimes curvatures \cite{ZhanChen:2010a,ZhanChen:2010b,ZhanChen:2012} for the physical modeling of biomolecular systems.  {Indeed, the importance of geometric modeling for structural biology \cite{XFeng:2012a}, biophysics \cite{XFeng:2013b,KLXia:2014a} and bioengineering \cite{Boileau:2013,Sazonov:2012,Sazonov:2012b,Sohn:2012,Ramalho:2012,Manzoni:2012,Mikhal:2013} cannot be overemphasized. However, geometry based models are often  inundated with too much  structural detail and computationally extremely expensive.} In many biological problems, such as the  open or close of ion channels, the association or disassociation of ligands, and the assembly or disassembly of proteins,  there exists an obvious topology-function relationship. In fact, just qualitative topological information, rather than quantitative geometric information is needed to understand many physical and biological functions. To state it differently, there is a {\it topology-function relationship} in many biomolecular systems. Topology is exactly the branch of mathematics that deals with the connectivity of different components in a space and is able to classify independent entities,  rings and higher dimensional faces within the space. Topology captures geometric properties that are independent of metrics or coordinates. Topological methodologies, such as homology and persistent homology, offer new strategies for analyzing biological functions from biomolecular data, particularly the point clouds of atoms in macromolecules.  

In the past decade, persistent homology has been developed as a new multiscale representation of topological features \cite{Edelsbrunner:2002,Zomorodian:2005,Zomorodian:2008}.  In general,  persistent homology characterizes the geometric features with persistent topological invariants by defining  a scale parameter relevant  to topological events. Through  filtration and persistence, persistent homology can capture topological structures continuously over a range of spatial  scales. Unlike commonly used computational homology which results in  truly metric free or coordinate free representations, persistent homology is able to embed geometric information to topological invariants so that ``birth"  and ``death" of  isolated components, circles, rings, loops, pockets, voids and  cavities at all geometric scales  can be monitored by topological measurements.  The basic concept was introduced by Frosini and Landi~\cite{Frosini:1999}, and in a general form by Robins~\cite{Robins:1999}, Edelsbrunner et al.~\cite{Edelsbrunner:2002}, and Zomorodian and Carlsson \cite{Zomorodian:2005}, independently. Efficient computational algorithms  have been proposed to track topological variations during the  filtration process \cite{Bubenik:2007, edelsbrunner:2010,Dey:2008,Dey:2013,Mischaikow:2013}. Usually, the persistent diagram is visualized through barcodes \cite{Ghrist:2008}, in which various horizontal line segments or bars are the homology generators lasted over filtration scales.  { It has been applied to a variety of domains, including image analysis \cite{Carlsson:2008,Pachauri:2011,Singh:2008,Bendich:2010}, image retrieval \cite{Frosini:2013}, chaotic dynamics verification \cite{Mischaikow:1999,Kaczynski:2004}, sensor network \cite{Silva:2005}, complex network \cite{LeeH:2012,Horak:2009}, data analysis \cite{Carlsson:2009,Niyogi:2011,BeiWang:2011,Rieck:2012,XuLiu:2012}, computer vision \cite{Singh:2008}, shape recognition \cite{DiFabio:2011} and computational biology \cite{Kasson:2007,Gameiro:2013,Dabaghian:2012}.} Compared with computational topology \cite{YaoY:2009,ChangHW:2013}  and/or computational homology, persistent homology  inherently has an additional dimension, the filtration parameter, which can be utilized to embed some crucial geometric or quantitative information into the topological invariants. The importance of retaining  geometric information in topological analysis has been recognized in a survey \cite{Biasotti:2008}.  However, most successful applications of persistent homology have been reported for qualitative characterization or classification. To our best knowledge,  persistent homology has hardly been employed for quantitative analysis,  mathematical modeling, and physical prediction. In general, topological tools often incur too much reduction of the original geometric/data information, while geometric tools frequently get lost in the geometric  detail  or are computationally too expensive to be practical in many situations. Persistent homology is able to bridge between geometry and topology. Given the big data challenge in biological science,  persistent homology ought to be more efficient for many biological problems. 

The objective of the present work is to  explore the utility of persistent homology for protein structure characterization, protein flexibility quantification and protein folding stability prediction. We introduce the molecular topological fingerprint (MTF) as a unique topological feature for protein characterization, identification and classification, and for the  understanding of the topology-function relationship of biomolecules. We  also  introduce all-atom and coarse-grained representations of protein topological fingerprints so as to utilize them for appropriate modeling. To analyze the topological fingerprints of alpha helices and beta sheets in detail, we propose  the method of slicing, which allows a clear tracking of geometric origins contributing to topological invariants.  Additionally, to understand the optimal cutoff distance in the GNM, we introduce a new distance based filtration matrix to recreate the cutoff effect in persistent homology. Our findings shed light on the topological interpretation of the optimal cutoff distance in GNM.  Moreover, based on the protein topological fingerprints, we propose accumulated bar lengths to  characterize protein topological evolution and quantitatively model protein rigidity based on protein topological connectivity. This approach gives rise to an accurate prediction of optimal characteristic distance used in the FRI method for protein flexibility analysis. Finally the proposed  accumulated bar lengths are also employed to predict the total energies of a series of protein folding  configurations generated by steered molecular dynamics.

The rest of this paper is organized as follows. Section \ref{sec:Theory} is devoted to fundamental concepts and algorithms for persistent homology, including simplicial complex, homology, persistence, $\check{\rm C}$ech  complex, Rips complex, filtration process, reduction algorithm, Euler characteristic, etc. To offer a pedagogic description, we discuss and illustrate the definition, generation and calculation of simplicial homology in detail. The persistent homology analysis of protein structure, flexibility and folding is developed in  Section \ref{sec:PHProtein}. Extensive examples, including alpha helices, beta sheets and beta barrel, are used to demonstrate the generation and analysis of protein topological fingerprints. Additionally, we utilize MTFs to explore the topology-function relationship of proteins.  Protein flexibility and rigidity is quantitative modeled by MTFs. We further explore protein topological evolution by analyzing the trajectory of protein topological invariants during the protein unfolding.  The quantitative prediction of protein folding stability is carried out over a series of protein configurations. This paper ends with some concluding remarks. 


\section{Theory and algorithm }\label{sec:Theory}

In general, homology utilizes a topological space with an algebraic group representation  to characterize topological features, such as isolated components, circles, holes and  void. For a given topological space $\mathbb{T}$, a $p$-dimensional hole in $\mathbb{T}$  induces the corresponding homology group $H_p(\mathbb{T})$. For a point set of data, such as atoms in a protein, one wishes to extracted the original topological invariants in its continuous description. Persistent homology plays an important  role in resolving this problem. By associating each point with an ever-increasing radius, a multi-scale representation can be systematically generated.  The corresponding series of homology groups is capable of characterizing the intrinsic topology in the point set.  Additionally, efficient computational algorithms have been proposed. The resulting persistent diagrams provide detailed information of the birth and death of topological features, namely, different dimensional  circles or holes.  In order to facilitate the detailed analysis of biomolecular systems, we briefly review  basic concepts and algorithms relevant to persistent homology, including simplicial complex, $\check{\rm C}$ech complex, Rips complex, filtration process, reduction algorithm, paring algorithm, etc. in this section. We illustrate several aspects including the definition, the generation and the computation of the simplicial homology with many simple examples.

\subsection{Simplicial homology and persistent homology}\label{sec:SimplicialHomology}

Simplicial complex is a topological space consisting of vertices (points), edges (line segments), triangles, and their high dimensional counterparts. Based on simplicial complex, simplicial homology can be defined and further used to analyze topological invariants.

\paragraph{Simplicial complex}
The essential component of simplicial complex $K$ is a $k$-simplex, $\sigma^k$, which can be defined as the convex hall of $k+1$ affine independent points in $\mathbb{R}^N$ ($N>k$). If we let $v_0,v_1,v_2,\cdots,v_k$ be $k+1$ affine independent points, a $k$-simplex $\sigma^k=\{v_0,v_1,v_2,\cdots,v_k\}$ can be expressed as
\begin{eqnarray}\label{eq:couple_matrix1}
\sigma^k=\left\{\lambda_0 v_0+\lambda_1 v_1+ \cdots +\lambda_k v_k \mid \sum^{k}_{i=0}\lambda_i=1;0\leq \lambda_i \leq 1,i=0,1, \cdots,k \right\}.
\end{eqnarray}
Moreover, an $i$-dimensional face of $\sigma^k$ is defined as the convex hall formed by the nonempty subset of $i+1$ vertices from $\sigma^k$ ($k>i$). Geometrically, a 0-simplex is a vertex, a 1-simplex is an edge, a 2-simplex is a triangle, and a 3-simplex represents a tetrahedron. We can also define the empty set as a (-1)-simplex.

To combine these geometric components, including vertices, edges, triangles, and tetrahedrons together under certain rules, a simplicial complex is constructed. More specifically, a simplicial complex $K$ is a finite set of simplicies that satisfy two conditions. The first is that any face of a simplex from  $K$  is also in  $K$. The second is that the intersection of any two simplices in  $K$ is either empty or  shared faces. The dimension of a simplicial complex is defined as the maximal dimension of its simplicies. The underlying space $|K|$ is a union of all the simplices of $K$, i.e.,  $|K|=\cup_{\sigma^k\in K} \sigma^k$.  In order to associate the topological space with algebra groups, we need to introduce the concept of chain.

\paragraph{Homology}
A $k$-chain $[\sigma^k]$ is a linear combination $\sum^{k}_{i}\alpha_i\sigma^k_i$ of $k$-simplex $\sigma^k_i$. The coefficients $\alpha_i$ can be chosen from different fields such as, rational field $\mathbb{Q}$, integer field $\mathbb{Z}$, and prime integer field  $\mathbb{Z}_p$ with prime number $p$. For simplicity, in this work the coefficients $\alpha_i$ is chosen in the field of $\mathbb{Z}_2$, for which  the addition operation between two chains is the modulo 2 addition for the coefficients of their corresponding simplices. The set of all $k$-chains of simplicial complex $K$ together with addition operation forms an Abelian group $C_k(K, \mathbb{Z}_2)$. The homology of a topological space is represented by a series of Abelian groups.

Let  us  define the boundary operation $\partial_k$ as $\partial_k: C_k \rightarrow C_{k-1}$. With no consideration of the orientation, the boundary of a $k$-simplex $\sigma^k=\{v_0,v_1,v_2,\cdots,v_k\}$ can be denoted as,
\begin{eqnarray}
\partial_k \sigma^k = \sum^{k}_{i=0} \{ v_0, v_1, v_2, \cdots, \hat{v_i}, \cdots, v_k \}.
\end{eqnarray}
Here $\{v_0, v_1, v_2, \cdots ,\hat{v_i}, \cdots, v_k\}$ means that  the $(k-1)$-simplex is generated by the elimination of vertex $v_i$ from the sequence. A key property of the boundary operator is that applying the boundary operation twice, any $k$-chain will be mapped to a zero element as $\partial_{k-1}\partial_k= \emptyset$. Also we have $\partial_0= \emptyset$. With the boundary operator, one can define the cycle group and boundary group. Basically, the $k$-th cycle group $Z_k$ and the $k$-th boundary group $B_k$ are the subgroups of $C_k$ and can be defined as,
\begin{eqnarray}
&& Z_k={\rm Ker}~ \partial_k=\{c\in C_k \mid \partial_k c=\emptyset\}, \\
&&  { B_k={\rm Im} ~\partial_{k+1}= \{ c\in C_k \mid \exists d \in C_{k+1}: c=\partial_{k+1} d\}.}
\end{eqnarray}
Element in the $k$-th cycle group $Z_k$ or the $k$-th boundary group $B_k$ is called the $k$-th cycle or the $k$-th boundary.  As the boundary of a boundary is always empty $\partial_{k-1}\partial_k= \emptyset$,  one has $B_k\subseteq Z_k \subseteq C_k$. Topologically, the $k$-th cycle is a $k$ dimensional loop or hole.

With all the above definitions, one can introduce the homology group. Specifically, the $k$-th homology group $H_k$ is the quotient group generated by the $k$-th cycle group $Z_k$ and $k$-th boundary group $B_k$: $H_k=Z_k/B_k$. Two $k$-th cycle elements are then called homologous if they are different by a $k$-th boundary element. From the fundamental theorem of finitely generated abelian groups, the $k$-th homology group $H_k$ can be expressed as a direct sum,
\begin{eqnarray}
H_k= {Z}\oplus \cdots \oplus {Z} \oplus {Z}_{p_1}\oplus \cdots \oplus {Z}_{p_n}= {Z}^{\beta_k} \oplus {Z}_{p_1}\oplus \cdots \oplus {Z}_{p_n},
\end{eqnarray}
where $\beta_k$, the rank of the free subgroup, is the $k$-th Betti number. Here  $ {Z}_{p_i}$ is torsion subgroup with torsion coefficients $\{p_i| i=1,2,...,n\}$, the power of prime number.   Therefore, whenever $H_k$ is torsion free. The Betti number can be simply calculated by  
\begin{eqnarray}
\beta_k = {\rm rank} ~H_k= {\rm rank }~ Z_k - {\rm rank}~ B_k.
\end{eqnarray}
Topologically, cycle element in $H_k$ forms a $k$-dimensional loop or hole that is not from the boundary of a higher dimensional chain element. The geometric meanings of  Betti numbers in $\mathbb{R}^3$ are the follows: $\beta_0$ represents the number of isolated components, $\beta_1$ is the number of one-dimensional loop or circle, and $\beta_2$ describes the number of two-dimensional voids or holes. Together, the Betti number sequence  { $\{\beta_0,\beta_1,\beta_2,\cdots \}$} describes the intrinsic topological property of the system.

\paragraph{Persistent homology}
For a simplicial complex $K$, the filtration is defined as a nested sub-sequence of its subcomplexes,
\begin{eqnarray}
\varnothing = K^0 \subseteq K^1 \subseteq \cdots \subseteq K^m=K.
\end{eqnarray}
The introduction of filtration is of essential importance and directly leads to the invention of persistent homology. Generally speaking,    abstract simplicial complexes generated from a filtration give a multiscale representation of the corresponding topological space, from which related homology groups can be evaluated to reveal  topological features. Furthermore, the concept of persistence is introduced for long-lasting topological features.  The $p$-persistent $k$-th homology group $K^i$ is
\begin{eqnarray}
H^{i,p}_k=Z^i_k/(B_k^{i+p}\bigcap Z^i_k).
\end{eqnarray}
Through the study of the persistent pattern of these topological features, the so called persistent homology is capable of capturing the intrinsic properties of the underlying space solely from the discrete point set.

\subsection{Simplicial complex construction and filtration}\label{sec:ConstructionHomology}

\paragraph{$\check{\rm C}$ech complex, Rips complex and alpha complex}
The concept of nerve is essential to the construction of simplicial complex from a given topological space. Basically,   given an index set $I$ and  open set ${\bf U}=\{U_i\}_{i\in I}$ which is a cover of  a point set $X \in \mathbb{R}^N$, i.e., $X \subseteq \{U_i\}_{i\in I}$, the nerve {\bf N} of {\bf U} satisfies two basic conditions. One  of the conditions is that $\emptyset \in {\bf N}$. The other states that if $\cap_{j \in J} U_j \neq \emptyset $ for $J \subseteq I $, then one has $J \in {\bf N}$. In general, for a set of point cloud data, the simplest way to construct a cover is to assign a ball of certain radius around each point. If the set of  point data is dense enough, then the union of all the  balls has the capability to recover the underlying space.

The nerve of a cover constructed from the union of balls is a $\check{\rm C}$ech complex. More specifically, for a point set $X \in \mathbb{ R}^N$, one defines a cover of closed balls ${\bf B}=\{B (x,\epsilon)\mid x \in X \}$ with radius $\epsilon$ and centered at $x$. The $\check{\rm C}$ech complex of $X$ with parameter $\epsilon$ is denoted as $\mathcal{C}(X,\epsilon)$, which is the  nerve of the closed ball set {\bf B},
\begin{eqnarray}
\mathcal{C}(X,\epsilon) = \left\{ \sigma \mid \cap_{x \in \sigma} B (x,\epsilon) \neq \emptyset \right\}.
\end{eqnarray}
The condition for a $\check{\rm C}$ech complex can be relaxed to generate a Vietoris-Rips complex, in which, a simplex $\sigma$ is generated if the largest distance between any of its vertices is at most $2\epsilon$. Denote  $\mathcal{R}(X,\epsilon)$ the    Vietoris-Rips complex, or Rips complex \cite{Edelsbrunner:1994}. These two abstract complexes satisfy the relation,
\begin{eqnarray}\label{eq:SandwichRelation}
\mathcal{C}(X,\epsilon)\subset \mathcal{R}(X,\epsilon) \subset \mathcal{C}(X,\sqrt{2}\epsilon).
\end{eqnarray}
In practice, Rips complex is much more preferred, due to the above sandwich relation and its computational efficiency.

$\check{\rm C}$ech complex and Rips complex are abstract complexes. Derived from computational geometry, alpha complex is also an important geometric concept. To facilitate the introduction, we review some basic definitions.  Let $X$ be a point set in Euclidean space $\mathbb{R}^d$. The Voronoi cell of a point $x \in X$ is defined as
\begin{eqnarray}
V_x = \{ u\in R^d \mid |u-x|\leq |u-x'|, \forall x'\in X \}.
\end{eqnarray}
The collection of all Voronoi cells forms a Voronoi diagram. Further, the nerve of the Voronoi diagram generates a Delaunay complex.

We define $R(x,\epsilon)$ as the intersection of Voronoi cell $V_x$ with ball $B(x,\epsilon)$, i.e., $R(x,\epsilon)= V_x \cap B(x,\epsilon)$. The alpha complex $\mathcal{A}(X,\epsilon)$ of point set $X$ is defined as the nerve of cover $\cup_{x\in X} R(x,\epsilon)$,
\begin{eqnarray}
\mathcal{A}(X,\epsilon) = \left\{ \sigma \mid \cap_{x \in \sigma} R (x,\epsilon) \neq \emptyset \right\}.
\end{eqnarray}
It can be seen that an alpha complex is a subset of a Delaunay complex.

\begin{figure}
\begin{center}
\begin{tabular}{c}
\includegraphics[width=0.6\textwidth]{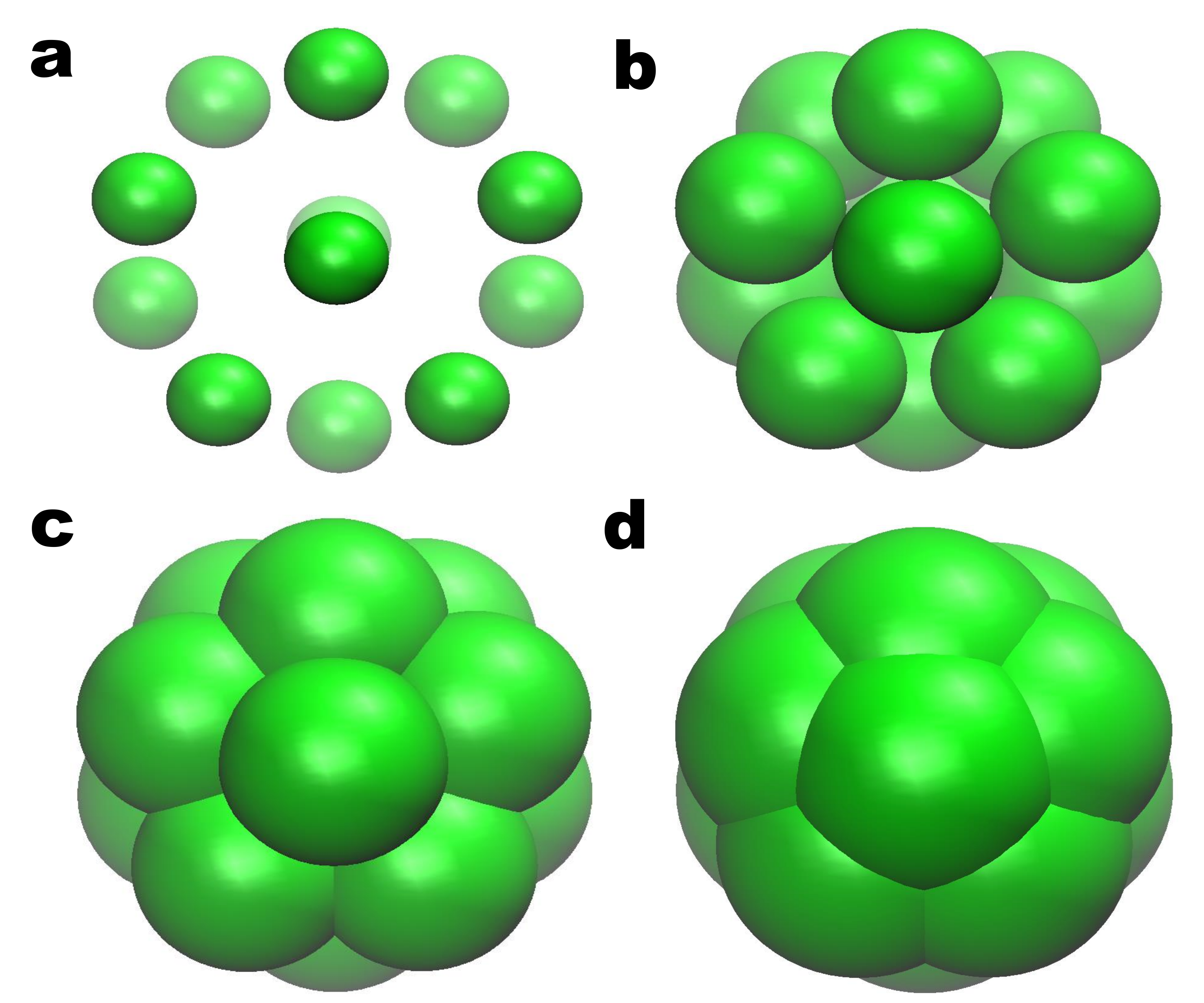}
\end{tabular}
\end{center}
\caption{The distance based filtration process of an icosahedron. Each icosahedron vertex is associated with an ever-increasing   radius to form a ball. With the increase of their radii, the balls overlap with each other to form higher simplexes. In this manner, the previously formed simplicial complex is included in the latter ones. 
}
\label{fig:icosahedron}
\end{figure}

\paragraph{General filtration processes}
To construct a simplicial homology from a set of point cloud data, a filtration process is required \cite{Bubenik:2007, edelsbrunner:2010,Dey:2008,Dey:2013,Mischaikow:2013}. For a specific system, the manner in which a suitable filtration is generated  is key to the persistent homology analysis. In practice, two filtration algorithms, Euclidean-distance based and the correlation matrix based ones, are commonly used.  These filtrations can be modified in many different ways to address physical needs as shown in the application part of this paper. 

The basic Euclidean-distance based filtration is straightforward. One associates each point with an ever-increasing  radius to form an ever-growing ball for each point. When these balls gradually overlap with each other, complexes can be identified from various complex construction algorithms described above. In this manner, the previously formed simplicial complex is an inclusion of latter ones and  naturally, a filtration process is created. One can  formalize this process by  the use of a distance matrix $\{d_{ij}\}$. Here the matrix element $d_{ij}$ represents the distance between atom $i$ and atom $j$. For diagonal terms, it is nature to assume $d_{ij}=0$.  Let us  denote the filtration threshold as a parameter $\varepsilon$. A 1-simplex is generated between vertices $i$ and $j$ if $d_{ij}\leq\varepsilon$. Similarly higher dimensional complexes can also be defined. Figure \ref{fig:icosahedron} demonstrates an Euclidean-distance based filtration  process of an icosahedron.

Sometimes, in order  to  explore the topology-function relationship or illustrate a physical concept, such as cutoff distance, a modification to the distance based filtration is preferred.  See Section \ref{Sec:PH_cutoff} for an example. 

\begin{figure}
\begin{center}
\begin{tabular}{c}
\includegraphics[width=0.7\textwidth]{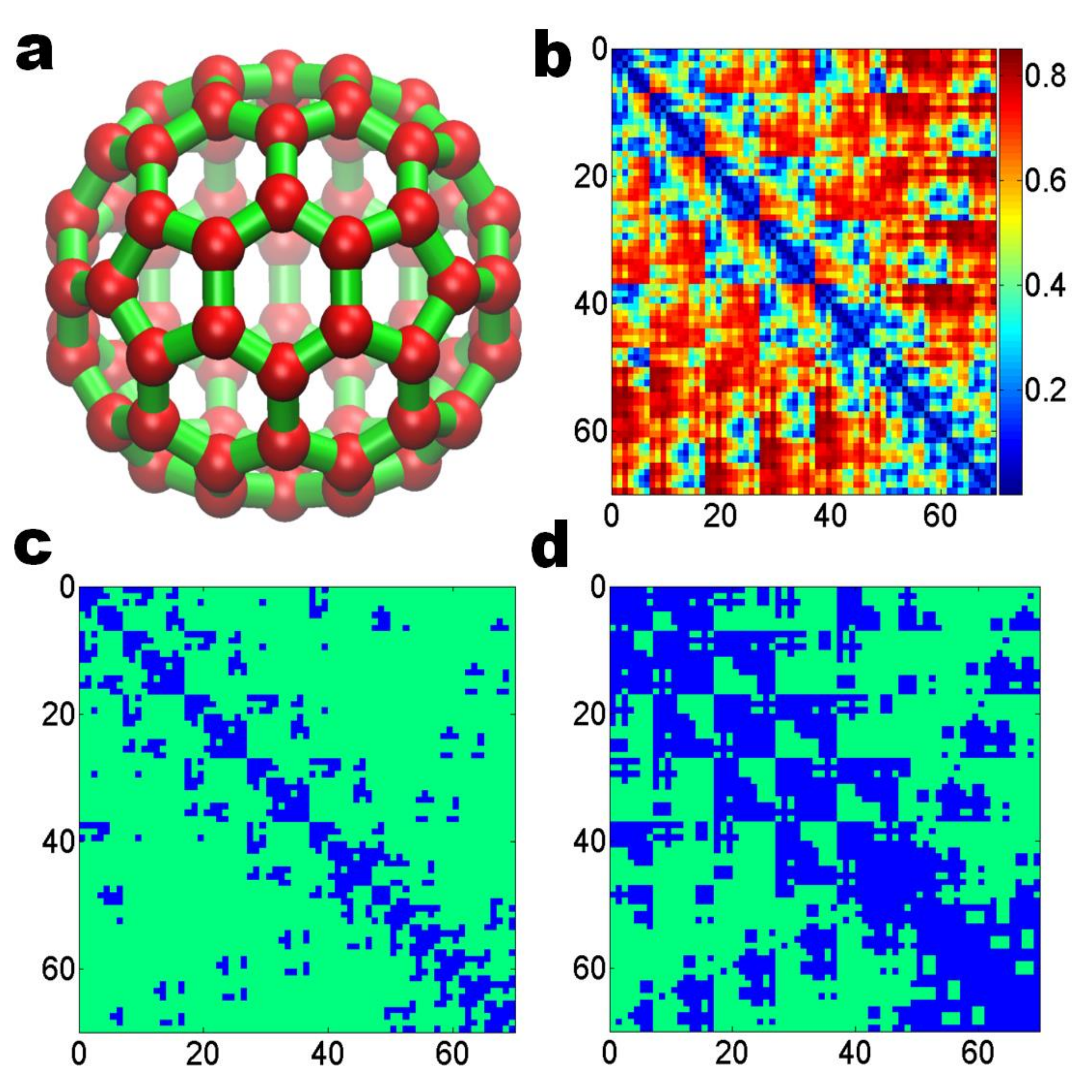}
\end{tabular}
\end{center}
\caption{Correlation matrix based filtration for fullerene C$_{70}$. The correlation matrix is constructed by using the  geometry to topology mapping \cite{KLXia:2013f,KLXia:2013d}. As the value of the filtration parameter increase,  Rips complex formed grows accordingly. {\bf a} is an image of fullerene C$_{70}$; {\bf b}, {\bf c} and  {\bf d} demonstrate the connectivity among C$_{70}$ atoms at filtration threshold $\epsilon=0.1$\AA, 0.3\AA~ and 0.5\AA, respectively. The blue color represents the atoms already formed simplicies.}
\label{fig:FiltrationMatrix}
\end{figure}

Usually, the physical properties are associated with geometric or topological features. However, these features, more often than not, cannot be used directly to predict the physical characteristics. Instead, correlation functions, either form fundamental laws or experimental observation, should be employed. Based on the correlation matrix generated by these functions, one can build another more abstract type of filtration. The resulting persistent homology groups can be simply understood as the topological intrinsic properties of the object. In this manner, one has a powerful tool to explore and reveal more interesting topology-function relationship and essential physical properties. Figure \ref{fig:FiltrationMatrix} demonstrates a correlation matrix based filtration process for fullerene C$_{70}$.  The correlation matrix is generated from the geometry to topology mapping discussed in Section \ref{sec:ExistingMethods}.

\subsection{Computational algorithms for homology}\label{sec:ComputationHomology}

For a given simplicial complex, we are interested in the related Betti numbers, which are topological invariants. In computational homology, the reduction algorithm is a standard method for Betti number evaluation. In this algorithm, the boundary operator is represented as a special matrix. Using invertible elementary row and column operations, this matrix can be further reduced to the Smith normal form. Finally, the Betti number can be expressed in terms of the rank of the matrix.

The matrix representation is essential to the reduction algorithm. For a boundary operator $\partial_i: C_i \rightarrow C_{i-1}$, under the chain group basis, it can be represented by an integer matrix $M_i$, which has $i$ columns and $i-1$ rows. Here entries in $M_i$ are directly related to the field  chosen. Elementary row and column operation, such as exchanging two rows (columns), multiplying a row (a column) with an invertible number, and replacing two rows (columns), can be employed to diagonalize the matrix $M_i$ to the standard Smith normal form $M_i={\rm diag}(a_1,a_2,\cdots,a_{i_{n}})$.  With this reduction algorithm, the ${\rm rank}~ M_i$ equals   parameter $i_{n}$. From the definition of cycle group and boundary group, one has ${\rm rank} ~Z_i= {\rm rank}~C_i - {\rm rank} ~M_i$ and ${\rm rank}~ B_i= {\rm rank} ~M_{i+1}$. Therefore the Betti number can be calculated as
\begin{eqnarray}\label{eq:CalculateBetti}
\beta_i = {\rm rank} ~C_i - {\rm rank}~ M_i - {\rm rank} ~M_{i+1}.
\end{eqnarray}
For a large simplicial complex, the constructed matrix may seem to be cumbersome. Sometimes, the topological invariant called Euler characteristic $\chi$ can be helpful in the evaluation of Betti numbers. More specifically, for the $k$-th simplicial complex K, $\chi(K)$ is defined as
\begin{eqnarray}
\chi(K) = \sum^{k}_{i=0}(-1)^i {\rm rank}~ C_i (K).
\end{eqnarray}
By using Eq. (\ref{eq:CalculateBetti}), the Euler characteristic can be also represented as
\begin{eqnarray}
\chi(K) = \sum^{k}_{i=0}(-1)^i \beta_i (K).
\end{eqnarray}
Since $K$ is the $k$-th simplicial complex, one has  ${\rm Im}~\partial_{k+1}= \emptyset$ and  ${\rm rank}~M_{k+1}= 0 $. As ${\rm rank}~ \partial_0=\emptyset $, one has ${\rm rank}~ M_{0}=0$. 

Proteins are often visualized by their surfaces of various definitions, such as van der Waals surfaces, solvent excluded surfaces, solvent accessible surfaces, minimal molecular surfaces \cite{Bates:2008} and Gaussian surfaces \cite{QZheng:2012} at some given van der Waals radii. Therefore,  another topological invariant, the genus number, is useful too. However, it is beyond the scope of the present work to elaborate this aspect.

\begin{figure}
\begin{center}
\begin{tabular}{cc}
\includegraphics[width=0.5\textwidth]{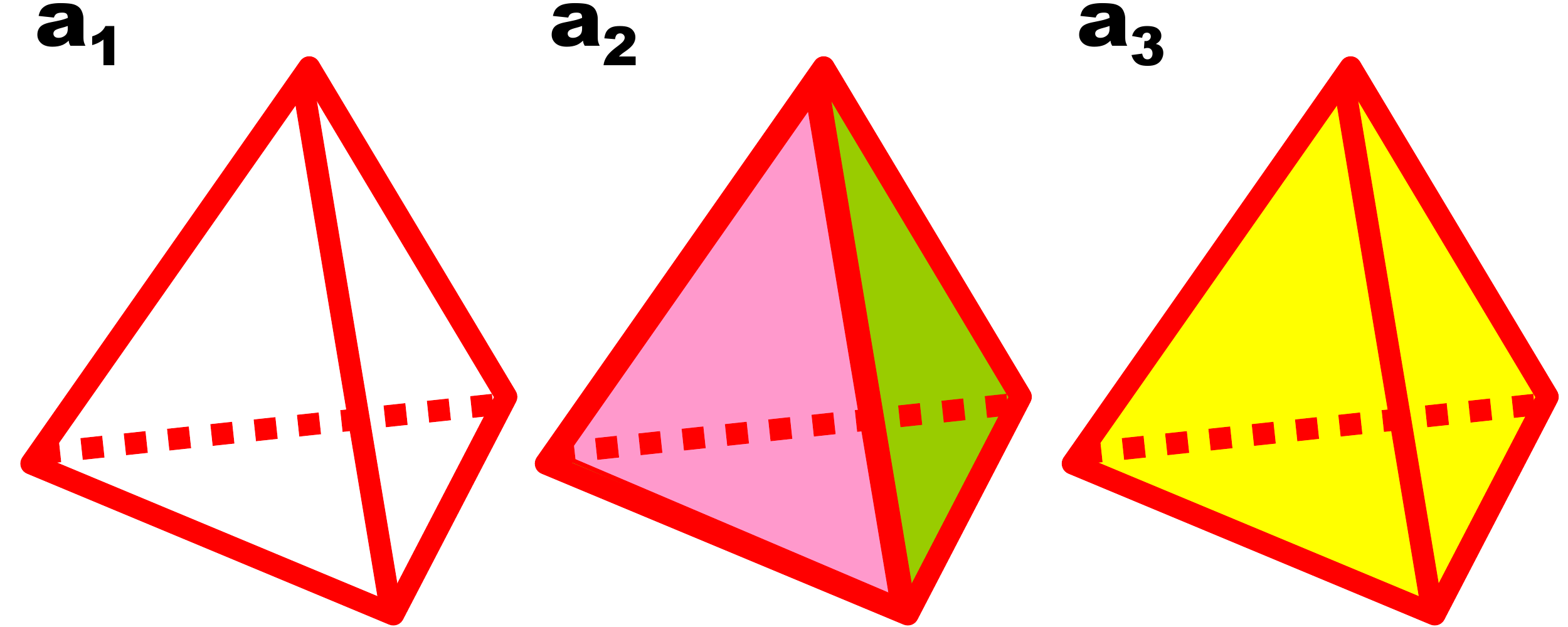}&
\includegraphics[width=0.5\textwidth]{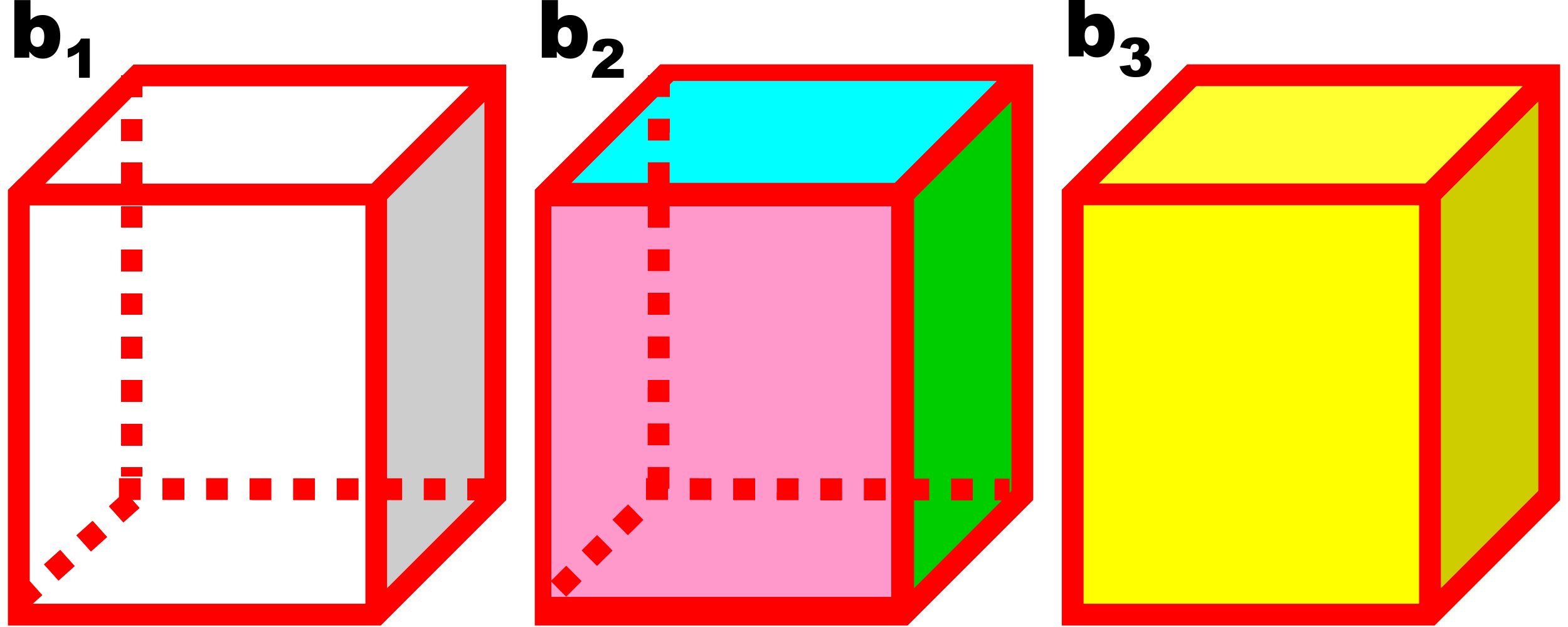}
\end{tabular}
\end{center}
\caption{ Illustration of simplicial complexes. The tetrahedron-shaped simplicial complexes are depicted in {\bf a}$_1$ to  {\bf a}$_3$, and the cube-shaped simplicial complexes are demonstrated in {\bf b}$_1$ to {\bf b}$_3$. In each shape, the leftmost simplicial complex has only 0-simplexes and 1-simplexes. As filtration processes, 2-simplexes emerge in the middle one. In the final stage, a 3-simplex is generated. Table \ref{tab:BettiEulerRelation} gives a description in terms of vertices, edges, faces and cells.}
\label{fig:BettiEulerRelation}
\end{figure}

To illustrate  Euler characteristic, Betti number and  reduction algorithm in detail, we have designed two toy models as shown in Fig. \ref{fig:BettiEulerRelation}. Let us discuss in detail of the first three charts of the figure, which are about three tetrahedron-like geometries. The first object is made of only points (0-simplex) and edge (1-simplex). Then face information (2-simplex) is added in the second chart. Further, a tetrahedron (3-simplex) is included in the third chart. This process resembles a typical filtration as the former one is the inclusion of the latter one. The same procedure is also used in the last three charts of Fig. \ref{fig:BettiEulerRelation} for a cube. Table \ref{tab:BettiEulerRelation} lists the basic properties of two simplicial complexes in terms of numbers of vertices, edges, faces and cells. Both Betti numbers and Euler characteristic are calculated for these examples.

\begin{table}
\caption{A summary of Betti number and Euler characteristic in Fig. \ref{fig:BettiEulerRelation}. Symbols V, E, F, and C stand for the number of vertices, edges, faces, and cells, respectively. Here $\chi$ is the Euler characteristic. }
\begin{center}
\begin{tabular}{|c|c|c| c|c|c|c|c|c|}
 \hline
  Simplex & $V$ & $E$ & $F$ & $C$ & $\beta_0$  & $\beta_1$ & $\beta_2$  & $\chi$ \\\hline
  Figure \ref{fig:BettiEulerRelation}{\bf a}$_1$  &4    &6     &0         &0          &1        &3        &0         &-2    \\\hline
  Figure \ref{fig:BettiEulerRelation}{\bf a}$_2$  &4    &6     &4         &0          &1        &0        &1         &2    \\\hline
  Figure \ref{fig:BettiEulerRelation}{\bf a}$_3$  &4    &6     &4         &1          &1        &0        &0         &1    \\\hline
  Figure \ref{fig:BettiEulerRelation}{\bf b}$_1$  &8    &12    &0         &0          &1        &5        &0         &-4    \\\hline
  Figure \ref{fig:BettiEulerRelation}{\bf b}$_2$  &8    &12    &6         &0          &1        &0        &1         &2    \\\hline
  Figure \ref{fig:BettiEulerRelation}{\bf b}$_3$  &8    &12    &6         &1          &1        &0        &0         &1    \\\hline
\end{tabular}
\label{tab:BettiEulerRelation}
\end{center}
\end{table}

For the filtration process of complicated point cloud data originated from a practical application, the calculation of the persistence of the Betti numbers is nontrivial. It is out of the scope of the present paper to discuss the Betti number calculation in  detail.   The interested reader is  referred to the literature \cite{Edelsbrunner:2002,Carlsson:2009}. In the past decade, many software packages  have been developed based on various algorithms, such as \href{https://code.google.com/p/javaplex/}{Javaplex}, \href{http://www.sas.upenn.edu/~vnanda/perseus/index.html}{Perseus}, \href{http://www.mrzv.org/software/dionysus/}{Dionysus} etc.  In this work, all the computations are carried out by using Javaplex \cite{javaPlex} and the persistent diagram is visualized through barcodes \cite{Ghrist:2008}.

\begin{figure}
\begin{center}
\begin{tabular}{c}
\includegraphics[width=0.7\textwidth]{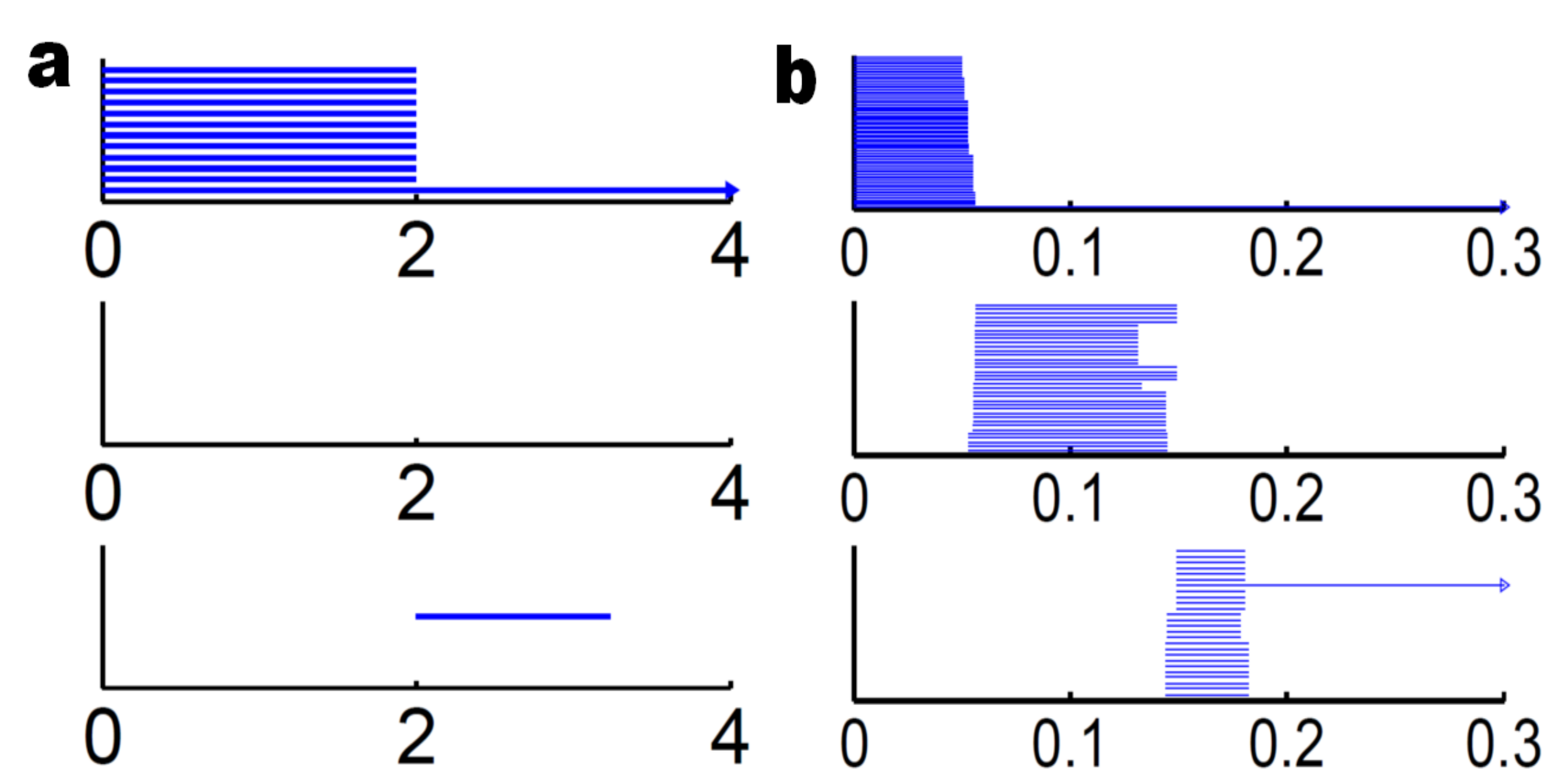}
\end{tabular}
\end{center}
\caption{Persistent homology analysis   of the icosahedron (left chart) and fullerene C$_{70}$ (right chart).
The horizontal axis is the filtration parameter, which has  the unit of  angstrom (\AA) in the distance based filtration for  icosahedron  and there is no unit in the correlation matrix based filtration for C$_{70}$.  From the top to the bottom, there are three panels corresponding to $\beta_0$, $\beta_1$ and $\beta_2$ bars, respectively. 
For the icosahedron barcode,  originally there are 12 bars due to 12 vertices in $\beta_0$ panel.  As the filtration continues, 11 of them terminate simultaneously with only one persisting to the end, indicating that all vertices are connected. Topologically, $\beta_0$ bars represent  isolated entities. At beginning, there are 12 individual vertices. They connect to each other simultaneously due to their structural symmetry. Nothing occurs at the $\beta_1$ panel, which means there is no one-dimensional circle ever formed. Finally, in the $\beta_2$ panel, the single bar represents the central void. 
For   fullerene C$_{70}$ barcode, there are 70 $\beta_0$ bars and  36 $\beta_1$ bars. The  $\beta_1$ bars are due to  12 pentagon rings and 25 hexagon rings.  The  hexagon rings further evolve into  two-dimensional holes, which are represented by 25 short-lived $\beta_2$ bars. The central void structure formed is captured by the persisting $\beta_2$ bar.}
\label{fig:barcodes}
\end{figure}

Figure \ref{fig:barcodes} illustrates the persistent homology analysis of icosahedron (left chart) and fullerene C$_{70}$ (right chart). Both distance based  filtration as illustrated  in Fig. \ref{fig:icosahedron}   and correlation matrix based filtration as depicted in Fig. \ref{fig:FiltrationMatrix} are employed. For the icosahedron chart, there exist three panels corresponding to $\beta_0$, $\beta_1$ and $\beta_2$ from top to bottom. For $\beta_0$ number, originally 12 bars coexist, indicating 12 isolated vertices. As the filtration continues, 11 of them disappear simultaneously with only one survived and persisting to the end. Geometrically, due to the high symmetry, these 12 vertices connect with each other simultaneously at 2\AA, i.e., the designed ``bond length''.  The positions where the bars terminate are exactly the corresponding bond lengths. Therefore, barcode representation is able to  incorporate certain geometric information. As no one-dimensional circle has ever formed, no $\beta_1$ bar is generated. Finally, in the $\beta_2$ panel, there is a single bar, which represents a two-dimensional void enclosed by the surface of the icosahedron.

In fullerene C$_{70}$ barcodes, there are 70 $\beta_0$ bars. Obviously, there are 6 distinct groups in  $\beta_0$ bars due to the factor that there are 6 types of bond lengths in  the C$_{70}$ structure.  Due to the formation of rings, $\beta_1$  bars emerge. There is a total of 36 $\beta_1$ bars corresponding to 12 pentagon rings and 25 hexagon rings. It appears that one ring is not accounted because  any individual ring can be represented as the linear combination of all other rings. Note that there are 6 types of rings. Additionally, 25 hexagon rings further evolve to two-dimensional holes, which are represented by 25  bars in the $\beta_2$ panel. The central void structure is captured by the persisting $\beta_2$ bar.

\section{Persistent homology analysis of proteins}\label{sec:PHProtein}

In this section, the method of persistent homology  is employed to study the topology-function relationship of proteins. Specifically,  the intrinsic features of protein structure, flexibility, and folding are investigated by using topological invariants. In protein structure analysis, we compare an all-atom representation with a coarse-grained (CG) model.  Two most important protein structural components, namely,  alpha helices and  beta sheets, are analyzed to reveal their  unique topological features, which can be recognized as their topological ``ID"s or fingerprints. A beta barrel is also employed as an example to further demonstrate the  potential of persistent homology in protein structure analysis.

In  protein flexibility and rigidity analysis, many elegant methods, such as normal mode analysis (NMA), Gaussian network model (GNM), elastic network model (ENM),    anisotropic network model (ANM), our molecular nonlinear dynamic (MND) and flexibility and rigidity index (FRI),   have been proposed. Although they differ in terms of theoretical foundations and computational realization, they share a similar parameter called cut off distance or characteristic distance. The physical meaning of this parameter is the relative influence domain of certain atoms. Usually, for the CG model with each amino acid represented by its C$_\alpha$ atom, the optimized cut off distance is about 7  to 8 \AA~ \cite{YangLei:2009}, based on fitting with a large number of experimental B-factors. To provide a different perspective and unveil the topological significance of the cutoff distance or characteristic distance, simplicial complexes and related filtration processes are built up. Different parameter values  induce dramatically distinguished topological patterns. Optimal characteristic distances are revealed from our persistent homology analysis. 

To study topological features of protein folding, a simulated unfolding process is considered. We use a constant velocity pulling algorithm in our steered molecule dynamics to generate a family of configurations.  Through the analysis of their topological invariants, we found that the accumulated bar length, which represents the total connectivity, continuously decreases in the protein unfolding process. As the relative stability of a protein is proportional to its topological connectivity, which is a topology-function relationship,  the negative accumulated bar length can be used to describe the stability of a protein.

\subsection{Topological  fingerprints of proteins}

\begin{figure}
\begin{center}
\begin{tabular}{c}
\includegraphics[width=0.8\textwidth]{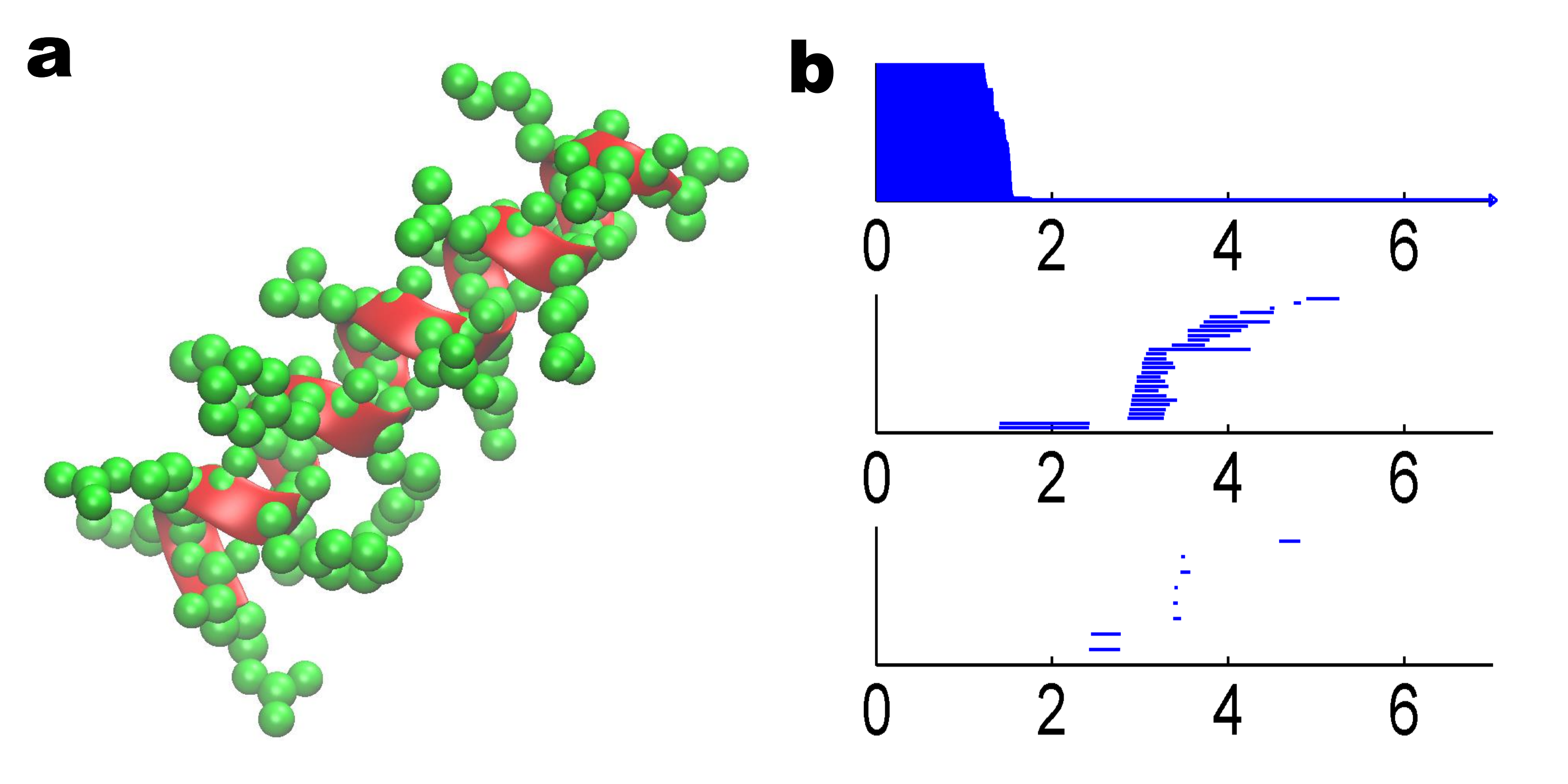}
\end{tabular}
\end{center}
\caption{Illustration of an alpha helix structure (PDB ID: 1C26) and its  topological  fingerprint obtained by  the distance filtration. In the left chart, atoms are demonstrated in green color and the helix structure of the main chain backbone is represent by the cartoon shape in red. The right chart is the corresponding barcode with the all-atom description. The horizontal axis is the filtration size (\AA). Although the alpha helix backbone has a loop-type structure, the corresponding barcode does not clearly demonstrate these patterns due to the fact that there are two many  atoms around the main chain. }
\label{fig:AlphaHelixAA}
\end{figure}

\begin{figure}
\begin{center}
\begin{tabular}{c}
\includegraphics[width=0.7\textwidth]{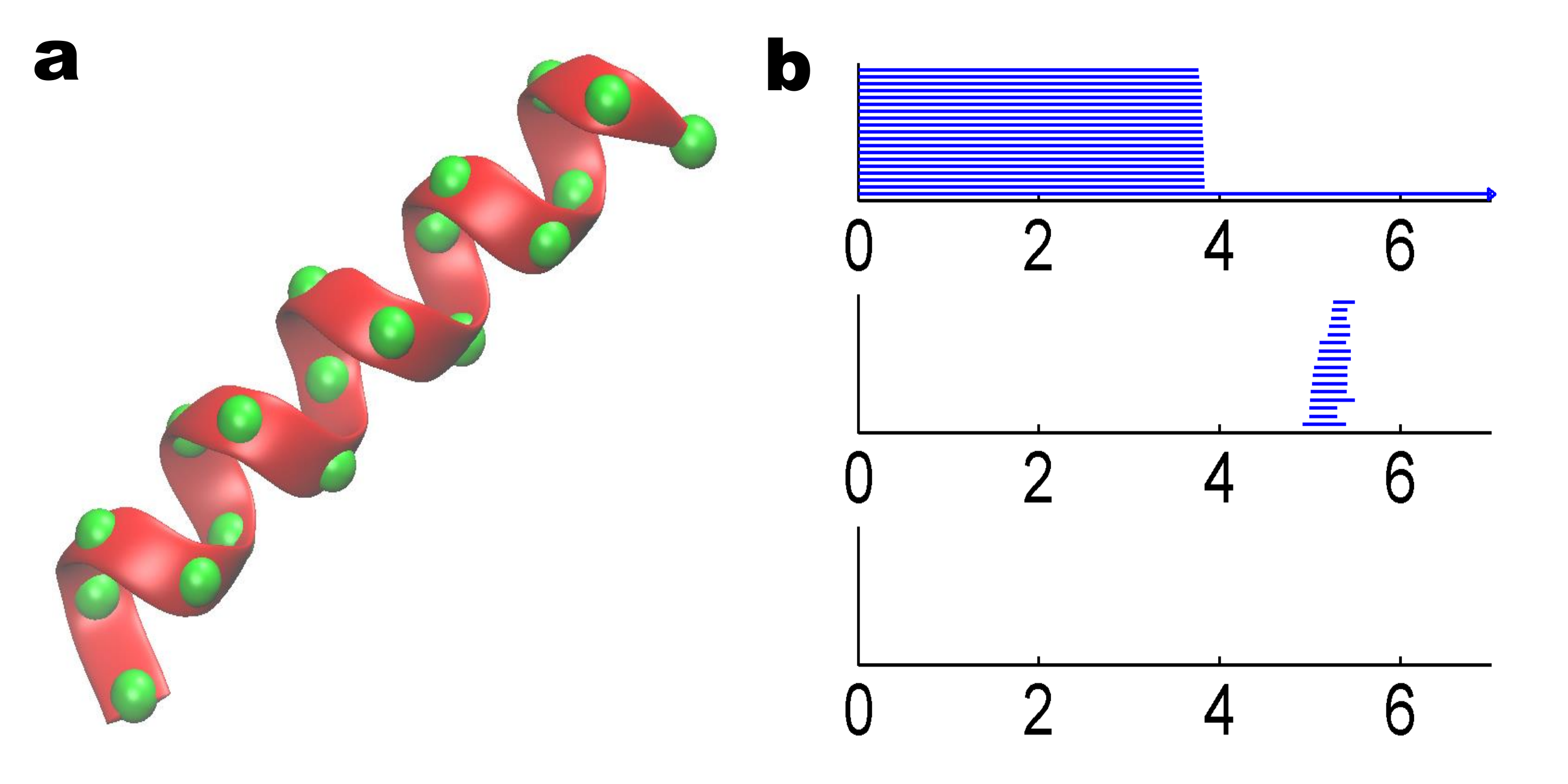}
\end{tabular}
\end{center}
\caption{The coarse-grained representation of an  alpha helix structure  (left chart) and its  topological  fingerprint  (right chart). The  alpha helix structure (PDB ID: 1C26) has 19 residues represented by  C$_{\alpha}$ atoms in green color. Each 4 C$_{\alpha}$ atoms contribute a $\beta_1$ loop and thus there are 16 short-lived bars in the $\beta_1$ panel. }
\label{fig:AlphaHelixCA}
\end{figure}

Protein molecules often consist of one or more coiled peptide chains and have highly complicated 3D structures.  Protein topological features include isolated entities, rings and cavities. However, each protein structure is unique. Our goal is to unveil protein intrinsic topologies and identify their molecular fingerprints.  Alpha helices and beta sheets are basic structural components of proteins. Biologically, alpha helix is a spiral conformation, stabilized through hydrogen bond formed between the backbone N-H group and the backbone C=O group of four residues earlier in the protein polymer. Usually, the backbone of an alpha helix is right-handedly coiled. Each amino acid residue contributes about a $100$ degree rotation in the helix structure. State differently, each spiral in the backbone is made of 3.6 amino acid residues. In contrast,   beta sheet  is a stretched polypeptide chain with about 3 to 10 amino acids.  Usually, beta strands connect laterally with each other through the backbone hydrogen bonds to form a pleated sheet. Two adjacent beta strands can be parallel or anti-parallel to each other with a slightly different pattern due to the relative position between N-H group and C=O group. Many amyloidosis related diseases, such as mad cow disease and Alzheimer's disease, are due to the insoluble  protein aggregates and fibrils made of  beta sheets.

\begin{figure}
\begin{center}
\begin{tabular}{c}
\includegraphics[width=0.8\textwidth]{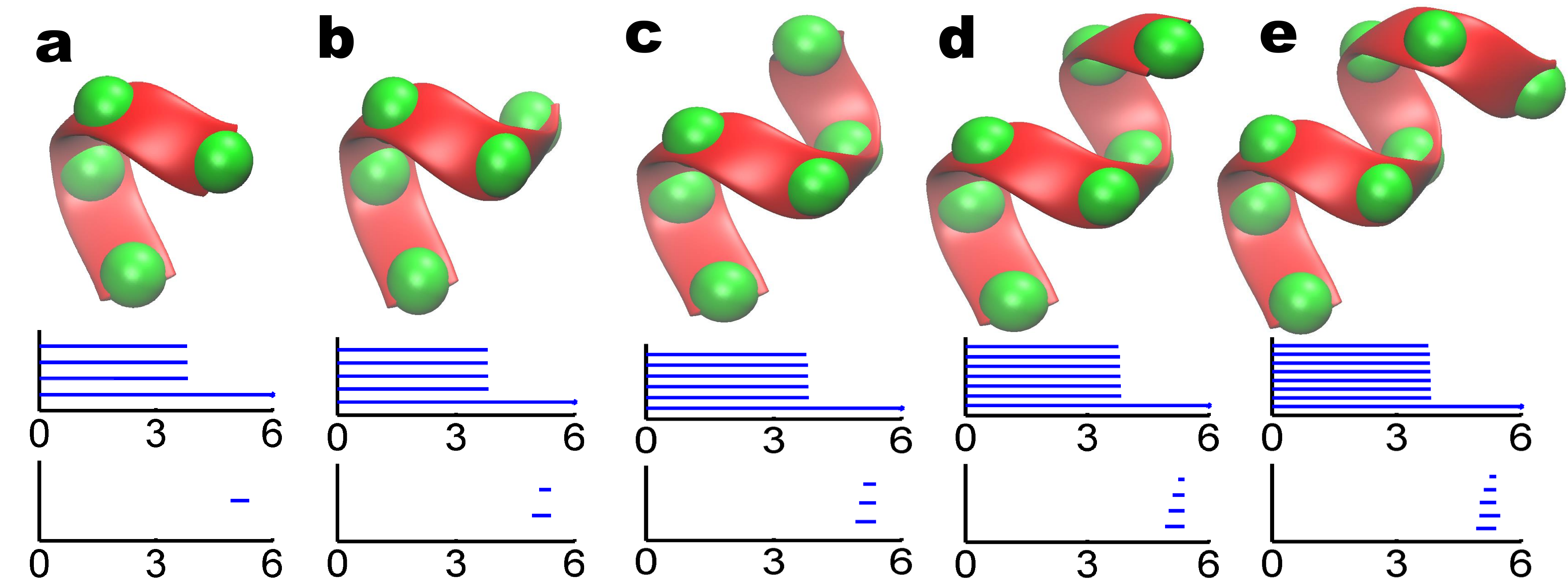}
\end{tabular}
\end{center}
\caption{Method of slicing for the analysis of alpha helix topological fingerprints. In the coarse-grain representation, each residue is represented by a C$_{\alpha}$ atom. In an alpha helix, each set of four C$_{\alpha}$ atoms forms a one-dimensional loop in the filtration process as depicted in  {\bf a}. By adding one more C$_{\alpha}$ atom, one more $\beta_1$ loop is generated and leads to an additional $\beta_1$ bar as shown in  {\bf b},  {\bf c}, {\bf d}, and {\bf e}. This explains the occurrence of 16 $\beta_1$ bars in Fig. \ref{fig:AlphaHelixCA}.}
\label{fig:slicing}
\end{figure}

In this section, two basic protein structure representations, i.e., an all-atom model and a CG model, are considered. For the all-atom model, various types of atoms, including  H,  O,  C, N, S, P, etc., are all included and regarded as equally important in our computation. The all-atom model gives an atomic description of the protein and is widely used in molecular dynamic simulation. In contrast, the CG model describes the protein molecule with a reduced number of degrees of freedom and is able to highlight important protein structure features. A standard coarse-grained representation of proteins is to represent each  amino acid by the corresponding C$_{\alpha}$ atom.   The CG model is efficient for describing large proteins and protein complexes.

\paragraph{Topological fingerprints of alpha helix and beta sheet}
Protein structure data are available through the Protein Data Bank (PDB). To analyze an alpha helix structure, we download a protein of PDB ID: 1C26 and adopt an alpha helix chain with 19 residues. In our all-atom model, we do not distinguish between different types of atoms. Instead, each atom is associated with the same radius in the distance based filtration. The persistent diagram is represented by the barcode as depicted in right chart of Fig. \ref{fig:AlphaHelixAA}. As discussed in Section \ref{sec:ComputationHomology}, the $\beta_0$ bars can reveal the bond length information. Physically, for protein molecule, the bond length is between 1   to 2 \AA, which is reflected in the distance based filtration. The occurrences of $\beta_1$ and $\beta_2$ bars are due to  the loop, hole and void type of structures. Because  the filtration process  generates a large number of  bars, it is difficult to directly decipher this high dimensional  topological information. From the left chart of Fig. \ref{fig:AlphaHelixAA}, it is seen that the alpha helix backbone has a regular spiral structure. However, residual atoms are quite crowd around the main chain and bury the spiral loop.  To extract more geometric and topological details of the helix structure, we utilize the CG with each amino acid represented by its C$_{\alpha}$ atom. The results are demonstrated in the left chart of Fig. \ref{fig:AlphaHelixCA}.  As there are  19 residues in the alpha helix structure, only 19 atoms are used in the CG model and the corresponding barcode is dramatically simplified. From the right chart of Fig.  \ref{fig:AlphaHelixCA}, it is seen that there are 19 bars  in $\beta_0$ panel and the bar length is around 3.8 \AA, which is the average length between two C$_{\alpha}$ atoms.  Additionally there are 16 bars in the $\beta_1$ panel. With similar birth time and persist length, these bars form a striking pattern. To reveal the topological meaning of these bars, we make use of a technique called slicing. Basically, we slice a piece of 4 C$_{\alpha}$ atoms from the back bone and study its persistent homology behavior. Then, one more C$_{\alpha}$ atom is added at a time. We repeat this process and generate the corresponding barcodes. The results are depicted in Fig. \ref{fig:slicing}. It can be seen clearly that each four C$_{\alpha}$ atoms in the alpha helix form a one-dimensional loop, corresponding to  a $\beta_1$ bar.  By adding  more C$_{\alpha}$ atoms, more loops are created and more $\beta_1$ bars are obtained. Finally, 19 residues in the alpha helix produce exactly 16 loops as seen in Fig. \ref{fig:AlphaHelixCA}.

\begin{figure}
\begin{center}
\begin{tabular}{c}
\includegraphics[width=0.8\textwidth]{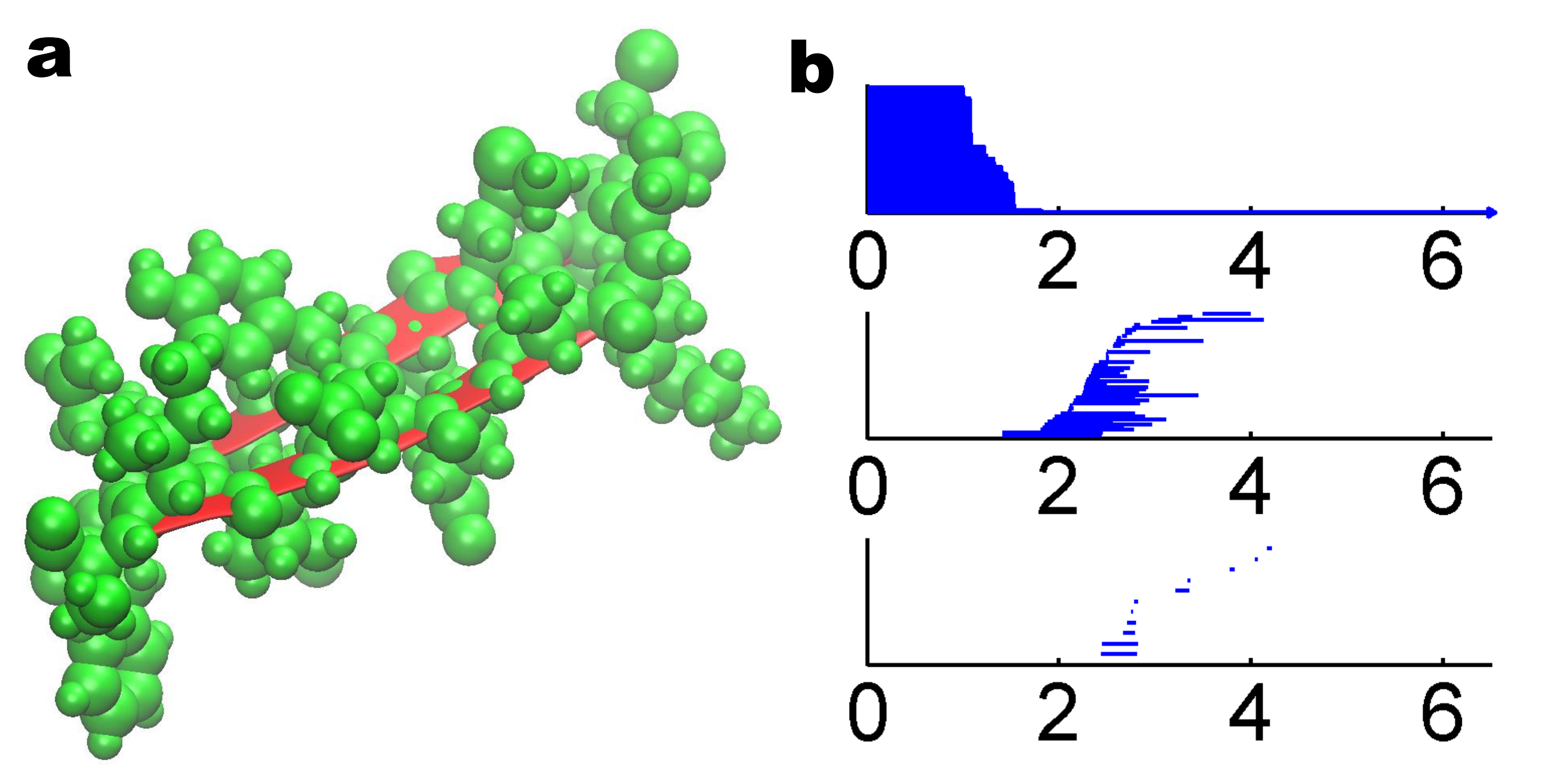}
\end{tabular}
\end{center}
\caption{The all-atom representation of the beta sheet structure generated from PDB 2JOX (left chart) and the related topological fingerprint (right chart). Each  beta sheet has 8 residues. The topological fingerprint generated from all-atom based filtration has a complicated  pattern due to excessively many residual atoms.}
\label{fig:BetaSheetAA}
\end{figure}

\begin{figure}
\begin{center}
\begin{tabular}{c}
\includegraphics[width=0.7\textwidth]{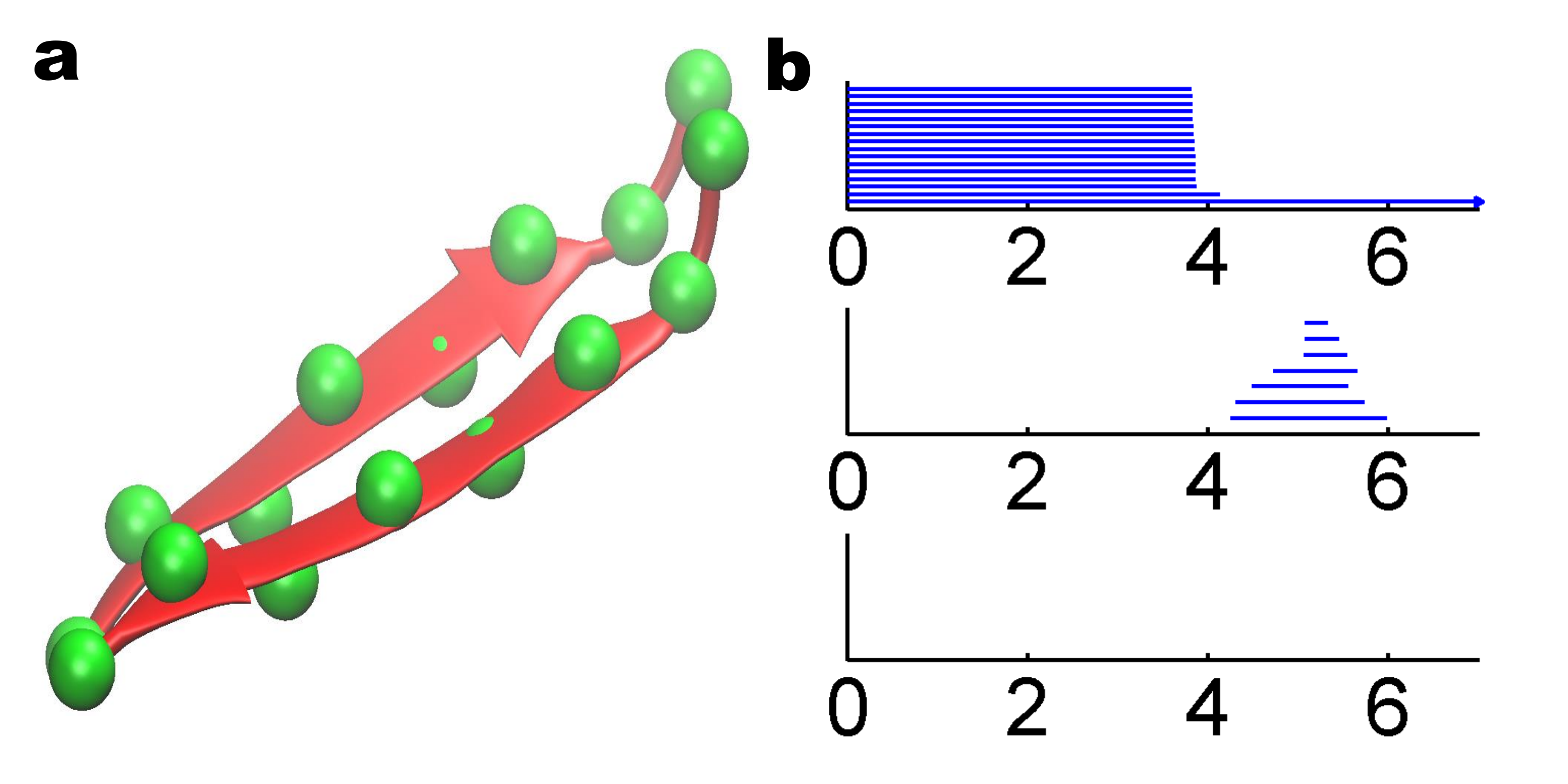}
\end{tabular}
\end{center}
\caption{The coarse-grained representation of two  beta sheet strands generated from protein 2JOX (left chart) and the corresponding topological fingerprint (right chart). There are 8 pairs of residues represented by 16 C$_{\alpha}$ atoms in the $\beta_0$ panel. Each 2 pairs of C$_{\alpha}$ atoms  contribute a $\beta_1$ loop to make up  7 bars in the $\beta_1$ panel. }
\label{fig:BetaSheetCA}
\end{figure}

To  explore the topological fingerprints of  beta sheet structures, we extract two parallel beta strands from protein 2JOX. Figure \ref{fig:BetaSheetAA} and Fig. \ref{fig:BetaSheetCA} demonstrate the persistent homology analysis of  all-atom model and CG model with the distance based filtration. Similar to the alpha helix case, in the all-atom representation, the generated barcode has a complicated pattern due to  excessively many  residual atoms. The barcode of the CG model, on the other hand, is much  simpler with only 7 individual $\beta_1$ bars. Each of these bars is formed by two adjacent residue pairs. From the $\beta_0$ panel we can see that the lengths of most bars  are still around 3.8 \AA, i.e., the average length between two adjacent C$_{\alpha}$ atoms as discussed above. These bars end when the corresponding atoms are connected. However, there exists a unique $\beta_0$ bar which has a length about 4.1 \AA. This bar reflects the shortest distance between  two closest adjacent two C$_{\alpha}$ atoms from two individual beta strands. With these geometric information, we can explain the mechanism of the birth and death of 7 individual $\beta_1$ bars. First, as the filtration begins, adjoined C$_{\alpha}$ atoms in the same strand  form 1-simplex. After that, adjacent C$_{\alpha}$ atoms in two different strands  connect with each other as the filtration continues, which leads to one-dimensional circles and  $\beta_1$ bars. The further  filtration terminates  all the $\beta_1$ bars. There is no  $\beta_1$ bar in the CG representation of beta sheet structures.

The persistent homology analysis of  alpha helices and  beta sheets reveals  their topological features which are useful for deciphering protein   fingerprints as demonstrated in the above example. Typically, CG model based filtration provides  more global scale properties because local characteristics  from  amino acids is ignored. However, all-atom model based filtration can preserve more local scale topological information. For instance,  there are two isolated bars around 2\AA~ in the $\beta_1$ panel of Fig. \ref{fig:AlphaHelixAA}. Meanwhile, there are also two individual bars around 2.5\AA ~ in the $\beta_2$ panel.  These bars are  the  fingerprints of ``PHE" or ``TYR" types of amino acid residues.  It turns out that there are two ``PHE" amino acid residues in the alpha-helix structure. Similar  fingerprints are of paramount importance for  the identification of protein structural motifs, topological modeling of biomolecules and  prediction of protein functions. However, these aspects are beyond the scope of the  present paper.

\begin{figure}
\begin{center}
\begin{tabular}{c}
\includegraphics[width=0.8\textwidth]{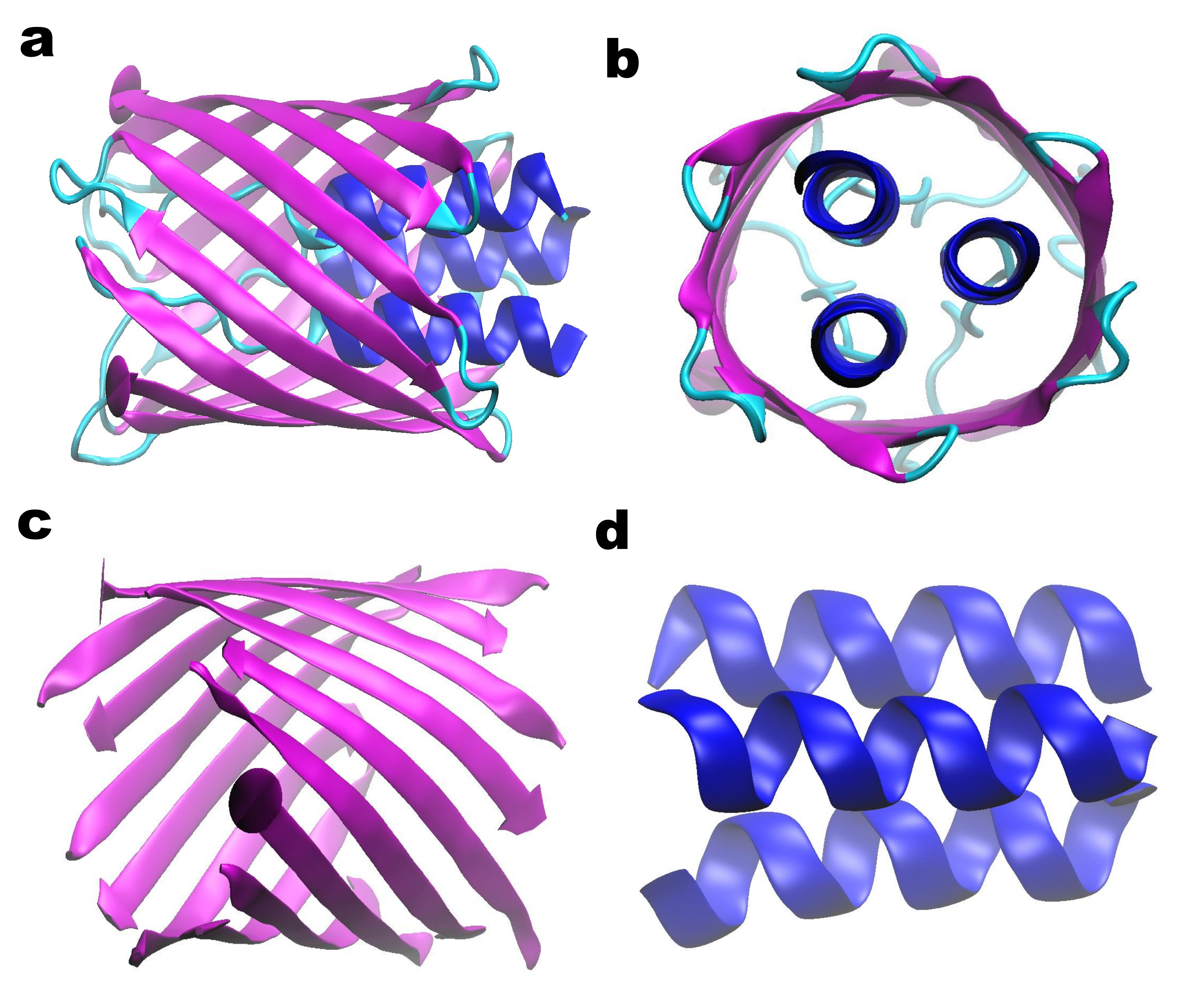}
\end{tabular}
\end{center}
\caption{The basic geometry of the beta barrel generated from protein 2GR8. Here {\bf a} and {\bf b} are cartoon representations from two views.  The  beta barrel structure is decomposed into  beta sheet and alpha helix  components for topological pattern recognition in {\bf c} and {\bf d}, respectively.    }
\label{fig:BetaBarrel}
\end{figure}

\begin{figure}
\begin{center}
\begin{tabular}{c}
\includegraphics[width=0.8\textwidth]{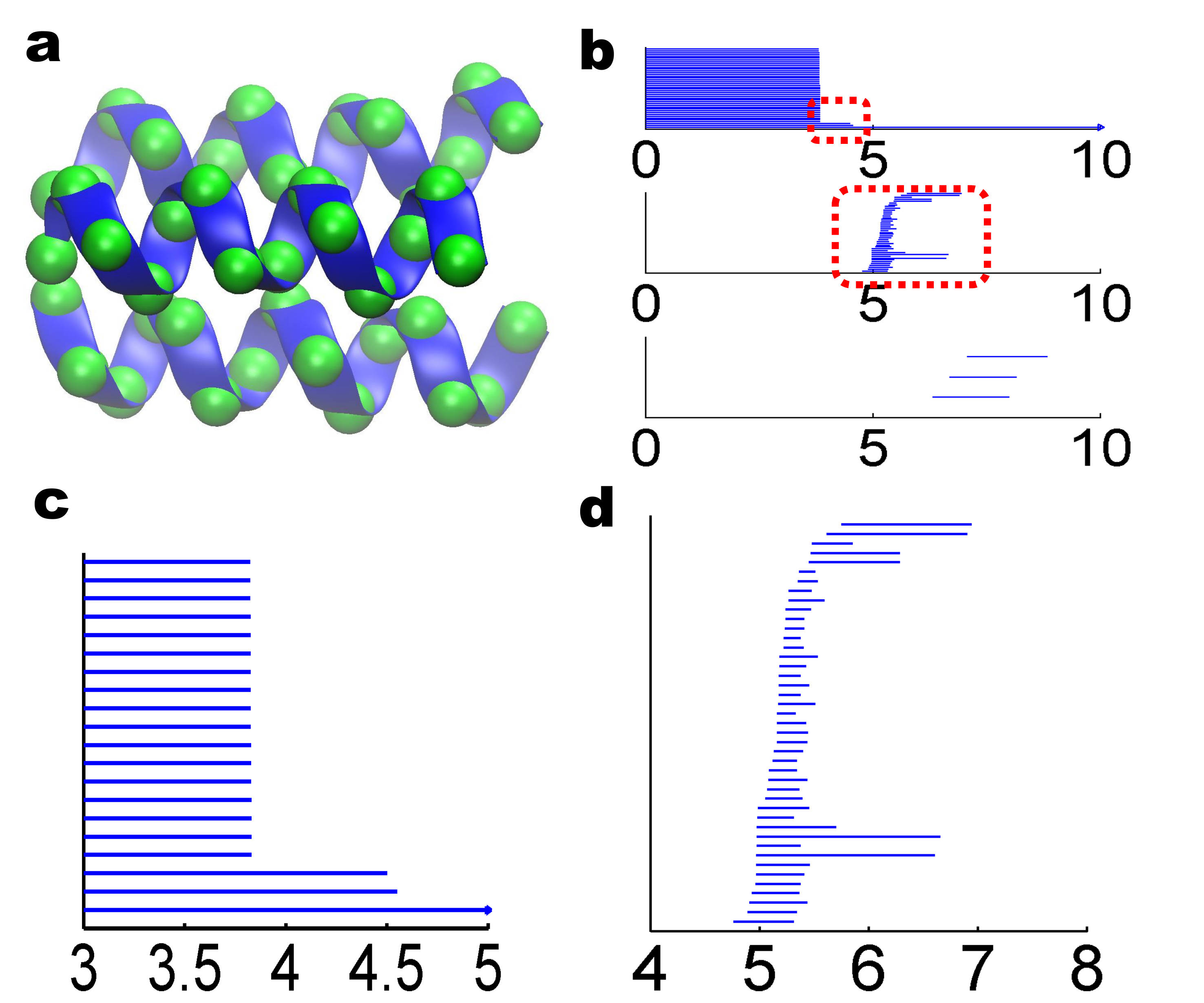}
\end{tabular}
\end{center}
\caption{Topological analysis of the alpha-helix structure  from beta barrel 2GR8.  The coarse-grained representation   is employed with  C$_{\alpha}$ atoms in the green color in {\bf a}.  The topological fingerprint  of the alpha-helix structure is depicted in {\bf b}. The details of the barcode  enclosed in  red boxes are demonstrated in {\bf c} and {\bf d}. It is seen from {\bf c} that in the $\beta_0$ panel, the length of most  bars  is around 3.8\AA ~ except for those of three bars. Two of these $\beta_0$ bars end around 4.5\AA, and the other lasts forever. These three bars represent   the remaining isolated alpha helices  when C$_{\alpha}$ atoms inside each alpha helix  become connected at about  3.8\AA. However, when the filtration parameter is increased to 4.5\AA,  which is exactly the smallest distance between alpha helices, three alpha helices are connected and the total number of independent entities becomes one. There are also three  bars in the $\beta_2$ panel as shown in {\bf b}. There is a total of 43 $\beta_1$ bars as shown in {\bf d}. 
}
\label{fig:BetaBarrel_helix}
\end{figure}

\paragraph{Topological  fingerprints of beta barrels}
Having analyzed  the topological fingerprints of alpha helix and beta sheet, we are interested in revealing the  topological patterns  of  protein structures. As an example, a beta barrel molecule (PDB ID: 2GR8) is used. Figures \ref{fig:BetaBarrel}{\bf a} and {\bf b} depict its basic structure viewed from two different perspectives. The sheet and helix  structures  in the protein  are then extracted and  demonstrated in magenta and bule colors in Figs. \ref{fig:BetaBarrel}{\bf c} and {\bf d}, respectively. The coarse grain model is used. We show the persistent homology analysis of alpha helices  in Fig.  \ref{fig:BetaBarrel_helix}. It can been seen from  the $\beta_0$ panel that nearly all the bar lengths are around 3.8\AA, except three.  Two of them persist to around 4.5\AA~  and the other forever. This pattern  reveals that there exist three isolated components when the filtration size is larger than 3.8\AA~ and less than 4.5\AA, which corresponds exactly to the number of individual alpha helices in the system. There are also 3 bars in the $\beta_2$ panel. Using our slicing technique, it can be found that each $\beta_2$ bar represents a void formed by one turn of  three helices due to the high symmetry of the structure. Each of symmetric turns also generates 3 $\beta_1$ bar. To be more specific, a circle is formed between each two alpha helices at the groove of the turning part. Moreover, it can be noticed that at the left end three alpha helices are more close to each other, with three C$_{\alpha}$ atoms symmetrically distributed to form a 2-simplex during the filtration. Therefore, no $\beta_1$ bar is generated. However, when we slice down the helices, this pattern of three C$_{\alpha}$ atoms forming a 2-simplex only happens once at the eighth C$_{\alpha}$ counted from the left end. When this occurs, a one-dimensional cycle is terminated. As it is well known that usually 3.6 residues form a turn in an alpha helix, we have three turning structures, which give rise  to 9 circles, i.e., 9 bars in the $\beta_1$ panel. Furthermore, a total of 44 (i.e., the number of  $\beta_0$ bars) atoms in three  alpha helices  contributes to 35 (i.e., $44-3*3$) circles. Together with the terminated one-dimensional cycle  at eighth C$_{\alpha}$ atom, we therefore  have 43 $\beta_1$ bars, which agrees with the persistent homology calculation.

\begin{figure}
\begin{center}
\begin{tabular}{c}
\includegraphics[width=0.8\textwidth]{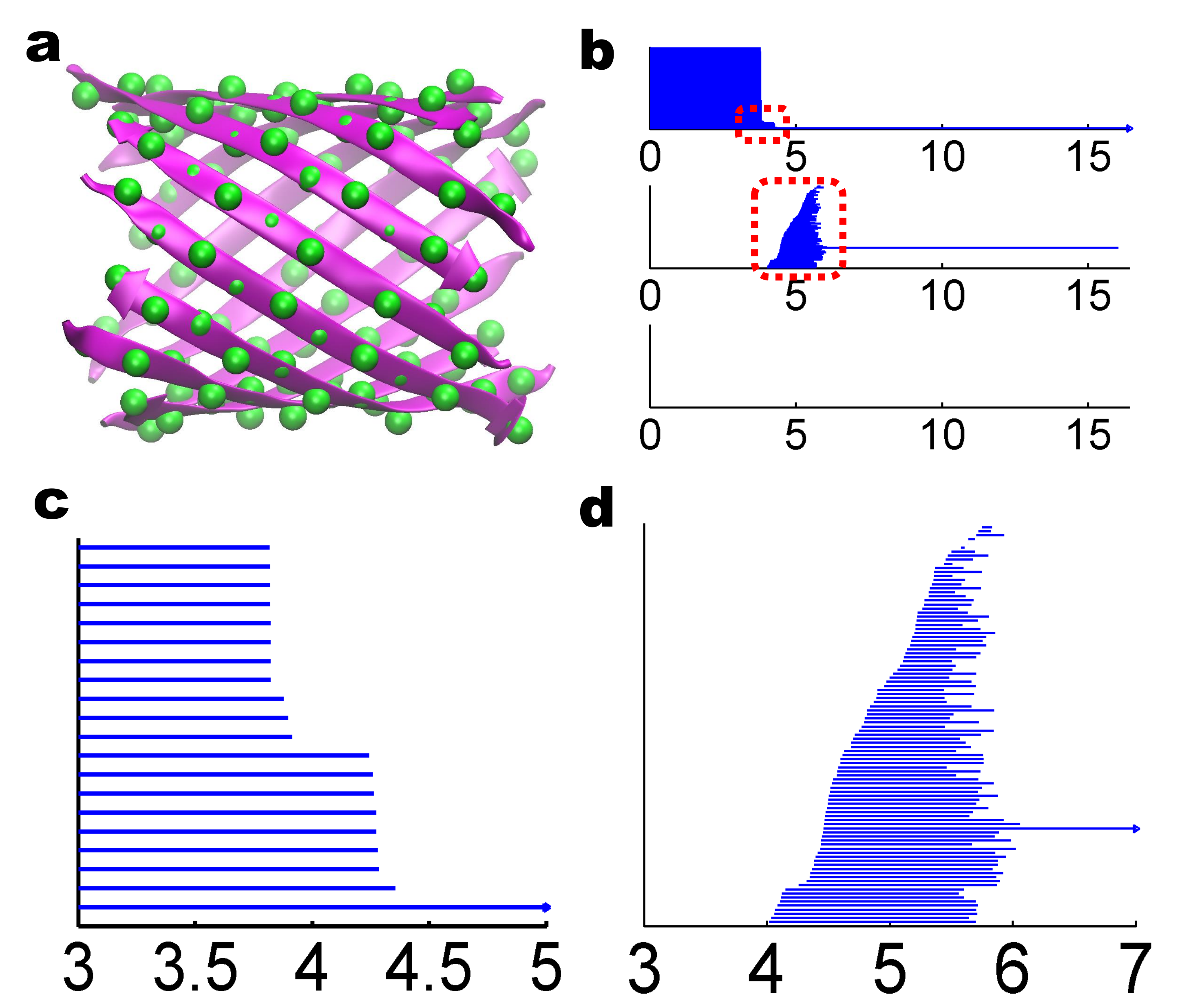}
\end{tabular}
\end{center}
\caption{The topological analysis of the  beta  sheet  structure extracted from beta barrel 2GR8.  The coarse-grained representation is employed with  C$_{\alpha}$ atoms in the green color in {\bf a}. There are 128 atoms organized in 12 beta sheets. The topological fingerprint of the beta sheet structure is depicted in {\bf b}. The red boxes  are zoomed in and demonstrated in {\bf c} and {\bf d}. There are 128 $\beta_0$ bars and 98 $\beta_1$ bars. It is seen from {\bf b} and {\bf c} that in the $\beta_0$ panel, 12 out of 128 bars persist beyond 3.8\AA, which corresponds to 12 isolated beta sheets.  The longest $\beta_1$ bar in {\bf b} is due to the large hole in beta barrel structure.  Other $\beta_1$ bars are formed from every adjacent 4 C$_{\alpha}$ atoms as discussed in the earlier analysis of parallel  beta sheet structure.  Due to the mismatch, adjacent two sheets contribute around 8 $\beta_1$ bars, which accounts for $12\times8=96$ short-lived $\beta_1$ bars. Together  with the bar for the global intrinsic ring, it is estimated that there are  97 $\beta_1$ bars. Although  computational result shows    98 $\beta_1$ bars,  one  of them (the fifth from top) is barely visible.   
}
\label{fig:BetaBarrel_sheet}
\end{figure}

The persistent homology analysis of the beta-sheet structure is shown in Fig. \ref{fig:BetaBarrel_sheet}. It is seen that in the $\beta_0$ panel, we have 12 bars that are longer than 3.8\AA. These bars correspond to 12 isolated beta sheets. In the  $\beta_1$ panel, there is a unique bar that lasts from around  4  to 16 \AA . Obviously, this   $\beta_1$ bar is due to the global hole of the  beta barrel. 

Except the longest $\beta_1$ bar, other  $\beta_1$ bars are generated from every adjacent 4 C$_{\alpha}$ atoms as discovered in the earlier analysis of parallel beta sheet structure. There are  12 near parallel beta sheets. Due to the mismatch in the structure as shown Fig. \ref{fig:BetaBarrel_sheet}({\bf a}) in detail, adjacent two sheets only contribute around 8 $\beta_1$ bars, which gives rise to $12\times8=96$ short-lived $\beta_1$ bars. Additionally, the global circle gives rise to another $\beta_1$ bar. Therefore,  we predict 97 bars. The persistent homology calculation shows 98 $\beta_1$ bars. However, one of these bars has an extremely short life and is hardly visible. Therefore,  there is very good consistency between our analysis and numerical computation.

\subsection{Persistent homology analysis of protein flexibility}

Rigidity and flexibility are part of protein functions. Theoretically, protein flexibility and rigidity can be studied based on fundamental laws of physics, such as molecular mechanics \cite{McCammon:1977}. However, the atomic representation and long time integration involve an excessively large number of degrees of freedom. To avoid this problem, normal mode based models \cite{Go:1983,Tasumi:1982,Brooks:1983,Levitt:1985}, such as elastic network model (ENM)  \cite{Tirion:1996}, anisotropic network model (ANM)  \cite{Atilgan:2001}, and Gaussian network model \cite{Flory:1976, Bahar:1997,Bahar:1998}  have been proposed. Combined with the coarse-grained representation, they are able to  access the flexibility of macromolecules or protein complexes.

Recently, we have proposed the molecular nonlinear dynamic (MND) model \cite{KLXia:2014b} and flexibility-rigidity index (FRI) theory \cite{KLXia:2013d} to analyze  protein flexibility. The fundamental assumption of our methods is that, protein structures are uniquely determined by various internal and external interactions, while the protein functions, such as stability and flexibility, are solely determined by the structure. Based on this assumption, we introduce a key element, the geometry to topology mapping to obtain protein topological connectivity from its geometry information. Furthermore, a correlation matrix is built up to transform the geometric information into functional relations.

In this section, we  provide a persistent homology analysis of protein flexibility.  We present a brief review to a few  techniques that are utilized in the present flexibility analysis.    Among them,  MND and FRI not only offer protein flexibility analysis, but also provide  correlation matrix based 
filtration for the persistent homology analysis of proteins.  Our results unveil the topology-function relationship of proteins. 

\subsubsection{Protein flexibility analysis}\label{sec:ExistingMethods}

\paragraph{Normal mode analysis}
Due to the limitation of computation power, MD or even coarse-grained MD sometimes falls short in the simulation of protein's real time dynamics, especially for macro-proteins or protein complexes, which have gigantic size and long-time-scale motions. However, if protein's relative flexibility and structure-encoded collective dynamics are of the major concern, MD simulation can be replaced by normal mode analysis, the related methods includes elastic network model, anisotropic network model, Gaussian network model, etc. In these methods, pseudo-bonds/pseudo-springs are used to connect atoms within certain cutoff distance. The harmonic potential induced by the pseudo-bond/pseudo-spring network dominates the motion of the protein near its equilibrium state. Through  low order eigenmodes obtained from diagonalizations of the  Hessian matrix of the interaction potential,  structure-encoded collective motion can be predicted along with the relative flexibility of the protein, which is measured experimentally by Debye-Waller factors. A more detailed description is available from review literature \cite{JMa:2005,LWYang:2008,Skjaven:2009,QCui:2010}.

\paragraph{Molecular nonlinear dynamics}
The key element in our MND model is the geometry to topology mapping  \cite{KLXia:2013d,KLXia:2014b}. Specifically,  we denote the coordinates of atoms in the molecule studied as ${\bf r}_{1}, {\bf r}_{2}, \cdots, {\bf r}_{j}, \cdots, {\bf r}_{N}$, where ${\bf r}_{j}\in \mathbb{R}^{3}$ is the position vector of the $j$th atom.  The Euclidean distance between $i$th and $j$th atom $r_{ij}$ can   be calculated. Based on these distances, topological connectivity matrix can be constructed with monotonically decreasing radial basis functions. The general form is,
\begin{eqnarray}\label{eq:couple_matrix0}
{C}_{ij} = w_{ij} \Phi( r_{ij},\eta_{ij}),   \quad i \neq j,
\end{eqnarray}
where $w_{ij}$ is associated with atomic types,  parameter $\eta_{ij}$ is the atom-type related characteristic distance, and $\Phi( r_{ij},\eta_{ij}) $ is a radial basis correlation kernel. 

The kernel definition is of great importance to the FRI model. From our previous experience,  both  exponential type and Lorentz type of kernels are  very efficient. A generalized exponential kernel has the form
\begin{eqnarray}\label{eq:ExpKernel}
\Phi(r,\eta) =    e^{-\left(r/\eta\right)^\kappa},    \quad \kappa >0
\end{eqnarray}
and the Lorentz type of kernels is 
\begin{eqnarray}\label{eq:PowerKernel}
 \Phi(r, \eta) =
 \frac{1}{1+ \left(r/\eta \right)^{\upsilon}}.
  \quad  \upsilon >0.
 \end{eqnarray}
The parameters $\kappa$, $\upsilon$, and $\eta$ are adjustable. We usually search over a certain reasonable range of parameters to find the best fitting result by comparing with  experimental B-factors.

Under physiological conditions, proteins experience ever-lasting thermal fluctuations. Although the whole molecule can have certain collective motions, each particle in a protein has its own  dynamics. It is speculated that each particle in a protein can be viewed as a nonlinear oscillator and its dynamics can be represented by a nonlinear equation \cite{KLXia:2014b}. The interactions between particles are represented by the correlation matrix \ref{eq:couple_matrix0}. Therefore, for the whole protein of  $N$ particles, we set a nonlinear dynamical system  as \cite{KLXia:2014b}
\begin{eqnarray}\label{eq:couple_matrix}
\frac{d{\bf u}}{dt} &=& {\bf F}({\bf u})+ {\bf E}{\bf u}, ~~~
\end{eqnarray}
where ${\bf u}=({\bf u}_1,{\bf u}_2,\cdots, {\bf u}_j,\cdots, {\bf u}_N )^T$ is an array of state functions for $N$ nonlinear oscillators ($T$ denotes the transpose), ${\bf u}_j=(u_{j1},u_{j2}, \cdots, u_{jn})^T$ is an $n$-dimensional nonlinear function for the $j$th oscillator,   $ {\bf F}({\bf u})=(f({\bf u}_1),f({\bf u}_2), \cdots, f({\bf u}_N))^T$ is an array of nonlinear functions of $N$  oscillators, and
\begin{eqnarray}
 {\bf E}=\varepsilon{\bf C}\otimes \Gamma.
\end{eqnarray}
Here,  $\varepsilon$ is the overall  coupling strength,  ${\bf C}=\{C_{ij} \}_{i,j=1,2,\cdots,N}$ is  { an $N\times N$ correlation matrix, and $\Gamma$ is an $n\times n$ linking matrix.}

It is found that, the transverse stability of the MND system gradually increases during the protein folding from disorder conformations to their well-defined natural structure. The interaction among particles leads to collective motions in a protein. The stronger the interaction is, the more unified dynamics will be. Eventually, the chaotic system assumes an intrinsically low dimensional manifold (ILDM) when the final folded state is reached, which indicates that protein folding tames chaos. To predict protein  Debye-Waller factors, we introduced a transverse perturbation to the dynamics of each particle and record the  relaxation time  defined as the time used to recover the original state within a factor of $1/e$,   which  measures the strength of  particle-particle and particle-environment
interactions. Therefore, the  relaxation time is associated with protein flexibility. This method has also been successfully applied to the prediction of Debye-Waller factors  \cite{KLXia:2013d}.

\paragraph{Flexibility rigidity index}
The FRI theory is very  simple. It can be directly derived from the correlation matrix.  Basically, we define the atomic  rigidity index  $\mu_i$ as \cite{KLXia:2013d}
\begin{eqnarray}\label{eq:rigidity1}
 \mu_i = \sum_{j}^N w_{ij} \Phi( r_{ij} ,\eta_{ij} ), \quad \forall i =1,2,\cdots,N.
\end{eqnarray}
The physical interpretation is straightforward. The stronger connectivity an atom has, the more rigid it becomes. After summarizing over all atoms, one arrives at the averaged molecular rigidity index (MRI),
\begin{eqnarray}\label{eq:Averagerigidity}
 \bar{\mu}_{\rm MRI} = \frac{1}{N}\sum_{i=1}^N {\mu}_{i}.
\end{eqnarray}
It has been pointed out that this index is related to molecular thermal stability, compressibility and  bulk modulus  \cite{KLXia:2013d}.

We also defined the atomic flexibility index as the inverse of the atomic rigidity index,
\begin{eqnarray}\label{eq:flexibility1}
f_i= \frac{1}{\mu_i}, \quad \forall i =1,2,\cdots,N.
\end{eqnarray}
An averaged molecular flexibility index (MFI) can be similar defined as the averaged molecular rigidity index
\begin{eqnarray}\label{eq:Averageflexibility}
 \bar{f}_{\rm MFI} = \frac{1}{N}\sum_{i=1}^N {f}_{i}.
\end{eqnarray}
 {We   set $\eta_{ij}=\eta$ and $w_{ij}=1$ for a CG model with one type of atoms}. 

The atomic rigidity index of each particle in a protein is associated with the particle's flexibility.   The FRI theory has been intensively validated by comparing with the experimental data, especially the Debye-Waller factors   \cite{KLXia:2013d}.  Although it is very simple, its application to B-factor prediction yields excellent results. The predicted results are proved to be very accurate and this method is highly efficient. FRI can also be used to analyze the protein folding behavior. 

\subsubsection{Topology-function relationship of protein flexibility}\label{Sec:PH_cutoff}

Topological connectivity is employed in the elastic network models, or GNM for protein flexibility analysis. To this end, a cutoff distance $r_c$ is often used to specify the spatial range of the connectivity in the protein elastic network. Each atom is assumed to be connected with all of its neighbor atoms within the designed cutoff distance by pseudo-springs or pseudo-bonds. Whereas atoms beyond the cutoff distance are simply ignored.  In contrast, our MND model and FRI theory do not use a cutoff distance, but utilize a characteristic distance $\eta$ to weight the distance effect in the geometry to topology mapping as shown in Eqs. (\ref{eq:ExpKernel}) and   (\ref{eq:PowerKernel}).  Atoms within the characteristic distance are assigned with relatively larger weights in the correlation matrix. It has been noted that the optimal cutoff distance can varies from protein to protein and from method to method.  For a given method, an optimal cutoff distance can be obtained by statistically averaging over a large number of proteins.  Such optimal cutoff distance is about 7 to 8\AA~ for GNM and around 13 to 15\AA~ for ANM \cite{YangLei:2009}. No rigorous analysis or explanation has been given for optimal cutoff distances. 
	
In this section, we explore the topology-function relationship of proteins. First, we present a persistent homology interpretation   of optimal cutoff distances in GNM. Additionally, we also provide a quantitative  prediction of optimal characteristic distances in MND and FRI based on persistent homology. To this end, we develop a new cutoff distance based filtration method to transfer a protein elastic network into a simplicial complex at each cutoff distance. The resulting patterns of topological invariants  shed light on the existence of optimal cutoff distances. We propose a new persistent homology based physical model to predict optimal characteristic distances in MND and FRI. 

\paragraph{Persistent homology analysis of optimal cutoff distance}
\begin{figure}
\begin{center}
\begin{tabular}{c}
\includegraphics[width=0.7\textwidth]{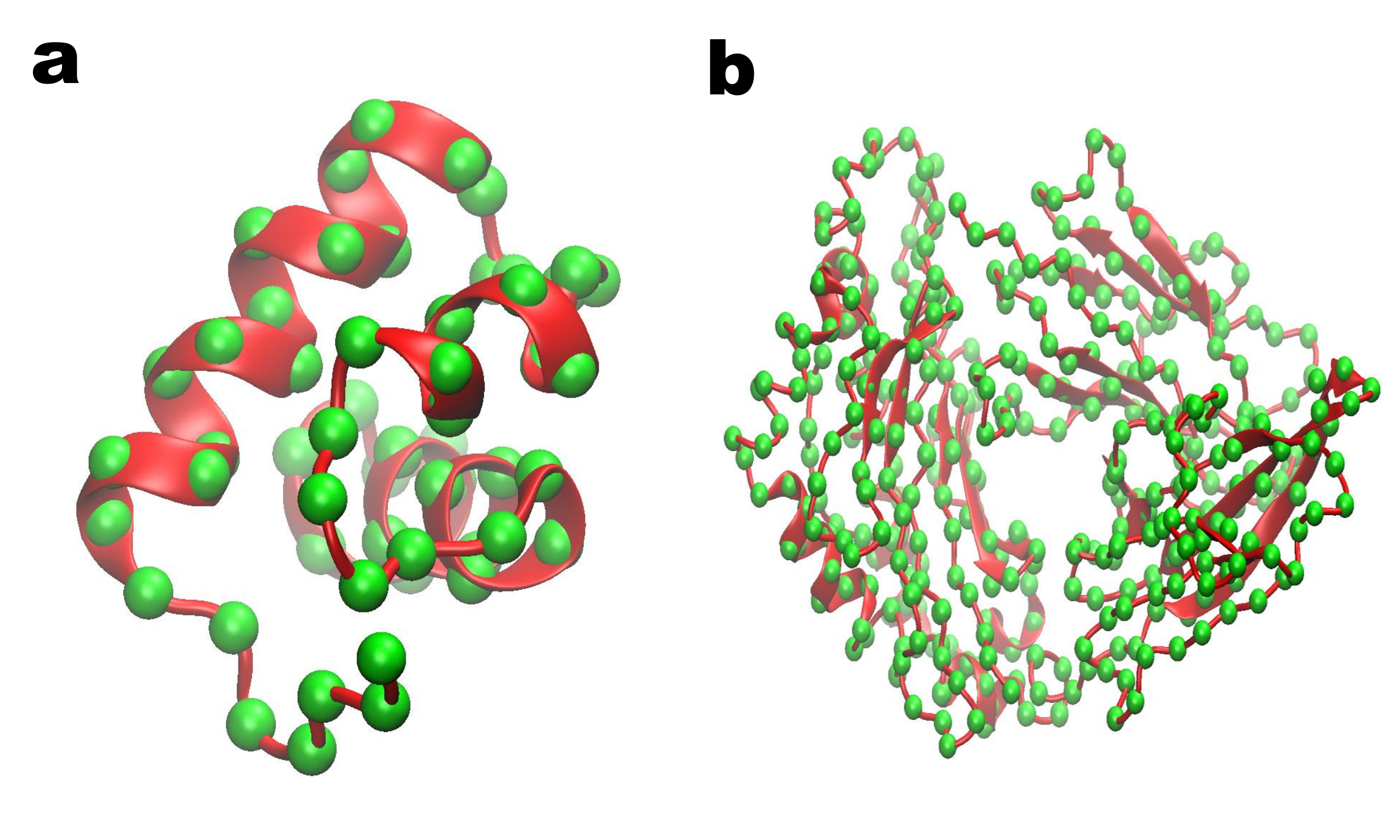}
\end{tabular}
\end{center}
\caption{Illustration of  proteins 1GVD (left chart) and 3MRE (right chart)  used in analyzing the optimal cutoff distance of the Gaussian
network model.  The coarse-grained model is employed with residues represented by their C$_{\alpha}$ atoms and displayed  as green balls.}
\label{fig:BetaBarrel2}
\end{figure}

Protein elastic network models usually employ the coarse-grained representation and do not distinguish between different residues.   We assume that the total number of  C$_{\alpha}$ atoms in the protein is $N$, and the distance between $i$th and $j$th C$_{\alpha}$ atoms is $d_{ij}$. To analysis the topological properties of protein elastic networks, we propose a new distance matrix ${\bf D} = \{d_{ij}|i=1,2,\cdots,N; j=1,2,\cdots,N\}$
\begin{eqnarray}\label{eq:rigidity12}
 d_{ij} = \begin{cases}
     d_{ij}, ~ d_{ij}\leq r_c; \\
     d_\infty, ~ d_{ij}>r_c,
\end{cases}
\end{eqnarray}
where $d_\infty$ is a sufficiently large value which is much larger than the final filtration size and  $r_c$ is a given cutoff distance. Here  $d_\infty$ is chosen to ensure  that atoms beyond the cutoff distance $r_c$ will never form any high order simplicial complex during the filtration process.  Consequently, the resulting persistent homology shares the same topological connectivity property with the elastic network model. With the barcode representation of topological invariants, the proposed persistent homology analysis gives rise to an effective visualization of topological connectivity.

\begin{figure}
\begin{center}
\begin{tabular}{cc}
\includegraphics[width=0.6\textwidth]{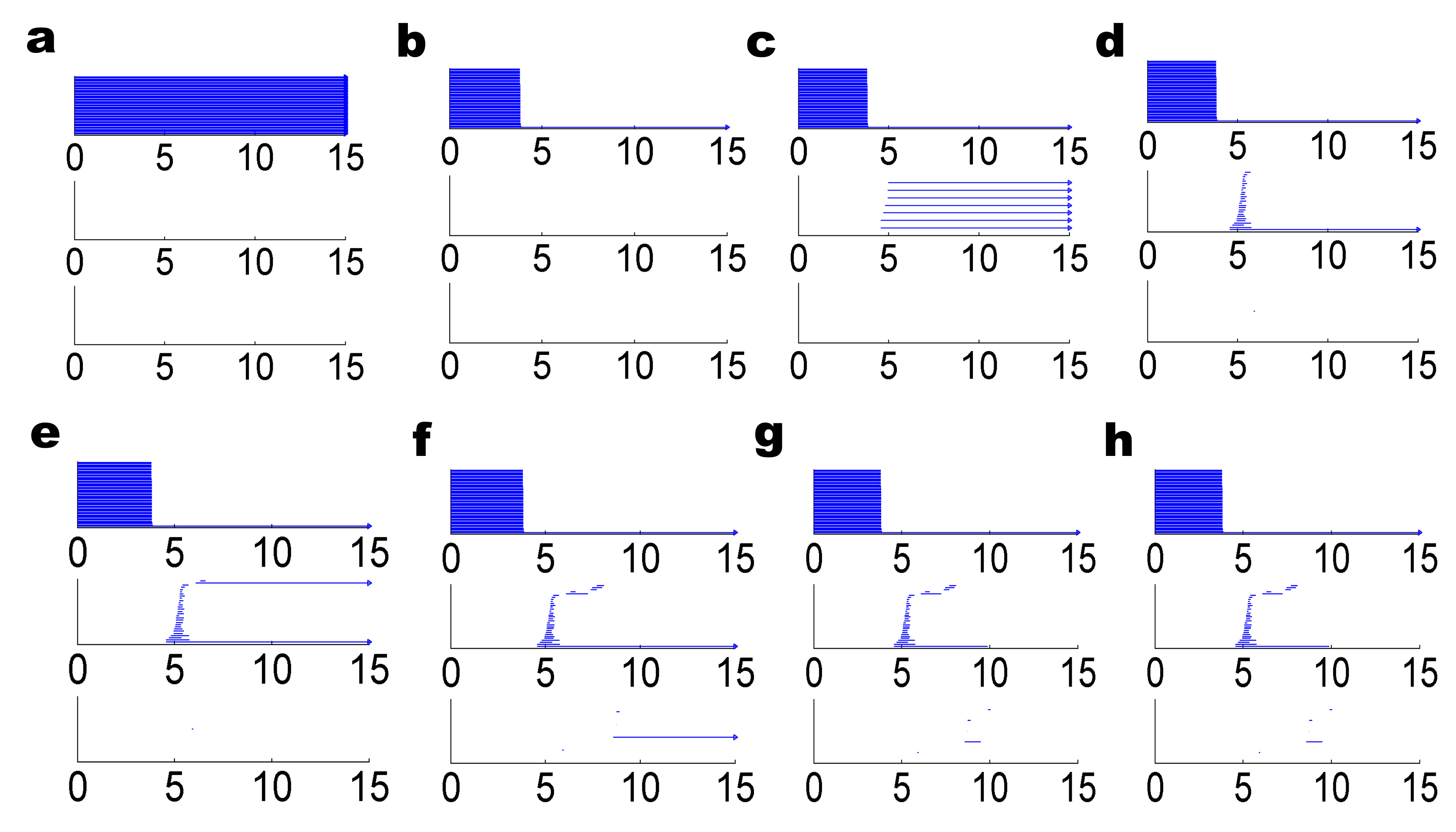} &
\includegraphics[width=0.4\textwidth]{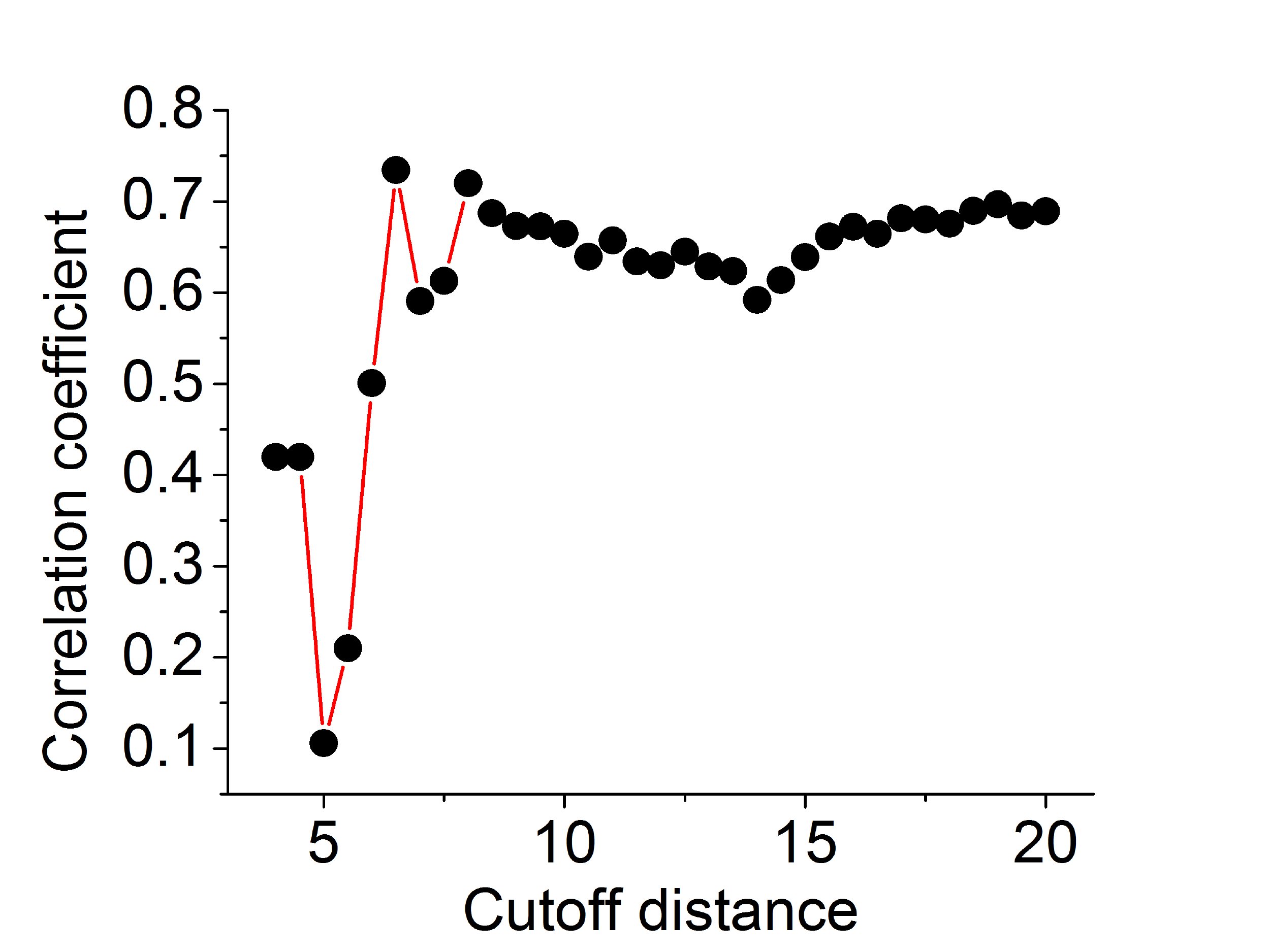}
\end{tabular}
\end{center}
\caption{Visualization of topological connectivity and  optimal cutoff distance of Gaussian network model for protein 1GVD.  The left charts from {\bf a} to {\bf h} 
are  barcodes generated based on the filtration given in Eq. (\ref{eq:rigidity12}).  From {\bf a} to {\bf h}, the cutoff distances used are 3\AA, 4\AA, 5\AA, 6\AA, 7\AA, 9\AA, 11\AA, and 13\AA,  respectively. The right chart is the correlation coefficient obtained from the  Gaussian network model under different cutoff distance (\AA).
}
\label{fig:cutoff_1GVD}
\end{figure}

\begin{figure}
\begin{center}
\begin{tabular}{cc}
\includegraphics[width=0.6\textwidth]{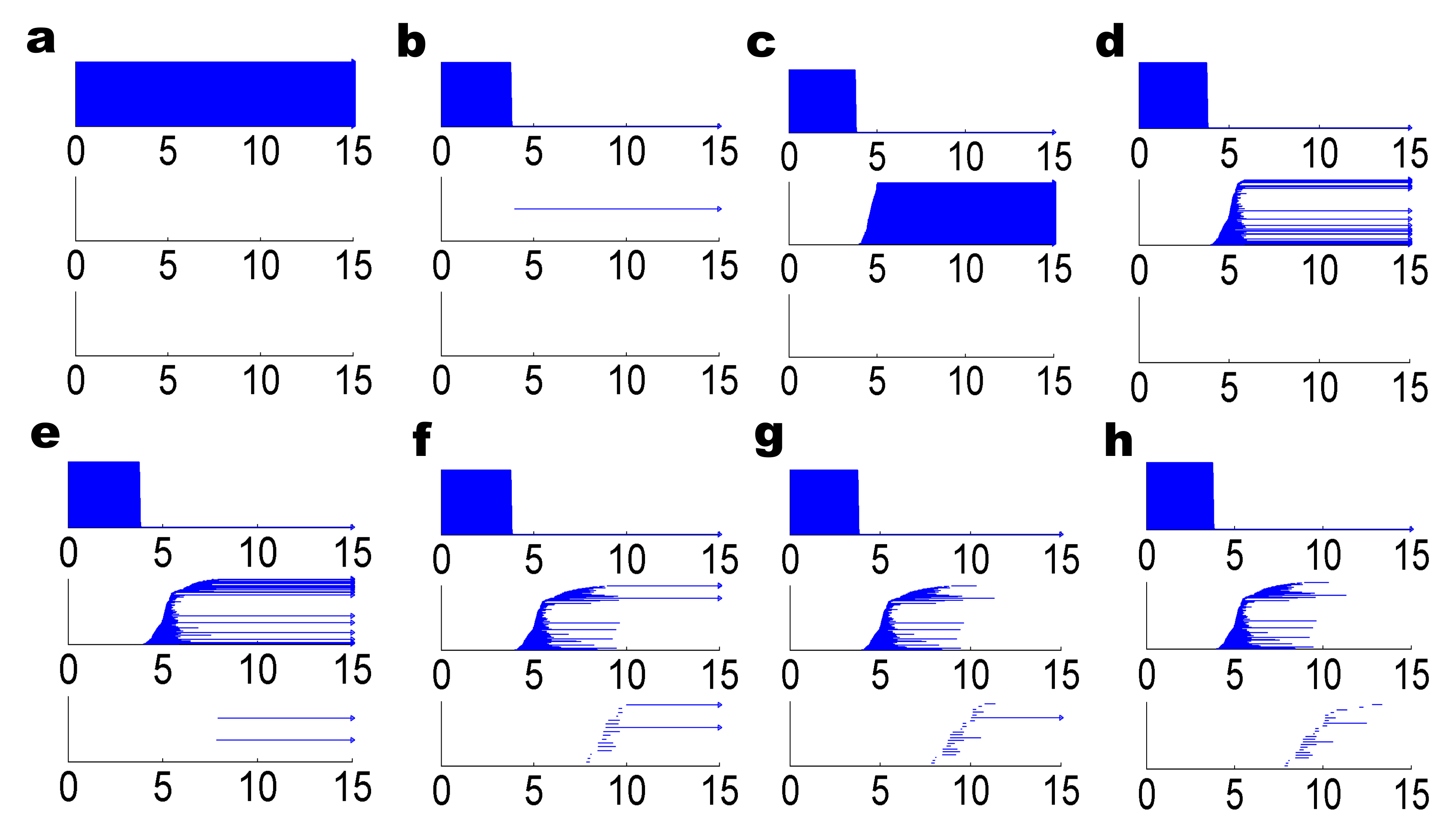}&
\includegraphics[width=0.4\textwidth]{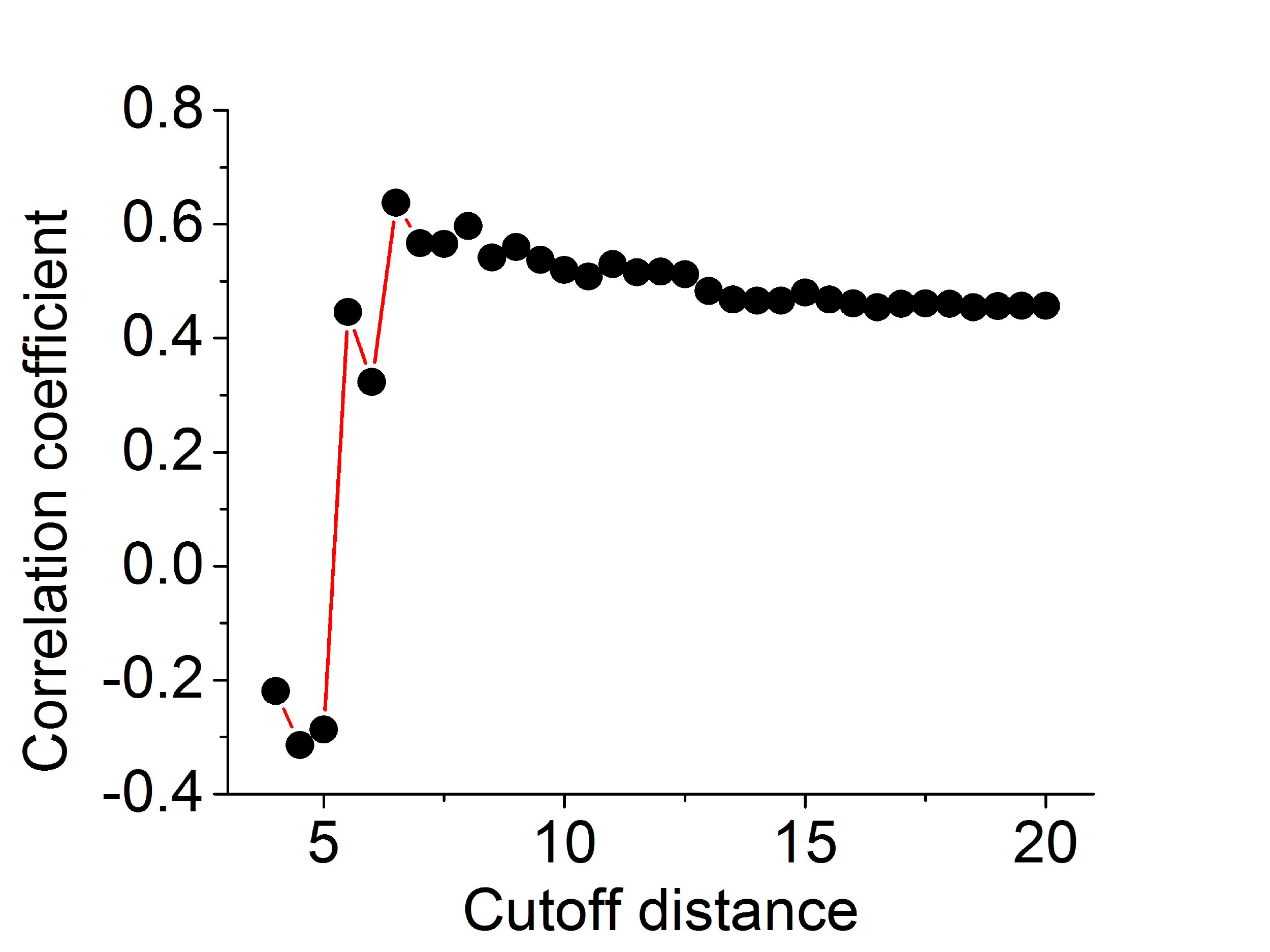}
\end{tabular}
\end{center}
\caption{Visualization of topological connectivity and  optimal cutoff distance of Gaussian network model for protein 3MRE.  The left charts from {\bf a} to {\bf h} are the   barcodes generated based on the filtration given in Eq. (\ref{eq:rigidity12}). From {\bf a} to {\bf h}, the cutoff distances used are 3\AA, 4\AA, 5\AA, 6\AA, 8\AA, 10\AA, 12\AA, and 14\AA, ~ respectively.  The right chart is the correlation coefficient obtained from the  Gaussian network model under different cutoff distance (\AA). }
\label{fig:cutoff_3MRE}
\end{figure}

To illustrate our persistent homology  analysis, we consider two proteins, 1GVD and 3MRE,  with 52 and 383 residues respectively, as shown in  Fig. \ref{fig:BetaBarrel2}.    With different cutoff distances, both the constructed elastic networks and persistent homology simplicial complexes demonstrate  dramatically different properties. The resulting persistent homology analysis using the proposed filtration (\ref{eq:rigidity12}) for protein 1GVD and 3MRE are illustrated in Figs. \ref{fig:cutoff_1GVD} and \ref{fig:cutoff_3MRE}, respectively. It can be seen that when the cutoff distance is 3\AA, all C$_{\alpha}$ atoms are isolated from each other, and $\beta_0$ bars persist forever. This happens because the average distance between two adjacent C$_{\alpha}$ atoms is about 3.8\AA, as discussed earlier. This particular distance also explains the filtration results in Figs. \ref{fig:cutoff_1GVD}{\bf b} and  \ref{fig:cutoff_3MRE}{\bf b} at  $r_c=4$ \AA. As the adjacent C$_{\alpha}$ atoms connect with each other, only a single $\beta_0$ bar survives. However, no nontrivial complex is formed at $r_c=4$ \AA. When the cutoff distance increases to 5\AA, a large  number of $\beta_1$ bars is produced and persists beyond the filtration size limit. Topologically, this means that almost all the generated loops never disappear during   the filtration. This happens because we artificially isolate atoms with distance larger than $r_c=5$\AA~ in our filtration matrix (\ref{eq:rigidity12}). Physically, without the consideration of  long distance interactions, local structural effects are over-amplified in this setting.  With the increase of the cutoff distance $r_c$, the number of  persistent $\beta_1$ bars drops. When $r_c$ is set  to about 12 \AA, almost no persistent $\beta_1$ bar can be found for both 1GVD and 3MRE. It should also note that for  a small protein like 1GVD, the number of persistent $\beta_1$ bars falls quickly. While a larger protein with a larger  number of $\beta_1$ bars, the reduction in the number of ring structures  is relatively slow. However, the further increase of the $r_c$ value does not change the persistent homology behavior anymore because all significant geometric features (i.e., isolated components, circles, voids, and holes) are already captured by the existing network at about $r_c \approx 12$\AA.

To understand the   physical impact of the topological connectivity found by persistent homology analysis, we analyze GNM predictions of protein B-factors at various cutoff distances. The accuracy of the GNM predictions is    quantitatively accessed by correlation coefficient (CC)
\begin{eqnarray}\label{correlation}
  {\rm CC}=\frac{\Sigma^N_{i=1}\left(B^e_i-\bar{B}^e \right)\left( B^t_i-\bar{B}^t \right)}
   { \left[\Sigma^N_{i=1}(B^e_i- \bar{B}^e)^2\Sigma^N_{i=1}(B^t_i-\bar{B}^t)^2\right]^{1/2}},
\end{eqnarray}
where $\{B^t_i,  i=1,2,\cdots,N\}$ are a set of predicted B-factors by using the proposed method and $\{B^e_i, i=1,2,\cdots, N\}$ are a set of experimental B-factors extracted from the PDB file. Here $\bar{B}^t$ and $\bar{B}^e$ the statistical averages of theoretical and experimental B-factors, respectively. Expression (\ref{correlation}) is used for the correlation analysis of other theoretical predictions as well.

We plot the CC with respect to the cutoff distance in the right charts of Figs. \ref{fig:cutoff_1GVD} and  \ref{fig:cutoff_3MRE} for proteins 1GVD and 3MRE, respectively. For both cases, when the cutoff distance is small than 3.8\AA,~ no CC is obtained because pseudo-spring/pseudo-bond is not constructed and GNM is not properly set. As $r_c$ increases to around 4\AA, we acquire  relatively small CC values. Further increase  of $r_c$ will lower  CC values until it reaches the bottom at around $r_c=5$\AA, where, as discussed in our persistent homology analysis, the influence of the local topological connectivity is over-estimated. Once the cutoff distance is larger than 5\AA, the CC values begin to increase dramatically until it reach the peak value around $r_c\approx 7$\AA, where the impact of local connectivity and global connectivity reaches an optimal balance.  The CC values fluctuate as the cutoff distance increases further, due to improper    balances in  local connectivity and global connectivity.

From the above analysis, it can seen that if the cutoff distance used in elastic network models is smaller than 5\AA, the constructed network is over-simplified without any global connectivity contribution. At the same time, if the cutoff distance is larger than 14\AA,  excessive global connections are included, which leads to a reduction in prediction accuracy,  especially for small proteins. To be specific, by excessive connections we mean that a given atom is connected by using elastic springs with too many remote atoms.  Even in the range of 6 to 14\AA,~ there  is always a tradeoff between excessive connection in certain part of the protein and lack of connection in the other regions. Also the relative size and the intrinsic topological properties of a protein should be considered when choosing the optimal cutoff distance. Although the selection of the optimal cutoff distance is complicated by many issues \cite{YangLei:2009}, one can make a suitable choice by the proposed persistent homology analysis. For instance, in the above two cases, when the cutoff distance is around 7 to 9\AA, the major global features in $\beta_1$ panel have already emerged and thus the selection of $r_c=7$\AA~to $r_c=9$\AA~ will generate a reasonable prediction. Therefore, our persistent homology analysis explains the optimal range of cutoff distances (i.e., $r_c=7$\AA~ to $8$\AA~) for GNM in the literature \cite{YangLei:2009} very well.

\paragraph{Persistent homology prediction of optimal characteristic distance}

\begin{figure}
\begin{center}
\begin{tabular}{c}
\includegraphics[width=0.7\textwidth]{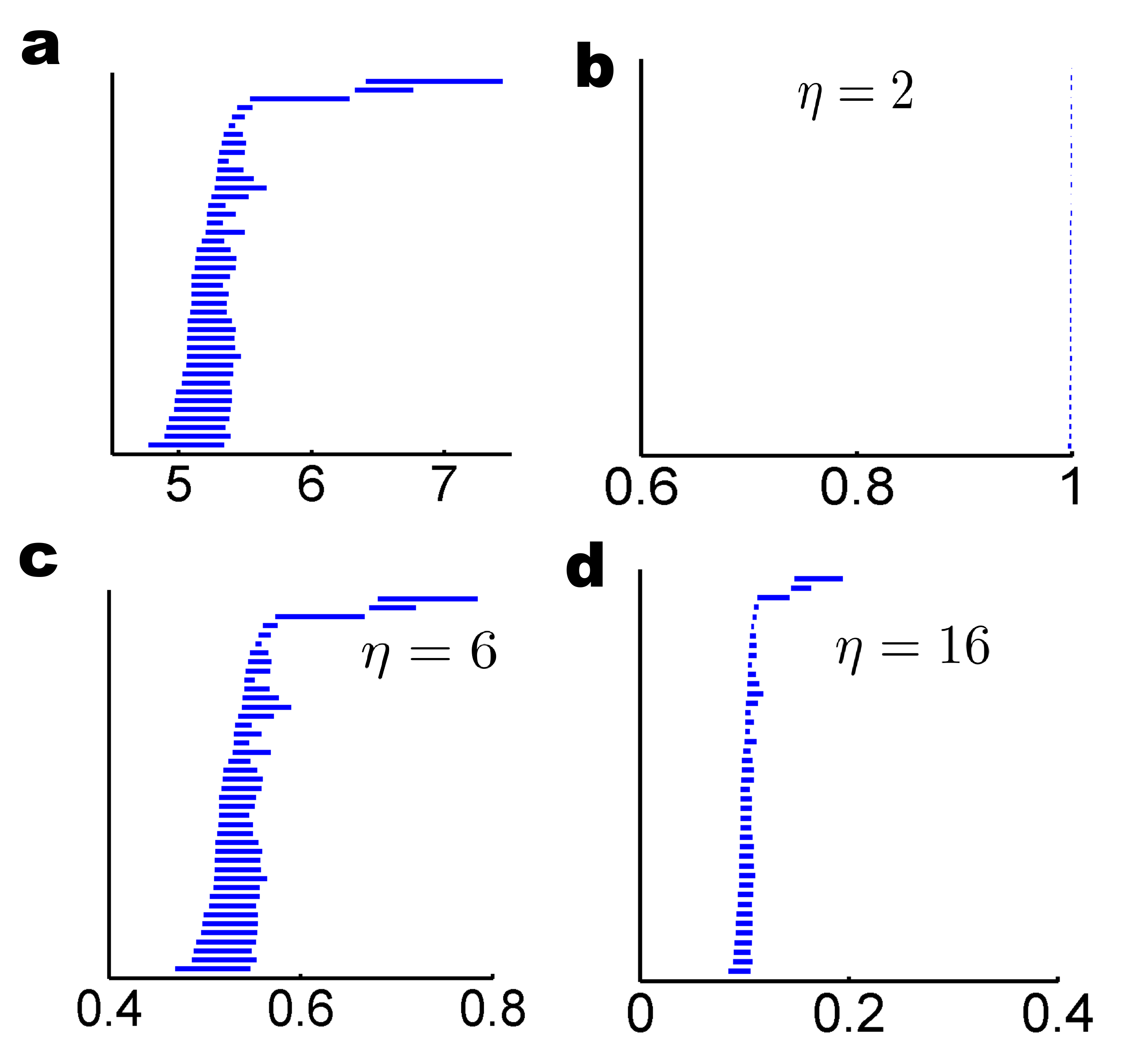}
\end{tabular}
\end{center}
\caption{Comparison of  $\beta_1$ behaviors in different filtration settings for protein  1YZM $C_\alpha$   point cloud data.  
Distance based filtration is shown in {\bf a}.  The correlation matrix based filtration with exponential kernel ($\kappa=2$)  is used in {\bf b}, {\bf c}  and {\bf d}.  The $\eta$ is chosen to be 2\AA, 6\AA~ and 16\AA ~ in  {\bf b}, {\bf c} and {\bf d}, respectively.  
The $\beta_1$ bar patterns   {\bf a}, {\bf c} and {\bf d} are very similar but have  different durations.
The $\beta_1$ bar pattern  in  {\bf b} differs much from the rest due to a small characteristic distance $\eta=2$\AA. 
 }
\label{fig:1YZMCompare}
\end{figure}

Unlike elastic network models which utilize  a  cutoff distance, the  MND method and  FRI theory employ   a characteristic distance $\eta$ in their  correlation kernel. 
The characteristic distance has a direct impact in the accuracy of protein B-factor prediction. 
Similar to the cutoff distance in GNM,  the optimal characteristic distance varies from protein to protein, although an optimal value can be found based on a statistical average over hundreds of proteins \cite{KLXia:2013d}.
In this section, we propose a persistent homology model to predict the optimal characteristic distance. 

Appropriate filtration process is of crucial importance to  the persistent homology analysis. To accurately predict optimal characteristic distance for MND and FRI models, we introduce a new filtration matrix $\{M_{ij}|i=1,2,\cdots,N; j=1,2,\cdots,N\}$  based on a modification of the correlation matrix (\ref{eq:couple_matrix0}) of MND and FRI 
\begin{eqnarray}\label{eq:couple_matrix01}
{M}_{ij} = \begin{cases}
      1-\Phi( r_{ij},\eta_{ij}),   \quad i \neq j, \\
      0, \quad i=j,
\end{cases}
\end{eqnarray}
where $0 \leq \Phi( r_{ij},\eta_{ij})\leq 1$ is defined in Eqs. (\ref{eq:ExpKernel}) or (\ref{eq:PowerKernel}). To avoid confusion, we simply use the exponential kernel with parameter $\kappa=2$ in the present work.  The visualization of  this new correlation matrix for fullerence C$_{70}$ is given in Fig. \ref{fig:FiltrationMatrix}.  

As the filtration continues, atoms with shorter distances and thus lower $M_{ij}$ values will form higher complexes first just like situation in the distance based filtration. However, when characteristic distance varies, the formation of simplicial complex or topological connectivity changes too. To illustrate this point, we use protein 1YZM as an example. The related persistent connectivity patterns in term of $\beta_1$ are depicted in Fig. \ref{fig:1YZMCompare}. From the  analysis of these topological invariants,  several interesting observations can be made. First, it can be seen that our persistent homology results obtained with $\eta=2$\AA, 6\AA, and $16$ \AA~ in  {\bf b}, {\bf c}, and {\bf d} share certain similarity with the $\beta_1$ pattern of the distance based filtration in {\bf a}. This similarity is most obvious when $\eta$ values are 6\AA~ and 16 \AA ~as shown in {\bf c} and {\bf d}. Second,  results differ much  in their  scales or persistent durations. The $\beta_1$ bars in {\bf b}, {\bf c} and {\bf d}  are located  around regions [0.996, 1], [0.4,0.8] and [0.05,0.2], respectively. Third, the global behavior is  captured in all cases and the local connectivity is not over-emphasized. This aspect  is different from that of the cutoff distance based filtration discussed previously and  implies the robustness of  the correlation matrix based filtration.  However, some bars are missing for certain $\eta$ values. Specifically,  the $\beta_1$ barcode in Fig. \ref{fig:1YZMCompare}{\bf b} does not have all the bars appeared in other cases, which leads to the  underestimation of certain protein connectivity. 

\begin{figure}
\begin{center}
\begin{tabular}{c}
\includegraphics[width=0.7\textwidth]{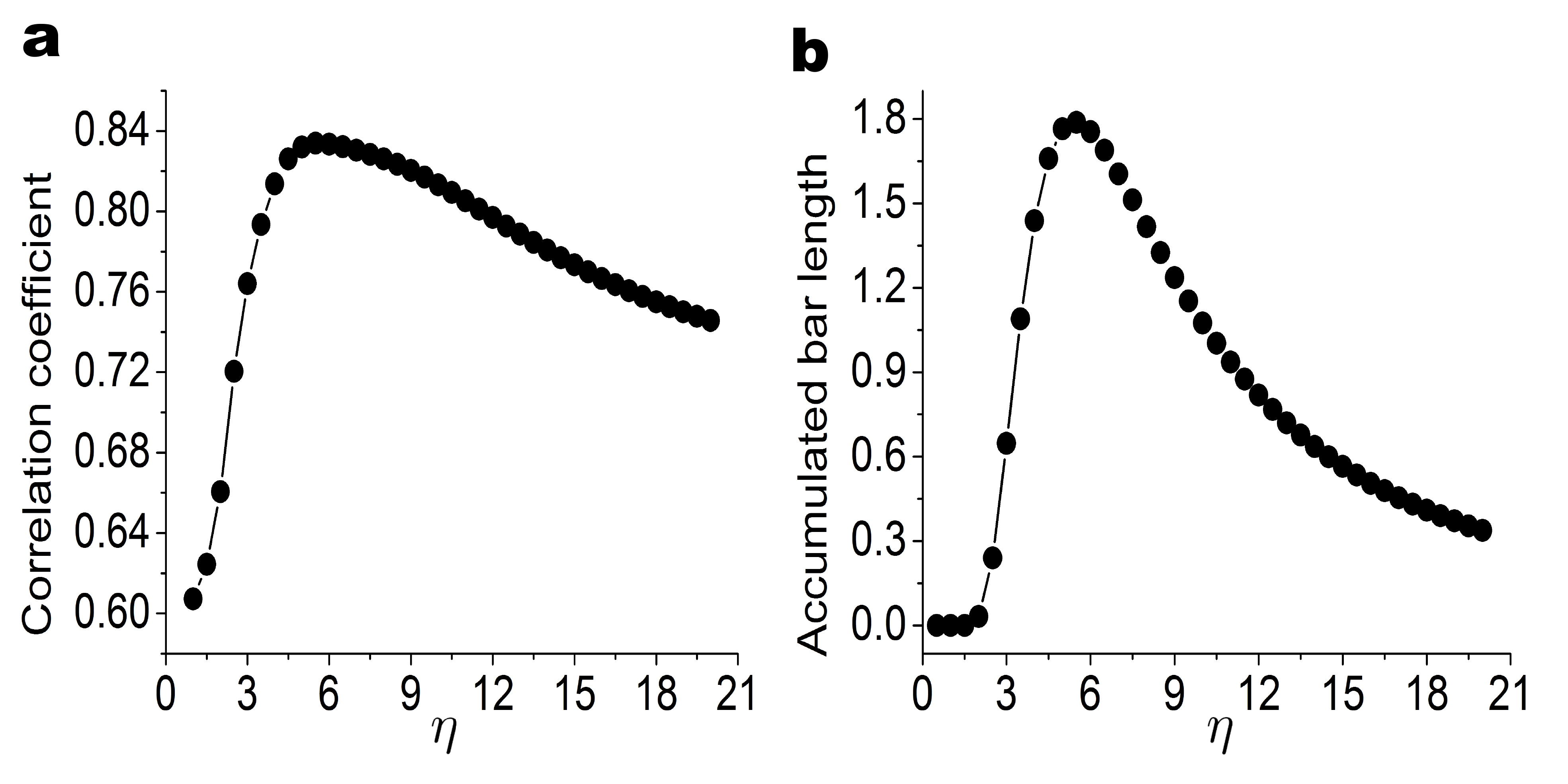}
\end{tabular}
\end{center}
\caption{The comparison between the correlation coefficient from the B-factor prediction by FRI (left chart) and accumulated bar length from persistent homology modeling (right chart) under various $\eta$ values in \AA~ for protein 1TZM. It is seen that correlation coefficient and accumulated bar length share a similar shape that their  values increase dramatically at first and then gradually decrease. The common  maximum near $\eta=6.0$\AA~ indicates the ability of persistent homology for the prediction of optimal characteristic distance.}
\label{fig:1YZMbetti}
\end{figure}

To quantitatively analyze protein connectivity and predict optimal characteristic distance, we propose a physical model based on persistent homology analysis. 
 We define  accumulation bar lengths $A_j$ as the summation of lengths of all  the bars for $\beta_j$,
\begin{equation}\label{AccumulationIndex}
A_j=\sum_{i=1}  L^{\beta_j}_{i}, ~j=0,1,2,
\end{equation}
where $L^{\beta_j}_{i}$ is the length of the $i$th bar of the $j$-th Betti number. To reveal the influence of the characteristic distance, we compute the accumulation bar length $A_1$ over a wide range of filtration parameter $\eta$. We vary the value of $\eta$ from 1\AA~ to 21\AA,~ for protein 1YZM  and compare the accumulated bar length $A_1$ with the CC values obtained with FRI over the same range of $\eta$. The exponential kernel with parameter $\kappa=2$ is used in both the FRI method and our persistent homology model. Results are displayed in  Fig. \ref{fig:1YZMbetti}.  It can be seen from the figure that two approaches  share the same general trend in their behavior as $\eta$ is increased.  Specifically,  when $\eta$ is less than 3\AA,~ the correlation coefficient is much lower than other values. Topologically, this is directly related to the absence of the certain bars in the persistent barcode as depicted in Fig. \ref{fig:1YZMbetti}{\bf b}. With the increase of $\eta$, both CC and $A_1$ reach their maximum around $\eta=6$ \AA. The further increase of $\eta$ leads to the decrease of both CC and $A_1$. 

We now analyze the aforementioned  behavior from the topological point of view.  The role of $\eta$ in the FRI model is to scale the influence of atoms at various  distances. An optimal $\eta$ in the FRI model balances the contributions from local atoms and nonlocal atoms, and offers the best prediction of protein flexibility.   In contrast, parameter $\eta$ in the correlation matrix based filtration is to impact the birth and death of each given $k$-complex. For example, a pair of 2-complexes that do not coexist at a given  cutoff distance in the distance based filtration  might coexist at an appropriate $\eta$ value in the correlation matrix based filtration. A maximum $A_1$  means the largest amount of coexisting 2-complexes (i.e., ring structures) at an appropriate $\eta$ value, which implies protein structural compactness and rigidity. Since the same kernel and the same $\eta$ are used in the FRI model and the persistent homology model (i.e., accumulation bar length), it is natural for the $\eta$ corresponding to the maximum $A_1$ to be the optimal characteristic distance in the FRI prediction. 


\begin{figure}
\begin{center}
\begin{tabular}{c}
\includegraphics[width=0.9\textwidth]{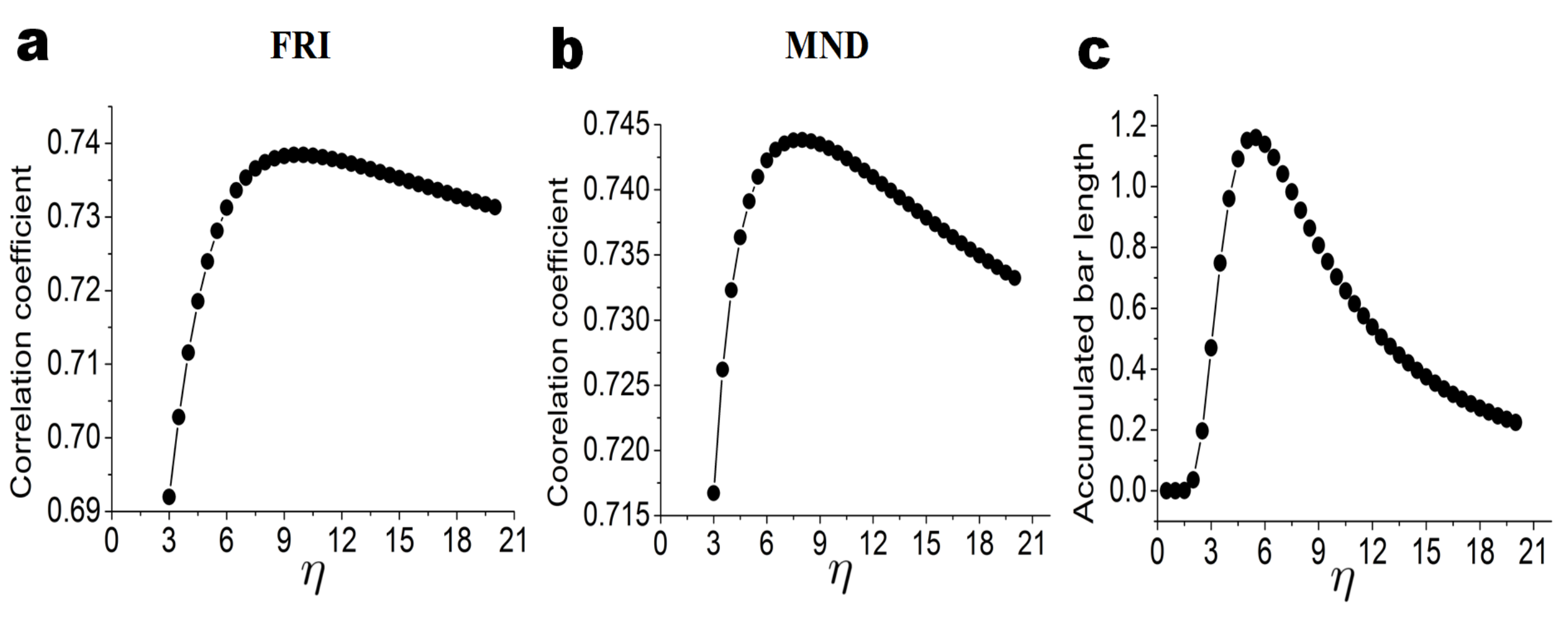}
\end{tabular}
\end{center}
\caption{Comparison between the patterns of average correlation coefficients for   B-factor predictions and the shape of average accumulated bar length for a set of 30 proteins listed in Table \ref{tab:30protein}. The average correlation coefficients obtained from FRI and MND are plotted over a range of characteristic distances $\eta$ (in \AA)  in {\bf a} and {\bf b}, respectively. The average accumulated bar length $A_1$ is shown in  {\bf c}. All patterns share a similar trend. The highest correlation coefficients are  reached around $\eta=$ 9\AA~ and 8\AA~ for MND and FRI, respectively.  While the highest accumulated bar length is found around $\eta=$6\AA. }
\label{fig:30ProteinSet}
\end{figure}

To further validate our topological analysis and prediction,  a set of 30 proteins are chosen and their PDB IDs are listed in Table \ref{tab:30protein}.  Two methods, namely FRI and MND, are employed for the flexibility  analysis via the B-factor prediction. The persistent homology analysis is carried out via the accumulation bar length $A_1$. We use the exponential kernel with parameter $\kappa=2$ for FRI, MND and $A_1$ calculations. The average correlation coefficients of MND and FRI methods are obtained  by averaging over 30 proteins at each given $\eta$ value. We compare these averaged CC values with the average accumulated length over the same set of 30 proteins. These results are illustrated in Fig. \ref{fig:30ProteinSet}. It can be seen that the average CC values obtained from FRI and MND  behave similarly as $\eta$ increases. They dramatically increase when $\eta$ goes beyond 3\AA, and reach the peak before decrease at large characteristic distances. The best correlation coefficient is achieved at about 9\AA~ and 8\AA~ for  MND and FRI models, respectively.   The average accumulation bar length $A_1$ behaves in a similar manner. However, its peak value is  around 6\AA, which is consistent with our earlier finding with protein 1YZM.  The deviation between optimal characteristic distance in protein the flexibility analysis and the ``optimal filtration parameter $\eta$'' is about 2\AA~ to 3\AA. We believe this deviation is due to several aspects. First, cofactors like metal ions, ligands, and small sugar molecules, play important role in protein stability and flexibility. Without any consideration of cofactors, our models may offer optimal  B-factor  prediction at wrong characteristic distances. 
Additionally, due to the limitation of our computational power, only relatively small sized proteins are considered in our test set. This may results in a lack of representation for relatively large proteins. Finally, the present persistent homology model is based only on $\beta_1$ numbers and may be improved by including $\beta_2$ as well. In spite of these issues, our persistent homology analysis successfully captures the basic correlation coefficient behavior. It provides an explanation for both the dramatic increase of correlation coefficients in small $\eta$ values and the slow decrease in large $\eta$ values. The predicted optimal characteristic distance  value is also in a reasonable  range.

\begin{table}
\caption{The 30   proteins used in our B-factor prediction and persistent homology analysis of optimal characteristic distance. }
{\small
\begin{center}
\begin{tabular}{|c|c|c|c|c|c|c|c|c|c|}
 \hline
  PDB ID & PDB ID & PDB ID & PDB ID & PDB ID & PDB ID  & PDB ID & PDB ID  & PDB ID & PDB ID \\\hline
    1BX7 & 1DF4   & 1ETL   &1FF4    &1GK7     &1GVD     &1HJE   &1KYC      &1NKD  &1NOT   \\
    1O06  &1OB4   &1OB7    &1P9I   & 1PEF    &1Q9B     &1UOY  & 1VRZ       &1XY1 &1XY2 \\
    1YZM  &2BF9   &2JKU    &2OL9   &2OLX   &3E7R  &3MD4 &3PZZ   &3Q2X   & 4AXY\\\hline
\end{tabular}
\label{tab:30protein}
\end{center}
}
\end{table}

\subsection{Persistent homology analysis of protein folding}

Protein folding produces characteristic and functional three-dimensional structures from unfolded polypeptides or disordered coils.  Although Anfinsen's dogma \cite{Anfinsen:1973} has been challenged due to  the existence of prions and amyloids, most functional proteins are well folded. The folding funnel hypothesis associates each folded protein structure with a global minimum of the Gibbs free energy. Unfolded conformations have higher energies and are thermodynamically unstable, albeit they can be kinetically favored.

\begin{figure}
\begin{center}
\begin{tabular}{c}
\includegraphics[width=0.9\textwidth]{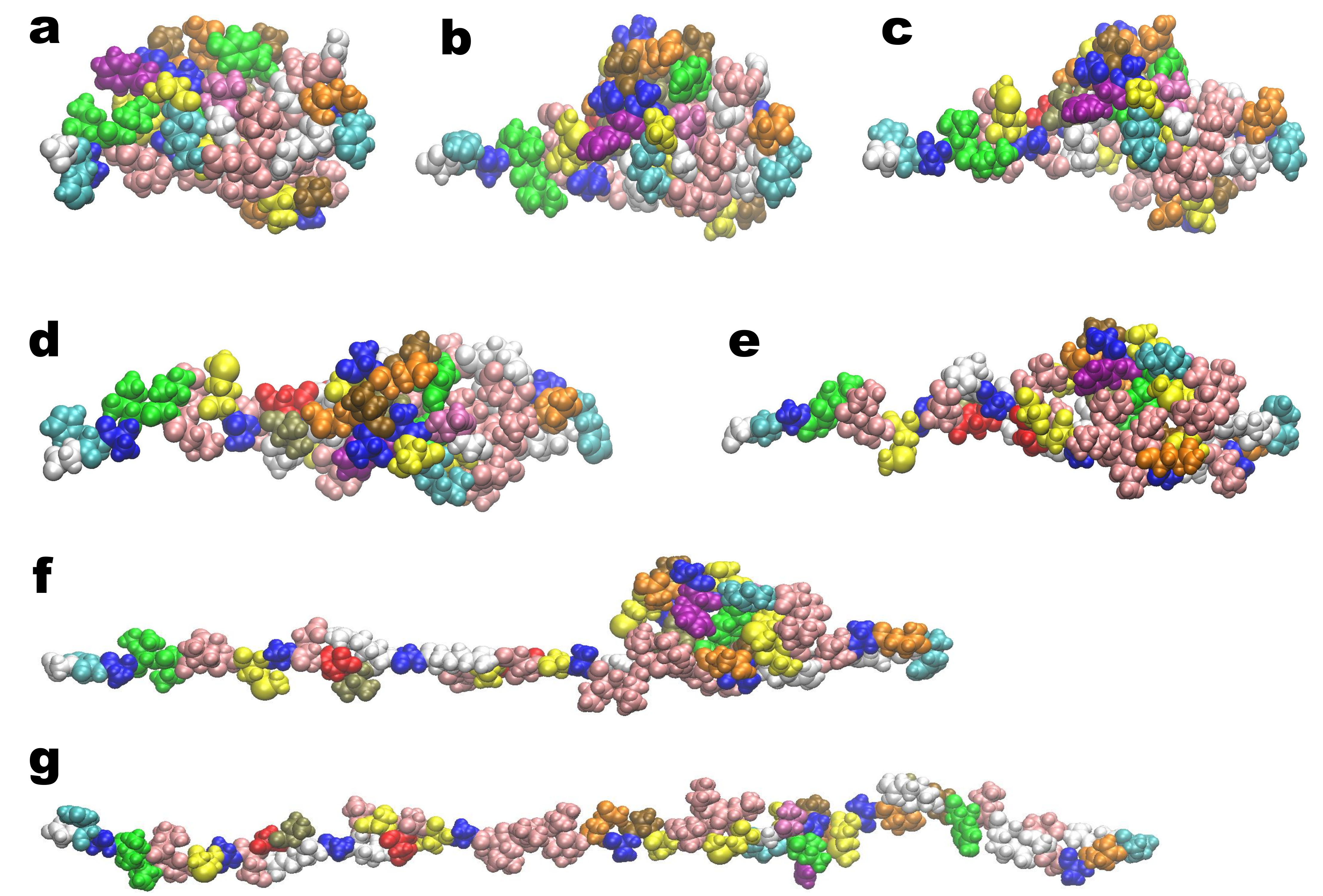}
\end{tabular}
\end{center}
\caption{ The unfolding configurations of protein 1I2T obtained from  the steered molecular dynamics with the constant velocity pulling algorithm. Charts {\bf a}, {\bf b}, {\bf c}, {\bf d}, {\bf e}, {\bf f} and {\bf g} are the corresponding configuration frames 1, 3, 5, 7, 10, 20, and 30. Amino acid residues are labeled with different colors. From {\bf a} to {\bf g}, protein topological connectivity decreases, while protein total energy increases.}
\label{fig:FoldingTotal}
\end{figure}

The protein folding process poses astonishing challenges to theoretical modeling and computer simulations. Despite the progress at the protein structure determination, the mechanism that how the polypeptide coils into its native conformation remains a puzzling issue, mainly due to the complexity and the stochastic dynamics involved in the process. Currently, experimental tools, such as atomic force microscopy, optical tweezers, and bio-membrane force probe, have been devised  to give information about unfolding force distribution, stable intermediates and transitional nonnative states. However, these approaches have a limited utility for unstable intermediate structures.  Using the steered molecular dynamics (SMD), more details such as some possible folding or unfolding pathway can be obtained.

Protein folding and unfolding involve massive changes in its local topology and global topology. Protein topological evolution can be tracked by the trajectory of protein topological invariants.   Typically, the folding of an amino acid polymer chain leads to dramatic increase in both 2-complexes and 3-complexes at appropriate filtration parameters. Therefore, persistent homology should provide an efficient tool for both qualitative characterization  and quantitative analysis of protein folding or unfolding.

In this section, we simulate the unfolding process with the constant velocity pulling algorithm of  SMD. Intermediate configurations are extracted from the trajectory. Then, we employ the persistent homology   to reveal the   topological features of intermediate configurations. Furthermore, we construct a quantitative model based on the accumulation bar length $A_1$ to predict the  energy and stability of protein configurations, which establishes a solid topology-function relationship of proteins.

\subsubsection{Steered molecular dynamics}\label{sec:SMD}

Usually, the SMD  is carried out   through one of three ways: high temperature, constant force pulling, and constant velocity pulling \cite{Paci:2000,Hui:1998,Srivastava:2013}. The study of the mechanical properties of protein FN-$III_{10}$ provides   information of how to carefully design the SMD. It is observed that the original implicit solvent models for SMD tend to miss  friction terms. For implicit solvent models, the design of the water environment is still nontrivial. If a water sphere is used, water deformation  requires additional artificial force. Further, the unfolding process may extend the protein out of the boundary of the initial sphere. Therefore, it is believed that using a large box which can hold the stretched protein is a more reliable way if one can afford the computational  cost \cite{Gao:2002}.

\begin{figure}
\begin{center}
\begin{tabular}{c}
\includegraphics[width=0.7\textwidth]{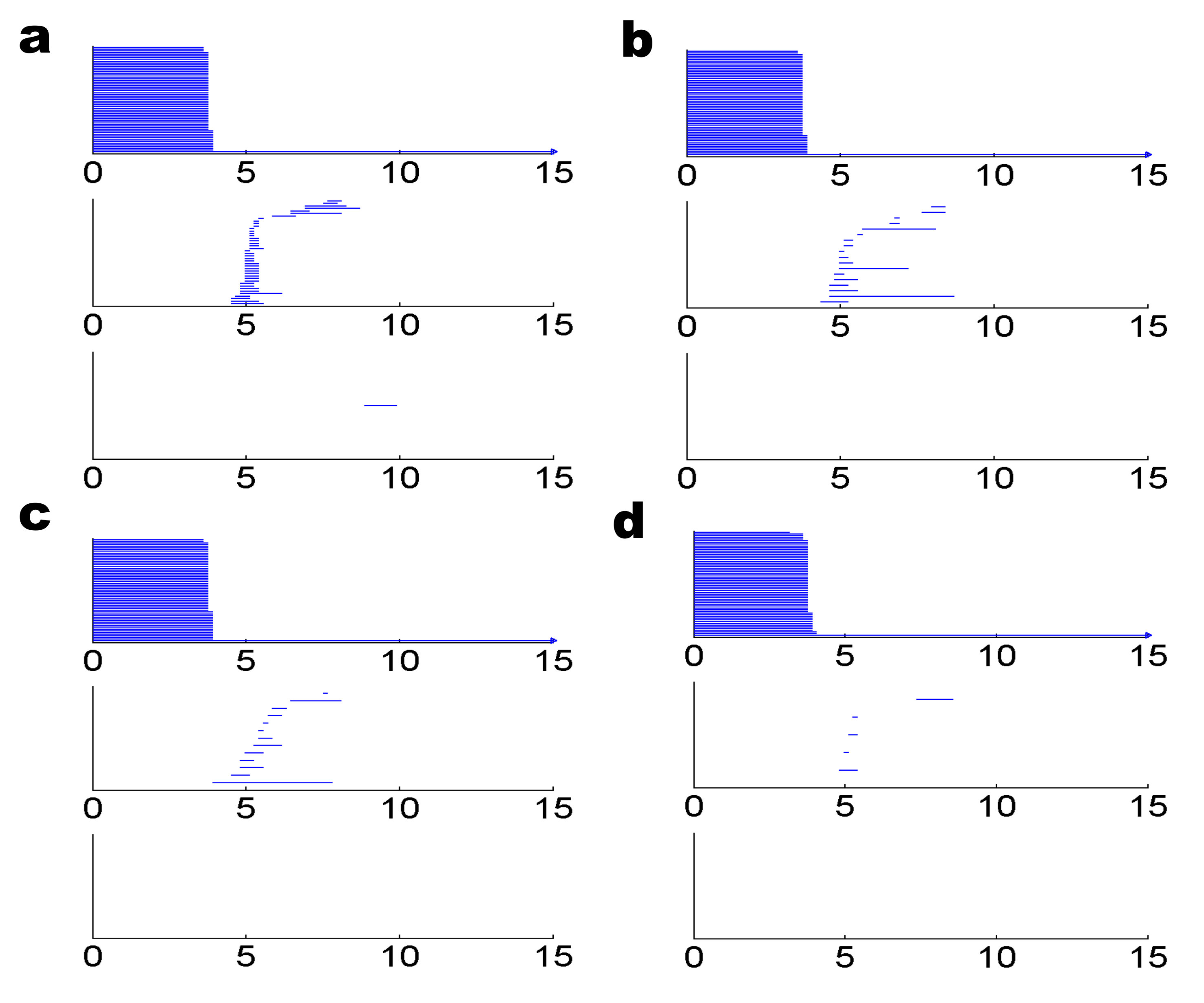}
\end{tabular}
\end{center}
\caption{ Topological fingerprints of four configurations of protein 1I2T generated by using the distance based filtration. Charts {\bf a}, {\bf b}, {\bf c}, and {\bf d} are   for frame 1, 10, 20 and 30, respectively (see Figs. \ref{fig:FoldingTotal}{\bf a}, \ref{fig:FoldingTotal}{\bf e}, \ref{fig:FoldingTotal}{\bf f} and \ref{fig:FoldingTotal}{\bf g} for their geometric shapes.). It can seen that as the protein unfolds, the $\beta_0$ bars are continuously decreasing, which corresponds to the reduction of topological connectivity among protein atoms. }
\label{fig:1i2tBarcodes}
\end{figure}

In the present work, the molecular dynamical simulation tool \href{http://www.ks.uiuc.edu/Training/Tutorials/namd/namd-tutorial-html/}{NAMD} is employed to generate the partially folded and unfolded protein conformations.  Two processes are involved, the relaxation of the structure and unfolding with constant velocity pulling. At first, the protein structure is downloaded from the PDB. Then, it is solvated with a water box which has an extra 5{\AA}~ layer, comparing with the minimal one that hold the protein structure. The standard minimization and equilibration process is employed. Basically, a total of 15000 time steps of equilibration iterations is carried out with the periodic boundary condition after 10000 time steps of initial energy minimization. The length of each time step  is 2fs in our simulation.

The setting of the constant velocity pulling is more complicated. First, pulling points where the force applied should be chosen. Usually, the first C$_\alpha$ atom is fixed and a constant pulling velocity is employed on the last C$_\alpha$ atom along the direction connecting these two points. The identification of the pulling points can be done by assigning special values to B-factor term and occupancy term in PDB data. Additionally, the pulling parameters should be carefully assigned. As  proteins  used in our simulation are relatively  small with about 80 residues. The spring constant is set as 7 kcal/mol{\AA}$^2$, with 1 kcal/mol{\AA}$^2/$  equaling 69.74 pN {\AA}. The constant velocity is 0.005{\AA}~ per time step. As many as  30000 simulation steps for protein 1I2T and 40000 for 2GI9 are employed for their pulling processes.  We extract 31 conformations from the simulation results at an equal time interval.  {The    total energies (kcal/mol) are computed for all configurations. For each pair of configurations, their relative values of total energies determine their relative stability.}  A few representative conformations of unfolding 1I2T are depicted in Fig. \ref{fig:FoldingTotal}. Obviously,  topological connectivities, i.e., 2-complexes and 3-complexes,  reduce dramatically from conformation Fig. \ref{fig:FoldingTotal}{\bf a} to conformation Fig. \ref{fig:FoldingTotal}{\bf g}.

\subsubsection{Persistent homology for protein folding analysis}\label{sec:SMD1}

\begin{figure}
\begin{center}
\begin{tabular}{cc}
\includegraphics[width=0.35\textwidth]{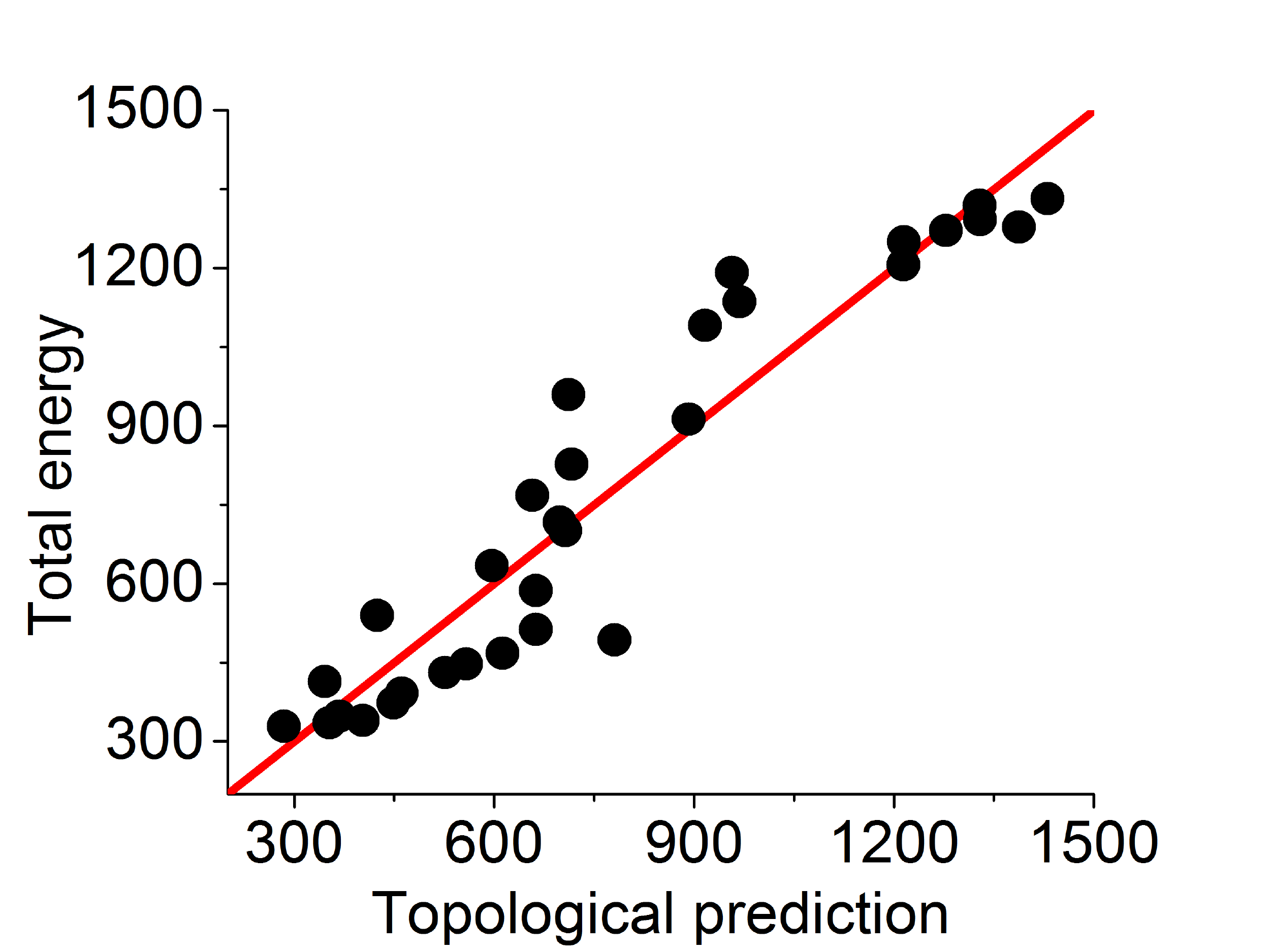}&
\includegraphics[width=0.35\textwidth]{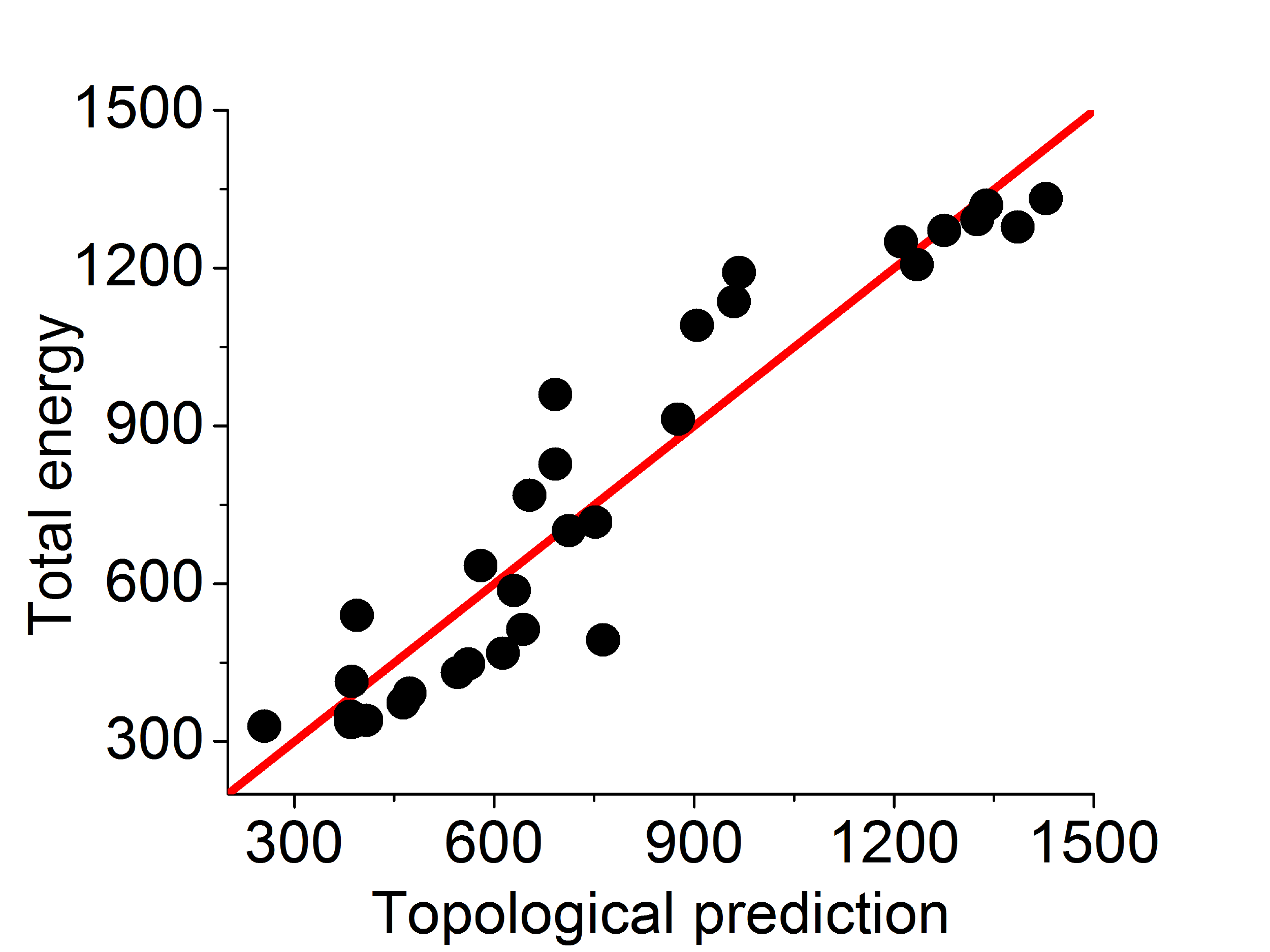}
\end{tabular}
\end{center}
\caption{Comparison between the total energies and the persistent homology prediction  for  31 configurations of protein 1I2T. The unfolding configurations are generated by using the SMD.  The negative accumulation bar length of $\beta_1$  ($A^-_1$)  is used in the persistent homology prediction with both distance based filtration (left chart) and correlation matrix based filtration (right chart). Their correlation coefficients are  0.947 and 0.944, respectively. Clearly, there is a linear correlation between the negative accumulation bar length of $\beta_1$ ($A^-_1$) and total energy.}
\label{fig:1i2t}
\end{figure}

\begin{figure}
\begin{center}
\begin{tabular}{cc}
\includegraphics[width=0.35\textwidth]{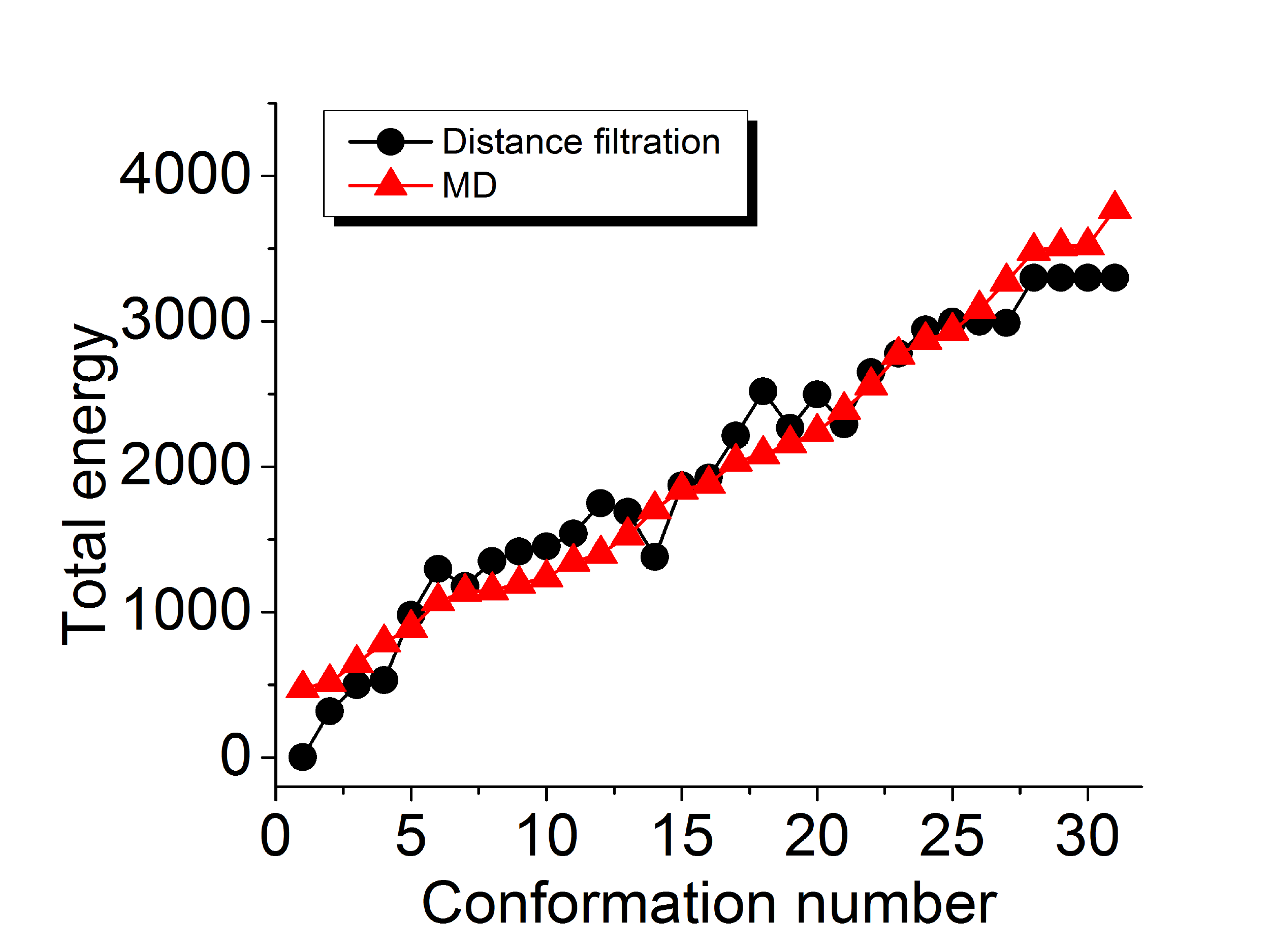}&
\includegraphics[width=0.35\textwidth]{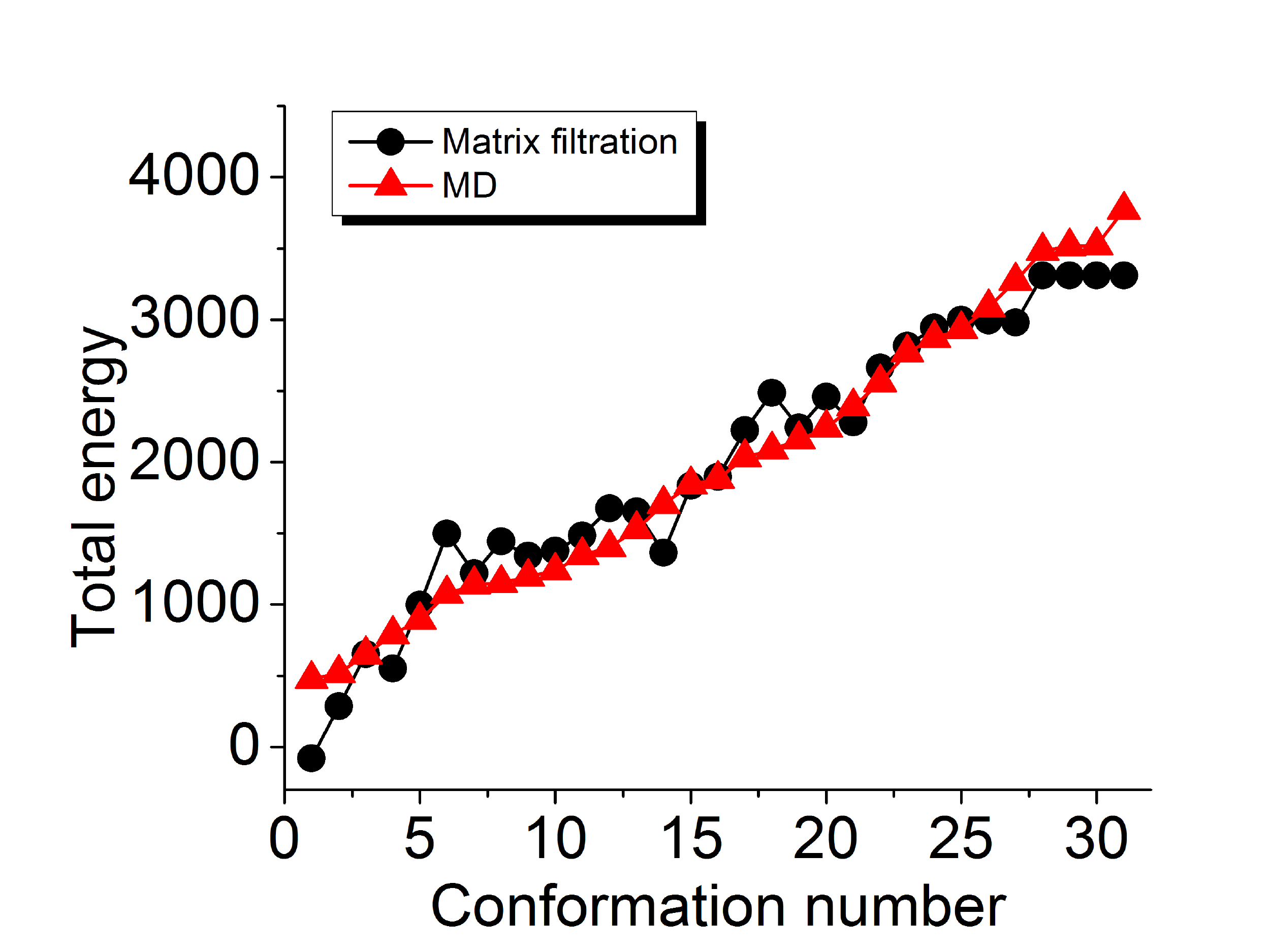}
\end{tabular}
\end{center}
\caption{Comparison between the total energy and the persistent homology prediction  for  31 configurations of protein 2GI9. The negative accumulation bar length of $\beta_1$ ($A^-_1$)  is used in the persistent homology prediction with both distance based filtration (left chart) and correlation matrix based filtration (right chart). Their correlation coefficients are  0.972 and 0.971, respectively. The linear correlation between the negative accumulation bar length and total energy is confirmed.}
\label{fig:2gi9}
\end{figure}

The  steered molecular dynamics (SMD) discussed in  Section \ref{sec:SMD} is used to generate the protein folding process. Basically, by pulling one end of the protein, the coiled structure is stretched   into a straight-line like shape. It is found that during the unfolding process, the hydrogen bonds that support the basic protein configuration are continuously broken. Consequently,  the number of high order complexes that may be formed during the filtration decreases because of protein unfolding. To validate our hypothesis, we  employ the coarse-grained model  in our persistent homology analysis although the all-atom model is used in our SMD calculations. The benefit of using the coarse-grained model is that the number of $\beta_1$ and $\beta_2$ is zero for a straight-line like peptide generated by SMD. We compute topological  invariants via the distance based filtration for all 31 configurations of protein 1I2T. The persistent homology barcodes for frames 1, 10, 20 and 30 are illustrated in Fig. \ref{fig:1i2tBarcodes}. We found that as the protein unfolded, their related $\beta_1$ value decreases. For  four protein configurations in Fig. \ref{fig:1i2tBarcodes}, their $\beta_1$ values are 47, 19, 13 and 5, respectively. Also there is a cavity in configuration 1. A comparison between configurations in Fig. \ref{fig:FoldingTotal} and their corresponding barcodes in Fig. \ref{fig:1i2tBarcodes} shows an obvious correlation between protein folding/unfolding and its topological trait. Therefore, topology and persistent homology are potential  tools for protein folding characterization.

Additionally, as a protein unfolds, its stability decrease. State differently, protein  becomes more and more unstable during its unfolding process, and its   total energy becomes higher during the SMD simulation.  As discussed above, the first Betti number  decreases as protein unfolds. Therefore, there is a strong anti-correlation between protein total energy and its first Betti number  during the protein unfolding process. However, using the least square fitting, we  found that this linear relation is not highly accurate with a correlation coefficient about 0.89 for 31 configurations of 1I2T.  A more robust quantitative  model is to correlate protein total energy with the negative accumulation bar length of $\beta_1$ ($A_1^-=-A_1$).  Indeed,  a striking linear relation between  the total energy and $A^-_1$ can be found. We demonstrate our results in the left chart of Fig. \ref{fig:1i2t}. Using a linear regression algorithm, a correlation coefficient about 0.947 can be obtained for 31    configurations of protein 1I2T. To further validate the relation between the negative accumulation bar length and total energy, the correlation matrix based filtration process is employed. We choose the exponential kernel with optimized parameter $\kappa=2$ and $\eta=7$\AA~. The results are illustrated in the right chart of Fig. \ref{fig:1i2t}. A linear correlation is found with the CC value of 0.944.

To further validate our persistent homology based quantitative model for the stability analysis of protein folding, we consider protein 2GI9. The same procedure described above is utilized to create 31  configurations. We use both distance based filtration and correlation matrix based filtration to compute $A_1^-$ for   all the extracted intermediate structures. Our results are depicted in Fig. \ref{fig:2gi9}. Again the linear correlation between the energy prediction using  negative accumulation bar length of the first Betti number and total energy is confirmed. The CC values are as high as 0.972 and 0.971, for distance based filtration and correlation matrix based filtration, respectively.

\section{Concluding remarks }

Persistent homology is a relatively new tool for topological characterization of signal,  image and data, which are often corrupted by noise. Compared with the commonly used  techniques in computational topology and computational homology, persistent homology incorporates a unique filtration process, through which, a sequence of nested simplicial complexes are generated by continuously enlarging a filtration parameter. In this manner, a multi-scaled representation of the underlying topological space can be constructed and further utilized to reveal the intrinsic topological properties against noise. Like other topological approaches, persistent homology  is able to dramatically reduce the complexity of the underlying problem and offer topological insight for geometric structures and/or intrinsic features that last over multiple length scales.  Additionally,  thanks to the introduction of the filtration process, persistent homology is capable of reintroducing the geometric information associated with topological features like isolated components, loops, rings, circles, pockets, holes and cavities. The most successful applications of persistent homology in the literature  have been about qualitative characterizations or classifications in the past. Indeed, there is hardly any successful quantitative model based on topology in the literature, because topological invariants preclude geometric description. It is interesting and desirable to develop quantitative models based on persistent  homology analysis. This work introduces persistent homology to protein structure characterization, flexibility prediction and folding analysis.  Our goal is to develop a mathematical tool that is able to dramatically simplify geometric complexity while incorporate sufficient geometric information in topological invariants for both qualitative characterization, identification  and classification and quantitative  understanding of the  topology-function relationship of proteins.  

To establish notation and facilitate our quantitative modeling, we briefly summarize the background of persistent homology.  Simplicial complex, homology, persistent homology,  filtration and  their computational algorithms are briefly reviewed. To be pedagogic, we illustrate persistent homology  concepts with a few carefully designed toy models, including a tetrahedron, a cube, an  icosahedron and a C$_{70}$ molecule. The relation between topological invariants, namely  Betti numbers and Euler characteristic,  is discussed using some simple models.   For sophisticated biomolecular systems,   we utilize both the all-atom model and coarse-grained model to deliver a multi-representational persistent homology analysis. Molecular topological fingerprints (MTFs) based  on the persistence of molecular topological invariants are extracted. 

Proteins are the most important molecules for living organism. The understanding of the molecular mechanism of protein structure, function, dynamics and transport is one of the most challenging tasks of modern science.  In order to understand  protein topological properties, we study the topological fingerprints of two most important protein structural components, namely alpha helices and beta sheets. To understand the geometric origin of each topological invariant and its persistence, we develop a method of slicing to systematically divide a biomolecule into small pieces and study their topological traits.   The topological fingerprints of alpha helices and beta sheets are further utilized to decipher the topological property and fingerprint of a beta barrel structure. Uncovering the connection between topological invariants and geometric features deepens our understanding of  biomolecular topology-function relationship.

Traditionally, long-lived   persistent bars in the barcode  have been celebrated as intrinsic topological features that  persistent homology was invented for, while other short-lived bars are generally regarded as useless noise. However, from our analysis, it is emphasized that all features generated from persistent homology analysis are equally important because they represent the topological fingerprint of a given protein structure. Therefore, the birth and death of each  $k$-complex uniquely represent  either a local or a global geometric feature. It is all of these topological features that make the MTF unique for each biomolecule. Just like nuclear magnetic resonance (NMR) signals or x-ray crystallography data, topological fingerprints are a new class of  biometrics  for the identification, characterization and classification of biomolecules. 

The rigidity of a protein is essentially determined by its non-covalent interactions and manifests in the compactness of its three-dimensional (3D) structure. Such a compactness can be measured by the topological connectivity of the protein polymer chain elements, i.e., amino acid residues. Consequently, topological invariants, such as  the first Betti number, give rise to a natural description of protein rigidity and flexibility. We therefore are able to reveal the topology-function relationship. Elastic network models, such as Gaussian network model, have been widely used for protein flexibility analysis. However, the performance of these methods depends on the cutoff distance which determines whether a given pair of atoms can be connected by an elastic spring in the model. In practical applications, the selection of a cutoff distance is empirical. We propose a new cutoff distance type of filtration matrix. The resulting topological diagrams shed light on the optimal cutoff distance used in the protein B-factor prediction with the Gaussian network model. 

The aforementioned persistent homology analyses are descriptive and qualitative. One of major goals of the present work is to exploit the geometric information  embedded in topological invariants for quantitative modeling.  To this end, we propose a correlation matrix based filtration to incorporate geometric information in topological invariants. We define accumulated bar length by   summing over  all the bars of the first Betti number.  We assume that the accumulated bar length correlates linearly to protein rigidity due to hydrogen bond strength, hydrophobic effects, electrostatic and van der Waals interactions.  In particular, we  investigate the dependence of the accumulated bar length on the characteristic distance used in our filtration process. It is found that the location of the maximum  accumulated bar length gives an accurate prediction of the optimal characteristic distance for the flexibility and rigidity index (FRI) analysis \cite{KLXia:2013d} of protein temperature factors.

To further exploit  persistent homology for  quantitative modeling, we consider protein folding which is an essential process for proteins to assume well-define structure and function. Our basic observation is that well-folded proteins, especially well-folded globular proteins, have abundant non-covalent bonds due to hydrogen bonds and van der Waals interactions, which translates into higher numbers of topological invariants, particularly, a large measurement of the first  Betti number. Additionally,  the funnel theory of protein folding states that a well-folded protein, i.e., the native structure, has the lowest free energy. In contrast, unfolded protein structures have less  numbers of topological invariants   and higher free energies. Based on this analysis, we propose a persistent homology based model to characterize protein topological evolution and  predict protein folding stability. We correlate the negative accumulated bar length of the first Betti number to the protein total energy for a series of protein configurations generated by the steered molecular dynamics. As such the evolution of topological invariants in the protein folding/unfolding process is tracked. Our persistent homology based model is found to provide an excellent quantitative prediction of protein total energy during the protein folding/unfolding process.

\section*{Acknowledgments}

 This work was supported in part by NSF grants   IIS-1302285 and DMS-1160352,   NIH grant R01GM-090208 and MSU Center for Mathematical Molecular Biosciences Initiative. The authors acknowledge the Mathematical Biosciences Institute for hosting valuable workshops.

\vspace{0.6cm}


\begin{thebibliography}{10}

\bibitem{Anfinsen:1973}
C.~B. Anfinsen.
\newblock Einfluss der configuration auf die wirkung den.
\newblock {\em Science}, 181:223 -- 230, 1973.

\bibitem{McCammon:1977}
J.~A. McCammon, B.~R. Gelin, and M.~Karplus.
\newblock Dynamics of of folded proteins.
\newblock {\em Nature}, 267:585--590, 1977.

\bibitem{Go:1983}
N.~Go, T.~Noguti, and T.~Nishikawa.
\newblock Dynamics of a small globular protein in terms of low-frequency
  vibrational modes.
\newblock {\em Proc. Natl. Acad. Sci.}, 80:3696 -- 3700, 1983.

\bibitem{Tasumi:1982}
M.~Tasumi, H.~Takenchi, S.~Ataka, A.~M. Dwidedi, and S.~Krimm.
\newblock Normal vibrations of proteins: Glucagon.
\newblock {\em Biopolymers}, 21:711 -- 714, 1982.

\bibitem{Brooks:1983}
B.~R. Brooks, R.~E. Bruccoleri, B.~D. Olafson, D.J. States, S.~Swaminathan, and
  M.~Karplus.
\newblock Charmm: A program for macromolecular energy, minimization, and
  dynamics calculations.
\newblock {\em J. Comput. Chem.}, 4:187--217, 1983.

\bibitem{Levitt:1985}
M.~Levitt, C.~Sander, and P.~S. Stern.
\newblock Protein normal-mode dynamics: Trypsin inhibitor, crambin,
  ribonuclease and lysozyme.
\newblock {\em J. Mol. Biol.}, 181(3):423 -- 447, 1985.

\bibitem{Tirion:1996}
M.M. Tirion.
\newblock Large amplitude elastic motions in proteins from a single-parameter,
  atomic analysis.
\newblock {\em Phys. Rev. Lett.}, 77:1905 -- 1908, 1996.

\bibitem{Flory:1976}
P.J. Flory.
\newblock Statistical thermodynamics of random networks.
\newblock {\em Proc. Roy. Soc. Lond. A,}, 351:351 -- 378, 1976.

\bibitem{Bahar:1997}
I.~Bahar, A.R. Atilgan, and B.~Erman.
\newblock Direct evaluation of thermal fluctuations in proteins using a
  single-parameter harmonic potential.
\newblock {\em Folding and Design}, 2:173 -- 181, 1997.

\bibitem{Bahar:1998}
I.~Bahar, A.R. Atilgan, M.C. Demirel, and Erman B.
\newblock Vibrational dynamics of proteins: Significance of slow and fast modes
  in relation to function and stability.
\newblock {\em Phys. Rev. Lett}, 80:2733 -- 2736, 1998.

\bibitem{Atilgan:2001}
A.R. Atilgan, S.R. Durrell, R.L. Jernigan, M.~C. Demirel, O.~Keskin, and
  I.~Bahar.
\newblock Anisotropy of fluctuation dynamics of proteins with an elastic
  network model.
\newblock {\em Biophys. J.}, 80:505 -- 515, 2001.

\bibitem{Cui:2002}
Q.~Cui.
\newblock Combining implicit solvation models with hybrid quantum
  mechanical/molecular mechanical methods: A critical test with glycine.
\newblock {\em Journal of Chemical Physics}, 117(10):4720, 2002.

\bibitem{YZhang:2009a}
Y.~Zhang, H.~Yu, J.~H. Qin, and B.~C. Lin.
\newblock A microfluidic dna computing processor for gene expression analysis
  and gene drug synthesisn.
\newblock {\em Biomicrofluidics}, 3(044105), 2009.

\bibitem{WFTian:2014}
W.~F. Tian and Shan Zhao.
\newblock A fast {ADI} algorithm for geometric flow equations in biomolecular
  surface generations.
\newblock {\em International Journal for Numerical Methods in Biomedical
  Engineering}, 30:490--516, 2014.

\bibitem{Geng:2011}
W.~Geng and G.~W. Wei.
\newblock Multiscale molecular dynamics using the matched interface and
  boundary method.
\newblock {\em J Comput. Phys.}, 230(2):435--457, 2011.

\bibitem{Wei:2009}
G.~W. Wei.
\newblock Differential geometry based multiscale models.
\newblock {\em Bulletin of Mathematical Biology}, 72:1562 -- 1622, 2010.

\bibitem{Wei:2012}
Guo-Wei Wei, Qiong Zheng, Zhan Chen, and Kelin Xia.
\newblock Variational multiscale models for charge transport.
\newblock {\em SIAM Review}, 54(4):699 -- 754, 2012.

\bibitem{DuanChen:2012a}
Duan Chen, Zhan Chen, and G.~W. Wei.
\newblock Quantum dynamics in continuum for proton transport {II: Variational}
  solvent-solute interface.
\newblock {\em International Journal for Numerical Methods in Biomedical
  Engineering}, 28:25 -- 51, 2012.

\bibitem{Wei:2013}
Guo-Wei Wei.
\newblock Multiscale, multiphysics and multidomain models {I: Basic} theory.
\newblock {\em Journal of Theoretical and Computational Chemistry},
  12(8):1341006, 2013.

\bibitem{KLXia:2013d}
K.~L. Xia, K.~Opron, and G.~W. Wei.
\newblock Multiscale multiphysics and multidomain models --- { Flexibility} and
  rigidity.
\newblock {\em Journal of Chemical Physics}, 139:194109, 2013.

\bibitem{XFeng:2012a}
Xin Feng, Kelin Xia, Yiying Tong, and Guo-Wei Wei.
\newblock Geometric modeling of subcellular structures, organelles and large
  multiprotein complexes.
\newblock {\em International Journal for Numerical Methods in Biomedical
  Engineering}, 28:1198--1223, 2012.

\bibitem{QZheng:2012}
Q.~Zheng, S.~Y. Yang, and G.~W. Wei.
\newblock { Molecular surface generation using PDE transform}.
\newblock {\em International Journal for Numerical Methods in Biomedical
  Engineering}, 28:291--316, 2012.

\bibitem{Bates:2008}
P.~W. Bates, G.~W. Wei, and Shan Zhao.
\newblock Minimal molecular surfaces and their applications.
\newblock {\em Journal of Computational Chemistry}, 29(3):380--91, 2008.

\bibitem{Bates:2009}
P.~W. Bates, Z.~Chen, Y.~H. Sun, G.~W. Wei, and S.~Zhao.
\newblock Geometric and potential driving formation and evolution of
  biomolecular surfaces.
\newblock {\em J. Math. Biol.}, 59:193--231, 2009.

\bibitem{ZhanChen:2010a}
Z.~Chen, N.~A. Baker, and G.~W. Wei.
\newblock Differential geometry based solvation models {I}: Eulerian
  formulation.
\newblock {\em J. Comput. Phys.}, 229:8231--8258, 2010.

\bibitem{ZhanChen:2010b}
Z.~Chen, N.~A. Baker, and G.~W. Wei.
\newblock Differential geometry based solvation models {II}: Lagrangian
  formulation.
\newblock {\em J. Math. Biol.}, 63:1139-- 1200, 2011.

\bibitem{ZhanChen:2012}
Z.~Chen, Shan Zhao, J.~Chun, D.~G. Thomas, N.~A. Baker, P.~B. Bates, and G.~W.
  Wei.
\newblock Variational approach for nonpolar solvation analysis.
\newblock {\em Journal of Chemical Physics}, 137(084101), 2012.

\bibitem{XFeng:2013b}
X.~Feng, K.~L. Xia, Y.~Y. Tong, and G.~W. Wei.
\newblock Multiscale geometric modeling of macromolecules {II:} lagrangian
  representation.
\newblock {\em Journal of Computational Chemistry}, 34:2100--2120, 2013.

\bibitem{KLXia:2014a}
K.~L. Xia, X.~Feng, Y.~Y. Tong, and G.~W. Wei.
\newblock Multiscale geometric modeling of macromolecules.
\newblock {\em Journal of Computational Physics}, 275:912--936, 2014.

\bibitem{Boileau:2013}
E.~Boileau, R.~L.~T. Bevan, I.~Sazonov, M.~I. Rees, and P.~Nithiarasu.
\newblock Flow-induced atp release in patient-specific arterial geometries - a
  comparative study of computational models.
\newblock {\em International Journal for Numerical Methods in Engineering},
  29:1038--1056, 2013.

\bibitem{Sazonov:2012}
Igor Sazonov and Perumal Nithiarasu.
\newblock Semi-automatic surface and volume mesh generation for
  subject-specific biomedical geometries.
\newblock {\em International Journal for Numerical Methods in Biomedical
  Engineering}, 28:133--157, 2012.

\bibitem{Sazonov:2012b}
I.~Sazonov, S.~Y. Yeo, R.~L.~T. Bevan, X.~H. Xie, R.~van Loon, and
  P.~Nithiarasu.
\newblock Modelling pipeline for subject-specific arterial blood flow -- a
  review.
\newblock {\em International Journal for Numerical Methods in Engineering},
  28:1868--1910, 2012.

\bibitem{Sohn:2012}
J.~S. Sohn, S.~W. Li, X.~F. Li, and J.~S. Lowengrub.
\newblock Axisymmetric multicomponent vesicles: A comparison of hydrodynamic
  and geometric models.
\newblock {\em International Journal for Numerical Methods in Engineering},
  28:346--368, 2012.

\bibitem{Ramalho:2012}
S~Ramalho, A~Moura, A.~M. Gambaruto, and A.~Sequeira.
\newblock Sensitivity to outflow boundary conditions and level of geometry
  description for a cerebral aneurysm.
\newblock {\em International Journal for Numerical Methods in Engineering},
  28:697--713, 2012.

\bibitem{Manzoni:2012}
Andrea Manzoni, Alfio Quarteroni, and Gianluigi Rozza.
\newblock Model reduction techniques for fast blood flow simulation in
  parametrized geometries.
\newblock {\em International Journal for Numerical Methods in Engineering},
  28:604--625, 2012.

\bibitem{Mikhal:2013}
J~Mikhal, D.~J. Kroon, C.~H. Slump, and B.~J. Geurts.
\newblock Flow prediction in cerebral aneurysms based on geometry
  reconstruction from 3d rotational angiography.
\newblock {\em International Journal for Numerical Methods in Engineering},
  29:777-- 805, 2013.

\bibitem{Edelsbrunner:2002}
H.~Edelsbrunner, D.~Letscher, and A.~Zomorodian.
\newblock Topological persistence and simplification.
\newblock {\em Discrete Comput. Geom.}, 28:511--533, 2002.

\bibitem{Zomorodian:2005}
A.~Zomorodian and G.~Carlsson.
\newblock Computing persistent homology.
\newblock {\em Discrete Comput. Geom.}, 33:249--274, 2005.

\bibitem{Zomorodian:2008}
Afra Zomorodian and Gunnar Carlsson.
\newblock Localized homology.
\newblock {\em Computational Geometry - Theory and Applications},
  41(3):126--148, 2008.

\bibitem{Frosini:1999}
Patrizio Frosini and Claudia Landi.
\newblock Size theory as a topological tool for computer vision.
\newblock {\em Pattern Recognition and Image Analysis}, 9(4):596--603, 1999.

\bibitem{Robins:1999}
Vanessa Robins.
\newblock Towards computing homology from finite approximations.
\newblock In {\em Topology Proceedings}, volume~24, pages 503--532, 1999.

\bibitem{Bubenik:2007}
Peter Bubenik and Peter~T. Kim.
\newblock A statistical approach to persistent homology.
\newblock {\em Homology, Homotopy and Applications}, 19:337--362, 2007.

\bibitem{edelsbrunner:2010}
Herbert Edelsbrunner and John Harer.
\newblock {\em Computational topology: an introduction}.
\newblock American Mathematical Soc., 2010.

\bibitem{Dey:2008}
T.~K. Dey, K.~Y. Li, J.~Sun, and C.~S. David.
\newblock Computing geometry aware handle and tunnel loops in 3d models.
\newblock {\em ACM Trans. Graph.}, 27, 2008.

\bibitem{Dey:2013}
Tamal~K. Dey and Yusu Wang.
\newblock Reeb graphs: Approximation and persistence.
\newblock {\em Discrete and Computational Geometry}, 49(1):46--73, 2013.

\bibitem{Mischaikow:2013}
K.~Mischaikow and V.~Nanda.
\newblock Morse theory for filtrations and efficient computation of persistent
  homology.
\newblock {\em Discrete and Computational Geometry}, 50(2):330--353, 2013.

\bibitem{Ghrist:2008}
R.~Ghrist.
\newblock Barcodes: {The} persistent topology of data.
\newblock {\em Bull. Amer. Math. Soc.}, 45:61--75, 2008.

\bibitem{Carlsson:2008}
G.~Carlsson, T.~Ishkhanov, V.~Silva, and A.~Zomorodian.
\newblock On the local behavior of spaces of natural images.
\newblock {\em International Journal of Computer Vision}, 76(1):1--12, 2008.

\bibitem{Pachauri:2011}
D.~Pachauri, C.~Hinrichs, M.K. Chung, S.C. Johnson, and V.~Singh.
\newblock Topology-based kernels with application to inference problems in
  alzheimer's disease.
\newblock {\em Medical Imaging, IEEE Transactions on}, 30(10):1760--1770, Oct
  2011.

\bibitem{Singh:2008}
G.~Singh, F.~Memoli, T.~Ishkhanov, G.~Sapiro, G.~Carlsson, and D.~L. Ringach.
\newblock Topological analysis of population activity in visual cortex.
\newblock {\em Journal of Vision}, 8(8), 2008.

\bibitem{Bendich:2010}
Paul Bendich, Herbert Edelsbrunner, and Michael Kerber.
\newblock Computing robustness and persistence for images.
\newblock {\em IEEE Transactions on Visualization and Computer Graphics},
  16:1251--1260, 2010.

\bibitem{Frosini:2013}
Patrizio Frosini and Claudia Landi.
\newblock Persistent betti numbers for a noise tolerant shape-based approach to
  image retrieval.
\newblock {\em Pattern Recognition Letters}, 34:863--872, 2013.

\bibitem{Mischaikow:1999}
K.~Mischaikow, M~Mrozek, J.~Reiss, and A.~Szymczak.
\newblock Construction of symbolic dynamics from experimental time series.
\newblock {\em Physical Review Letters}, 82:1144--1147, 1999.

\bibitem{Kaczynski:2004}
T.~Kaczynski, K.~Mischaikow, and M.~Mrozek.
\newblock {\em Computational homology}.
\newblock Springer-Verlag, 2004.

\bibitem{Silva:2005}
V.~D. Silva and R~Ghrist.
\newblock Blind swarms for coverage in 2-d.
\newblock In {\em In Proceedings of Robotics: Science and Systems}, page~01,
  2005.

\bibitem{LeeH:2012}
H~Lee, H.~Kang, M.~K. Chung, B.~Kim, and D.~S. Lee.
\newblock Persistent brain network homology from the perspective of dendrogram.
\newblock {\em Medical Imaging, IEEE Transactions on}, 31(12):2267--2277, Dec
  2012.

\bibitem{Horak:2009}
D.~Horak, S~Maletic, and M.~Rajkovic.
\newblock Persistent homology of complex networks.
\newblock {\em Journal of Statistical Mechanics: Theory and Experiment},
  2009(03):P03034, 2009.

\bibitem{Carlsson:2009}
G.~Carlsson.
\newblock Topology and data.
\newblock {\em Am. Math. Soc}, 46(2):255--308, 2009.

\bibitem{Niyogi:2011}
P.~Niyogi, S.~Smale, and S.~Weinberger.
\newblock A topological view of unsupervised learning from noisy data.
\newblock {\em SIAM Journal on Computing}, 40:646--663, 2011.

\bibitem{BeiWang:2011}
Bei Wang, Brian Summa, Valerio Pascucci, and M.~Vejdemo-Johansson.
\newblock Branching and circular features in high dimensional data.
\newblock {\em IEEE Transactions on Visualization and Computer Graphics},
  17:1902--1911, 2011.

\bibitem{Rieck:2012}
Bastian Rieck, Hubert Mara, and Heike Leitte.
\newblock Multivariate data analysis using persistence-based filtering and
  topological signatures.
\newblock {\em IEEE Transactions on Visualization and Computer Graphics},
  18:2382--2391, 2012.

\bibitem{XuLiu:2012}
Xu~Liu, Zheng Xie, and Dongyun Yi.
\newblock A fast algorithm for constructing topological structure in large
  data.
\newblock {\em Homology, Homotopy and Applications}, 14:221--238, 2012.

\bibitem{DiFabio:2011}
Barbara Di~Fabio and Claudia Landi.
\newblock A mayer-vietoris formula for persistent homology with an application
  to shape recognition in the presence of occlusions.
\newblock {\em Foundations of Computational Mathematics}, 11:499--527, 2011.

\bibitem{Kasson:2007}
P.~M. Kasson, A.~Zomorodian, S.~Park, N.~Singhal, L.~J. Guibas, and V.~S.
  Pande.
\newblock Persistent voids a new structural metric for membrane fusion.
\newblock {\em Bioinformatics}, 23:1753--1759, 2007.

\bibitem{Gameiro:2013}
M.~Gameiro, Y.~Hiraoka, S.~Izumi, M.~Kramar, K.~Mischaikow, and V.~Nanda.
\newblock Topological measurement of protein compressibility via persistence
  diagrams.
\newblock {\em preprint}, 2013.

\bibitem{Dabaghian:2012}
Y.~Dabaghian, F.~Memoli, L.~Frank, and G.~Carlsson.
\newblock A topological paradigm for hippocampal spatial map formation using
  persistent homology.
\newblock {\em PLoS Comput Biol}, 8(8):e1002581, 08 2012.

\bibitem{YaoY:2009}
Y.~Yao, J.~Sun, X.~H. Huang, G.~R. Bowman, G.~Singh, M.~Lesnick, L.~J. Guibas,
  V.~S. Pande, and G.~Carlsson.
\newblock Topological methods for exploring low-density states in biomolecular
  folding pathways.
\newblock {\em The Journal of Chemical Physics}, 130:144115, 2009.

\bibitem{ChangHW:2013}
H.~W. Chang, S.~Bacallado, V.~S. Pande, and G.~E. Carlsson.
\newblock Persistent topology and metastable state in conformational dynamics.
\newblock {\em PLos ONE}, 8(4):e58699, 2013.

\bibitem{Biasotti:2008}
S.~Biasotti, L.~De~Floriani, B.~Falcidieno, P.~Frosini, D.~Giorgi, C.~Landi,
  L.~Papaleo, and M.~Spagnuolo.
\newblock Describing shapes by geometrical-topological properties of real
  functions.
\newblock {\em ACM Computing Surveys}, 40(4):12, 2008.

\bibitem{Edelsbrunner:1994}
H.~Edelsbrunner and E.~P. Mucke.
\newblock Three-dimensional alpha shapes.
\newblock {\em Physical Review Letters}, 13:43--72, 1994.

\bibitem{KLXia:2013f}
K.~L. Xia and G.~W. Wei.
\newblock A stochastic model for protein flexibility analysis.
\newblock {\em Physical Review E}, 88:062709, 2013.

\bibitem{javaPlex}
Andrew Tausz, Mikael Vejdemo-Johansson, and Henry Adams.
\newblock Javaplex: A research software package for persistent (co)homology.
\newblock Software available at \url{http://code.google.com/p/javaplex}, 2011.

\bibitem{YangLei:2009}
L.~Yang, G.~Song, and R.~L. Jernigan.
\newblock Protein elastic network models and the ranges of cooperativity.
\newblock {\em Proceedings of the National Academy of Sciences},
  106(30):12347--12352, 2009.

\bibitem{KLXia:2014b}
K.~L. Xia and G.~W. Wei.
\newblock Molecular nonlinear dynamics and protein thermal uncertainty
  quantification.
\newblock {\em Chaos}, 24:013103, 2014.

\bibitem{JMa:2005}
J.~Ma.
\newblock Usefulness and limitations of normal mode analysis in modeling
  dynamics of biomolecular complexes.
\newblock {\em Structure}, 13:373 -- 180, 2005.

\bibitem{LWYang:2008}
L.~W. Yang and C.~P. Chng.
\newblock Coarse-grained models reveal functional dynamics--i. elastic network
  models--theories, comparisons and perspectives.
\newblock {\em Bioinformatics and Biology Insights}, 2:25 -- 45, 2008.

\bibitem{Skjaven:2009}
L.~Skjaerven, S.~M. Hollup, and N.~Reuter.
\newblock Normal mode analysis for proteins.
\newblock {\em Journal of Molecular Structure: Theochem.}, 898:42 -- 48, 2009.

\bibitem{QCui:2010}
Q.~Cui and I.~Bahar.
\newblock {\em Normal mode analysis: theory and applications to biological and
  chemical systems}.
\newblock Chapman and Hall/CRC, 2010.

\bibitem{Paci:2000}
E.~Paci and M.~Karplus.
\newblock Unfolding proteins by external forces and temperature: The importance
  of topology and energetics.
\newblock {\em Proceedings of the National Academy of Sciences}, 97:6521 --
  6526, 2000.

\bibitem{Hui:1998}
L.~Hui, B.~Isralewitz, A.~Krammer, V.~Vogel, and K.~Schulten.
\newblock Unfolding of titin immunoglobulin domains by steered molecular
  dynamics simulation.
\newblock {\em Biophysical Journal}, 75:662--671, 1998.

\bibitem{Srivastava:2013}
A.~Srivastava and R.~Granek.
\newblock Cooperativity in thermal and force-induced protein unfolding:
  integration of crack propagation and network elasticity models.
\newblock {\em Phys. Rev. Lett.}, 110(138101):1--5, 2013.

\bibitem{Gao:2002}
M.~Gao, D.~Craig, V.~Vogel, and K.~Schulten.
\newblock Identifying unfolding intermediates of $fn-iii_{10}$ by steered
  molecular dynamics.
\newblock {\em J. Mol. Biol.}, 323:939--950, 2002.

\end{thebibliography}

\end{document}